\documentstyle[psfig]{mn}

\begin{document}

\journal{To Appear in MNRAS}

\title[X-ray evidence for multiphase gas with solar abundances in the
brightest ellipticals] {X-ray evidence for multiphase hot gas with
solar abundances in the brightest elliptical galaxies}
\author[D.~A. Buote]{David A. Buote$^{1,2}$\\ $^1$ UCO/Lick
Observatory, University of California at Santa Cruz, Santa Cruz, CA
95064, U.S.A. \\ $^2$ {\sl Chandra} Fellow\\ }

\maketitle

\begin{abstract}

We examine whether isothermal models of the hot gas can successfully
describe the {\sl ASCA} and {\sl ROSAT} spectra of NGC 1399, NGC 4472,
NGC 4636, and NGC 5044, which are among the brightest elliptical
galaxies in X-rays. Broad-band spectral fitting of the {\sl ASCA} SIS
and GIS data accumulated within a radius of $\sim 5\arcmin$ for each
galaxy shows that isothermal models (which also include a component
for discrete sources) are unable to fit the SIS data near 1 keV,
although a marginal fit for NGC 4636 is obtained if the relative
abundances of several elements with respect to Fe are allowed to
depart substantially from their solar values. In addition, these
isothermal models typically fail to produce the large equivalent
widths of the K$\alpha$ line blends of Si and S which are measured
independently of the Fe L emission lines.

Two-temperature models provide substantially better broad-band fits to
both the SIS and GIS data of each galaxy with the relative abundances
(except for NGC 4636) fixed at their solar values. A simple multiphase
cooling flow model fits nearly as well as the two-temperature model
for NGC 1399, NGC 4472, and NGC 5044. The multiphase models also
predict more accurately the Si and S equivalent widths and the ratios
of Si XIV/XIII and S XVI/XV than the isothermal models.  From detailed
comparison of broad-band fits to the {\sl ASCA} data of these
ellipticals using the MEKAL and Raymond-Smith plasma codes we
determine that the MEKAL plasma code (as expected) is significantly
more accurate for the important energies $\sim 0.7-1.4$ keV, and the
small residuals in the Fe L region for the best-fitting multiphase
models imply that remaining inaccuracies in the MEKAL code are
insufficient to change qualitatively the results for data of the
present quality.

Using various approaches we find that the temperature gradients
inferred from the {\sl ROSAT} PSPC data of these galaxies, especially
for NGC 1399 and NGC 5044, are inconsistent with the isothermal models
obtained from fitting the {\sl ASCA} data within a single aperture but
are very consistent with the multiphase models.  Therefore, models
which assume isothermal gas within $r\sim 5\arcmin$ are inconsistent
with the {\sl ASCA} and {\sl ROSAT} PSPC data of these elliptical
galaxies. Simple two-temperature models and multiphase cooling flows
provide much better descriptions of these data sets with Fe abundances
of $\sim 1$-2 solar and (except for NGC 4636) relative abundances
fixed at their solar values. We discuss the implications of these
nearly solar abundances.

\end{abstract}

 
\section{Introduction}
\label{intro}

The elemental abundances inferred from X-ray observations of
elliptical galaxies are sensitive diagnostics of models of the
formation and evolution of the hot gas in these systems.  For example,
the relative abundances of the $\alpha$-process elements with respect
to Fe are affected by the relative contributions of Type Ia and Type
II supernovae to the enrichment of the hot gas (e.g. David, Jones, \&
Forman 1991). The standard models that rely on Type Ia supernova to
enrich the hot gas tend to predict Fe abundances well in excess of
solar (e.g. Ciotti et al. 1991). Smaller Fe abundances can be
accommodated by other models (e.g. Fujita et al. 1997; Brighenti \&
Mathews 1998b).

Most studies of ellipticals based on X-ray spectral data from the {\sl
ASCA} satellite \cite{tanaka} find very sub-solar Fe abundances $(\sim
0.3Z_{\sun})$. These low abundances are generally deduced from fitting
isothermal models of the hot gas to the accumulated {\sl ASCA} spectra
within circular apertures of typical radius $r\sim
3\arcmin$-$5\arcmin$. These findings have been christened as a
``Standard Model'' by Loewenstein \& Mushotzky \shortcite{lmushy} who
review the results obtained from the {\sl ASCA} spectra of
ellipticals.

In a recent study of the {\sl ASCA} data of 20 bright ellipticals,
Buote \& Fabian (1998; hereafter BF) concluded that the spectra of the
brightest galaxies in their sample require at least two temperature
components in the hot gas (in addition to any possible emission from
discrete sources). The multiphase models predict Fe abundances for
these galaxies that are generally consistent with solar.  These
results for the brightest ellipticals highlight the sensitivity of the
Fe abundance to the spectral model.

These results of BF also do not fit into the Standard Model and have
been criticized by Loewenstein \& Mushotzky (1998; also Loewenstein
1998) in two key respects. First, Loewenstein \& Mushotzky state that
the {\sl ASCA} data can generally be described by two-component models
consisting of one ``soft'' component of isothermal hot gas and another
``hard'' component due to discrete sources. Second, they claim that
the K$\alpha$ emission line ratios of Si XIV/XIII are very consistent
with these isothermal models but are only marginally consistent with
the multiphase models of BF.

In this paper we re-examine the {\sl ASCA} data of the brightest
ellipticals to address these criticisms and thus determine whether
indeed isothermal models with very sub-solar Fe abundances are
adequate descriptions of the X-ray spectra of elliptical galaxies. We
extend the work of BF in the following respects: (1) we examine in
detail the predictions of the Standard Model using two-component
models where one component is isothermal hot gas and the other is a
high-temperature bremsstrahlung component presumably due to discrete
sources; (2) the equivalent widths and ratios of the Si and S line
blends measured locally for each galaxy are used to constrain the
isothermal and multiphase models obtained from broad-band analysis;
and (3) we examine whether the radial temperature gradients found by
studies of the {\sl ROSAT} data of bright ellipticals are consistent
with the Standard Model.

We focus our analysis on NGC 1399, NGC 4472, NGC 4636, and NGC 5044
which are among the brightest ellipticals in X-rays and have {\sl
ASCA} data with the best signal-to-noise ratios (S/N) for the sample
studied by BF. Moreover, these galaxies were observed during the {\sl
ASCA} Performance-Verification (PV) phase at which time the
performance of the CCDs, in particular the energy resolution, had not
yet degraded from radiation damage (see Dotani et al. 1995). This is
especially important for our local measurements of the Si and S
lines. Data from a very deep exposure of NGC 4636 has also become
available to the public. These new data for NGC 4636 have the highest
S/N in our sample.

The paper is organized as follows. In section \ref{obs} we discuss the
observations and reduction of the {\sl ASCA} data. The broad-band
spectral fitting is presented in detail in section \ref{broad}. In
section \ref{lines} we summarize our measurements of the Si and S line
blends and compare them to the models obtained from the broad-band
fits. We analyze the {\sl ROSAT} data and compare the results to the
models derived from fitting the {\sl ASCA} data in section
\ref{rosat}. In section \ref{disc} we evaluate our results in light of
the Standard Model and discuss the implications of the derived metal
abundances. Finally, we present a detailed summary of our results in
section \ref{conc}.

\section{{\sl ASCA} observations and data reduction}
\label{obs}

\begin{table*}
\caption{Galaxy Properties}
\label{tab.prop}
\begin{tabular}{llclccc}
Name & Type & $z$ & $B_{\rm T}^0$  & $\sigma_0$
     & $N_{\rm H}$ & Group\\ 
     &      &     &       &   (km s$^{-1}$)    &
($10^{21}$cm$^{-2}$)\\  
NGC 1399 & E1P    & 0.00483 & 10.79  & 310 & 0.13 & Fornax\\
NGC 4472 & E2     & 0.00290 & 09.32  & 287 & 0.16 & Virgo\\
NGC 4636 & E$0^+$ & 0.00365 & 10.50  & 191 & 0.17 & Virgo\\
NGC 5044 & E0     & 0.00898 & 11.87  & 234 & 0.50 & WP 23\\
\end{tabular}

\medskip
\raggedright

Morphological types and redshifts are taken from RC3.  Total blue
magnitudes $(B_{\rm T}^0)$ and central velocity dispersions
$(\sigma_0)$ are taken from Faber et al. \shortcite{7s}. Galactic
Hydrogen column densities $(N_{\rm H})$ are from Stark et
al. \shortcite{stark}. 

\end{table*}

\begin{table*}
\caption{{\em ASCA} Observation Properties}
\label{tab.obs}
\begin{tabular}{llrrrrrrrr}
Name  & Sequence \# & \multicolumn{2}{c}{Exposure} &
      \multicolumn{2}{c}{Count Rate} & \multicolumn{2}{c}{Exposure} &
      \multicolumn{2}{c}{Count Rate}\\
      &             &  \multicolumn{2}{c}{($10^{3}$s)}  &
      \multicolumn{2}{c}{($10^{-2}$ ct s$^{-1}$)} &
      \multicolumn{2}{c}{($10^{3}$s)}  & \multicolumn{2}{c}{($10^{-2}$
      ct s$^{-1}$)} \\
      &             & SIS0 & SIS1 & SIS0 & SIS1 & GIS2 & GIS3 & GIS2 & GIS3\\
NGC 1399 & 80038000 & 17.0 & 17.6 & 38.0 & 39.1 & 19.6 & 19.6 & 18.7 & 17.9\\ 
      & 80039000 & 14.8 & 17.0 & 41.9 & 28.8 & 19.8 & 19.9 & 15.4 & 13.6\\ 
NGC 4472 & 60029000 & 14.3 & 12.5 & 43.8 & 34.0 & 22.7 & 22.7 & 15.4 & 16.0\\ 
      & 60030000 & 16.5 & 13.9 & 37.0 & 28.8 & 22.2 & 22.2 & 11.2 & 14.0\\
NGC 4636 & 64008000 & 241.2 & 241.4 & 37.0 & 30.0 & 191.2 & 191.2 & 6.4 & 7.7 \\
NGC 5044 & 80026000 + 80026010 & 18.4 & 13.1 & 109.2 & 72.0 & 17.9 &
      15.5 & 31.2 & 33.5\\
\end{tabular}

\medskip
\raggedright

The exposures include any time filtering. The count rates are given
for energies 0.55-9 keV for the SIS and 1-9 keV for the GIS. The count
rates are background subtracted within the particular aperture (see
text in section \ref{obs}).

\end{table*}

The {\sl ASCA} X-ray satellite consists of four detectors: two X-ray
CCD cameras (Solid State Imaging Spectrometers -- SIS0 and SIS1) and
two proportional counters (Gas Imaging Spectrometers -- GIS2 and
GIS3). Each SIS is a square array of four $420\times 422$ pixel CCD
chips with a total field of view of 22 arcmin$^2$. The usable field of
view of each GIS lies within a circle of radius 20 arcmin$^2$.  Each
detector is illuminated by its own X-ray telescope (XRT) consisting of
119 nested layers of thin foil. Although the point spread function
(PSF) of each XRT has a relatively sharp core, the wings of the PSF
are quite broad (half power diameter $\sim 3\arcmin$) and increase
markedly for energies above a few keV (e.g. Kunieda et
al. 1995). Since the angular sizes of the elliptical galaxies in our
sample are comparable to this PSF, we do not attempt to analyze the
spatial distribution of the {\sl ASCA} data of these sources. Rather,
(as in most previous studies) we analyze the {\sl ASCA} X-ray emission
within a single large aperture for each galaxy which encloses most of
the detectable emission; we address the spatial properties of the
X-ray emission with {\sl ROSAT} data in section \ref{rosat}.

We obtained {\sl ASCA} data for NGC 1399, NGC 4472, NGC 4636, and NGC
5044 from the public data archive maintained by the High Energy
Astrophysics Science Archive Research Center (HEASARC).  The basic
properties of these galaxies are listed in Table \ref{tab.prop} and
the observation sequence numbers are listed in Table \ref{tab.obs}.
The second sequence (80026010) for NGC 5044 began 2 minutes after the
first sequence (80026000) and the pointing for each sequence was
identical. Since the gain and response of the detectors do not change
significantly over such time scales, we directly combined the events
of these two sequences. The observation sequence for NGC 4636 listed
in Table \ref{tab.obs} corresponds to the deep exposure taken in
1995. (Note that we do find that the significantly shorter PV-phase
observation of NGC 4636 (see section \ref{1t}) gives results that are
consistent with this deep observation.)

The data were reduced with the standard {\sc FTOOLS} (v4.1) software
according to the procedures described in The {\sl ASCA} Data Reduction
Guide and the WWW pages of the {\sl ASCA} Guest Observer Facility
(GOF)\footnote{See http://heasarc.gsfc.nasa.gov/docs/asca/abc/abc.html
and http://heasarc.gsfc.nasa.gov/docs/asca/.}  For the observations of
NGC 1399, NGC 4472, and NGC 5044 (all taken in the PV
phase) we used the events files generated by the default screening
criteria processed under the Revision 2 Data Processing (REV2). As is
standard procedure for analysis of SIS data, only data in BRIGHT mode
taken in medium or high bit rate were used for these galaxies.  Since
the NGC 4636 observation was performed in 1995 its SIS data are
degraded to some extent because of radiation damage to the CCDs (see
Dotani et al. 1995).

One of the problems caused by radiation damage is the Residual Dark
Distribution (RDD) which is essentially an increase in the dark
current of the CCDs (see
http://www.astro.isas.ac.jp/$\sim$dotani/rdd.html). RDD degrades the
energy resolution and the detection efficiency of the data. We
corrected the NGC 4636 data for this effect following the procedure
outlined on the {\sl ASCA} GOF WWW pages which produces corrected SIS
event files in BRIGHT2 mode constructed using the default screening
criteria. (Again, only data with medium or high bit rate were used.)

The gain of the SIS varies from chip-to-chip, and the variation is a
function of time owing to radiation damage. We have corrected the SIS
data for these effects using the most up-to-date calibration files as
of this writing (sisph2pi\_110397.fits). Finally, the screened SIS and
GIS events files for each galaxy were further modified by excluding
time intervals of high background. We mention that no dead time
corrections were required for the GIS data because the count rates
were less than 1 ct s$^{-1}$ for each GIS detector for all the
observations in Table \ref{tab.obs}.

The final processed events were then extracted from a region centered
on the emission peak for each detector of each sequence.  We selected
a particular extraction region using the following general
guidelines. Our primary concern is to select a region that encloses
most of the X-ray emission yet is symmetrically distributed about the
origin of the region; as a result, we used circles for most of our
extraction apertures.  We limited the size of the aperture to ensure
that the entire aperture fit on the detector, an issue more important
for the SIS because of its smaller field-of-view. Moreover, for the
SIS0 and SIS1 we tried to limit the apertures to as small a number of
chips as possible to reduce the effects of residual errors in
chip-to-chip calibration. The GIS regions were chosen to be of similar
size to the corresponding SIS regions of a given sequence for
consistency. Since, however, the GIS+XRT PSF is somewhat larger than
that of the SIS+XRT the extraction regions for the GIS are usually
$\sim 20\%$ larger.

We extracted the events using regions defined in detector coordinates
as is recommended in the {\it ASCA} ABC GUIDE because the spectral
response depends on the location in the detector not the position on
the sky. (Also, this avoids the problem in the current software in
handling regions defined in sky coordinates.) For the GIS data of all
of the galaxies we chose circular extraction regions: $r=6\arcmin$ for
NGC 1399 and NGC 4636 and $r=6.5\arcmin$ for NGC 4472 and NGC
5044. Similarly, circles were used for the SIS observations of NGC
1399 ($r=4.7\arcmin$) and NGC 5044 ($r=5.3\arcmin$); note the radius
of NGC 1399 is slightly smaller to reduce contamination from NGC 1404.

For the SIS data of NGC 4472 and NGC 4636 we used apertures different
from circles. The resulting extracted spectra, however, are not very
sensitive to these differences in aperture shape since the S/N is
lowest near the aperture boundaries where these shape differences are
most pronounced.  The X-ray emission of NGC 4472 is elongated (see
Irwin \& Sarazin 1996) and thus we used a slightly flattened ellipse
for the SIS data of sequence 60029000: $(a,b)=
(4.6\arcmin,5.5\arcmin)$. The pointing of sequence 60030000 places the
center closer to the edge of one of the CCDs and thus we decided a box
with half-widths $(4.2\arcmin,4.7\arcmin)$ was a better choice to
maximize S/N yet still be symmetrically positioned about the
origin. NGC 4636 was observed in single-ccd mode and thus a box
(half-width $4.6\arcmin$) was used. (Note the aperture radii for the
SIS are averages of the SIS0 and SIS1 radii, where the SIS1 radii are
generally smaller by $\sim 10\%$ because of their lower count rates.)

We computed background spectra for each detector using the standard
deep observations of blank fields. There are important advantages to
using these background templates instead of a local background
estimate. First, the templates allow background to be extracted from
the same parts of the detector as the source and thus the vignetting
and other exposure effects are the same for each; this is not the case
for background taken from a different region of the detector as the
current software does not allow the required corrections to be made
for spectral analysis. Second, it is known from {\sl ROSAT}
observations of these galaxies that their emission extends at least
over the whole field of view of the SIS and probably more (Trinchieri
et al. 1994; David et al. 1994; Rangarajan et al. 1995; Irwin \&
Sarazin 1996; Jones et al. 1997). This is especially important for NGC
4636 as it was observed in single-ccd mode.  Also, since the
background templates are created from deep exposures ($>100$ks) the
statistical error is much less than for local background obtained from
the NGC 1399, NGC 4472, and NGC 5044 observations.  However, it should
be emphasized that for these sources the integrated spectra within the
relatively large apertures are not dominated by background and, as a
result, the deduced temperatures and abundances are hardly affected
whether the templates or local background are used (e.g. Buote \&
Canizares 1997).

The SIS background templates we obtained from the HEASARC data archive
were selected to have the same standard event screening criteria as
were the events files of the galaxies in our sample. We performed the
rise-time filtering (i.e. {\sc gisclean}) on the standard GIS
templates as required to match the standard screening. In Table
\ref{tab.obs} the background-subtracted count rates for each
observation in each detector are listed.

The instrument response matrix required for spectral analysis of {\sl
ASCA} data is the product of a spectral Redistribution Matrix File
(RMF) and an Auxiliary Response File (ARF). The RMF specifies the
channel probability distribution for a photon (i.e. energy resolution
information) while the ARF contains the information on the effective
area. The RMFs for the GIS2 and GIS3 are equivalent, and thus for all
GIS spectra we use the GIS RMFs gis2v4\_0.rmf and its twin
gis3v4\_0.rmf obtained from the HEASARC archive. An RMF needs to be
generated specifically for each SIS of each observation because, among
other reasons, each chip of each SIS requires its own RMF and the
spectral resolution of the SIS is degrading with time. We generated
the responses for each SIS using the {\sc FTOOL sisrmg} selecting for
the standard event grades (0234). Using the response matrix and
spectral PI (Pulse Invariant) file we constructed an ARF file with the
{\sc FTOOL ascaarf}.

The source apertures used for the SIS observations of NGC 1399, NGC
4472, and NGC 5044 overlapped more than one chip. In order to analyze
the spectra of such regions we followed the standard procedure of
creating a new response matrix that is the average of the individual
response matrices of each chip weighted by the number of source counts
of each chip (i.e. within the source aperture). (Actually, the current
software only allows the RMFs to be averaged and then an ARF is
generated using the averaged RMF. This is considered to be a good
approximation for most sources.)  Unfortunately, some energy
resolution is lost as a result of this averaging. For the observations
of our sources, however, this small energy broadening is rendered
undetectable by the statistical noise of the data.

To maximize the S/N of our data we desire to combine the spectra of
all available observation sequences for a given galaxy; e.g. as in NGC
5044 above. There are two observations each for NGC 1399 and NGC 4472
taken during the PV phase. Since these observations are separated by
only one day for NGC 1399 and six days for NGC 4472 the responses of
each observation for each galaxy can be considered the same. However,
the pointings for the consecutive observations in each case were
somewhat different; e.g., the center of NGC 1399 is positioned at the
midpoints of different SIS chips for each observation. Hence, we can
combine the spectra of the consecutive observations for NGC 1399 and
NGC 4472 and use averaged RMFs and ARFs for spectral analysis without
significant loss of energy resolution. However, we do not combine the
PV-phase (1993) observation of NGC 4636 (see BF) with the 1995
observation listed in Table \ref{tab.obs} because the responses are
sufficiently different to warrant separate analysis of the data.

To further improve the S/N we also consider combining the spectra from
each SIS and similarly for each GIS; i.e. SIS=SIS0+SIS1 and
GIS=GIS2+GIS3. For the GIS data this does not result in significant
loss of information because the RMFs of the GIS2 and GIS3 are the same
and time-independent; only the ARFs need to be averaged for the GIS2
and GIS3. However, the responses of the SIS0 and SIS1 detectors are
different and thus the required averaging of the RMFs and ARFs leads
again to loss of information. Because the gain in S/N is important for
our study, we examine the combined SIS and combined GIS data of each
galaxy where all of the sequences listed in Table \ref{tab.obs} for a
given galaxy have been summed. For each galaxy the source spectra,
RMF, ARF, and background for each detector of each observation are
summed and scaled appropriately using the {\sc FTOOL addascaspec}
(v1.27).

Overall we found that the results obtained from analysis of these
summed SIS spectra agree well with results obtained without summing
the data. We mention below (section \ref{calib}) the instances where
differences in the SIS0 and SIS1 data are significant.

\section{Broad-band spectral fitting of {\sl ASCA} data}
\label{broad}

\begin{figure*}
\parbox{0.49\textwidth}{
\centerline{\psfig{figure=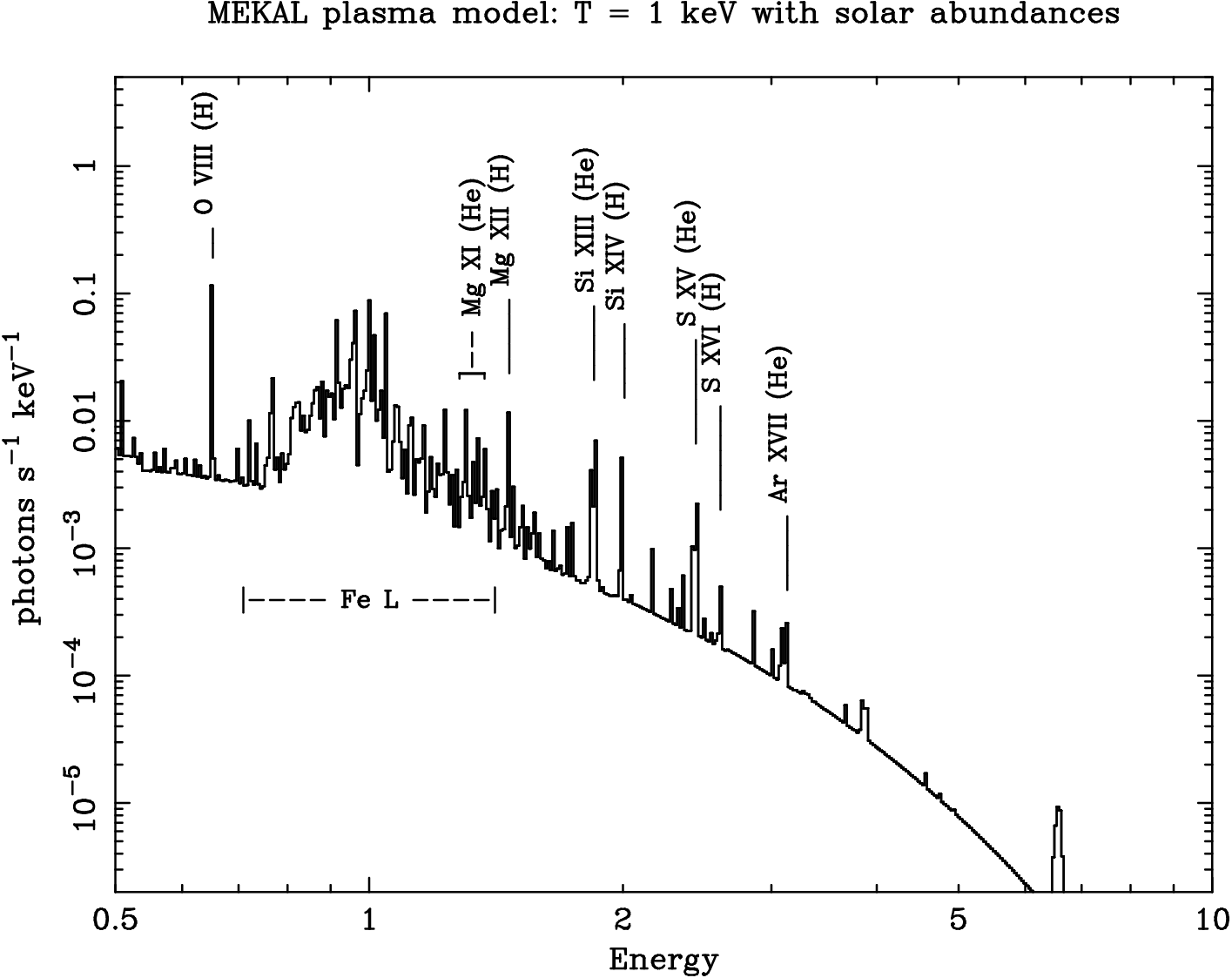,angle=-90,height=0.25\textheight}}
}
\parbox{0.49\textwidth}{
\centerline{\psfig{figure=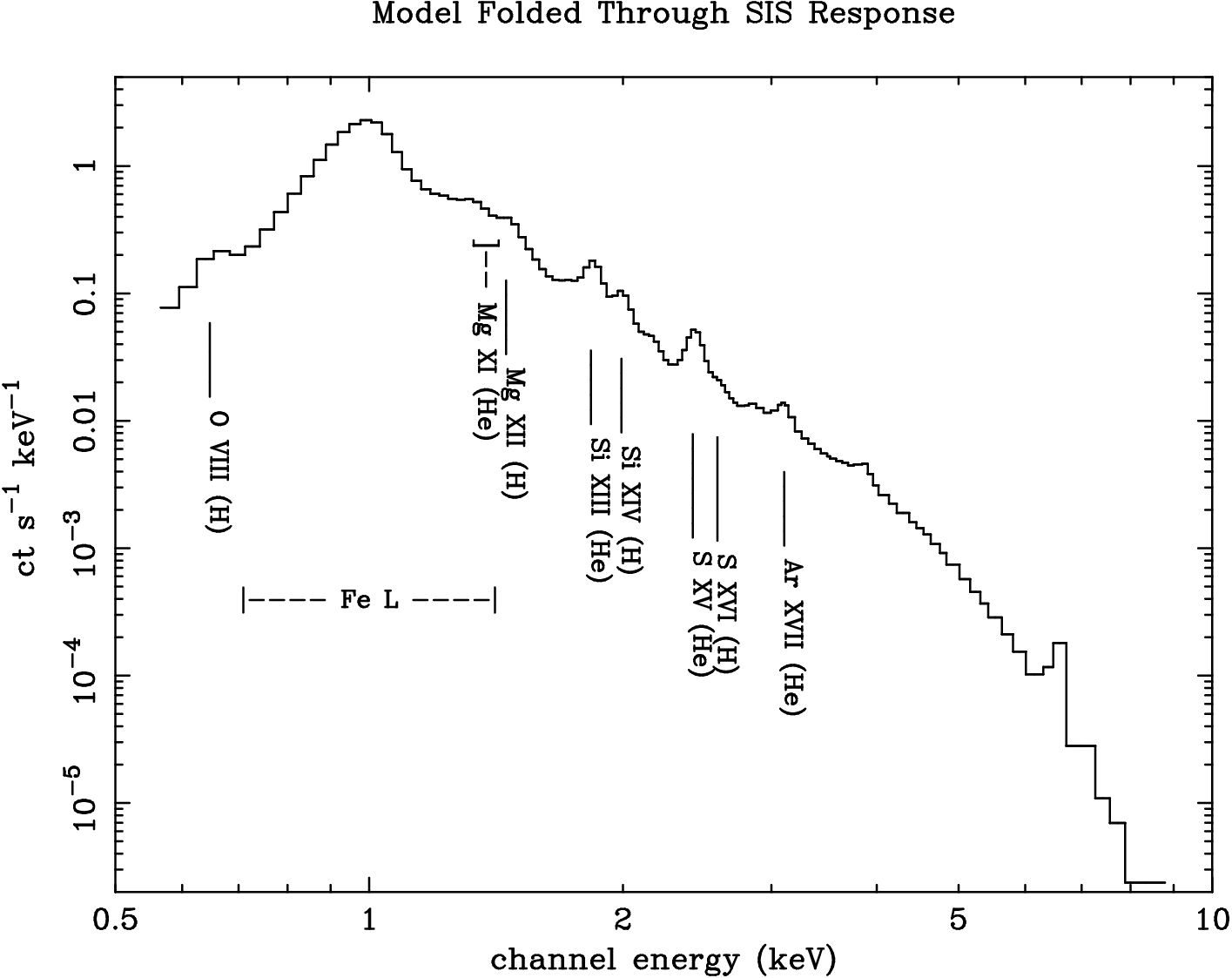,angle=-90,height=0.25\textheight}}
}

\vskip 0.4cm

\parbox{0.49\textwidth}{
\centerline{\psfig{figure=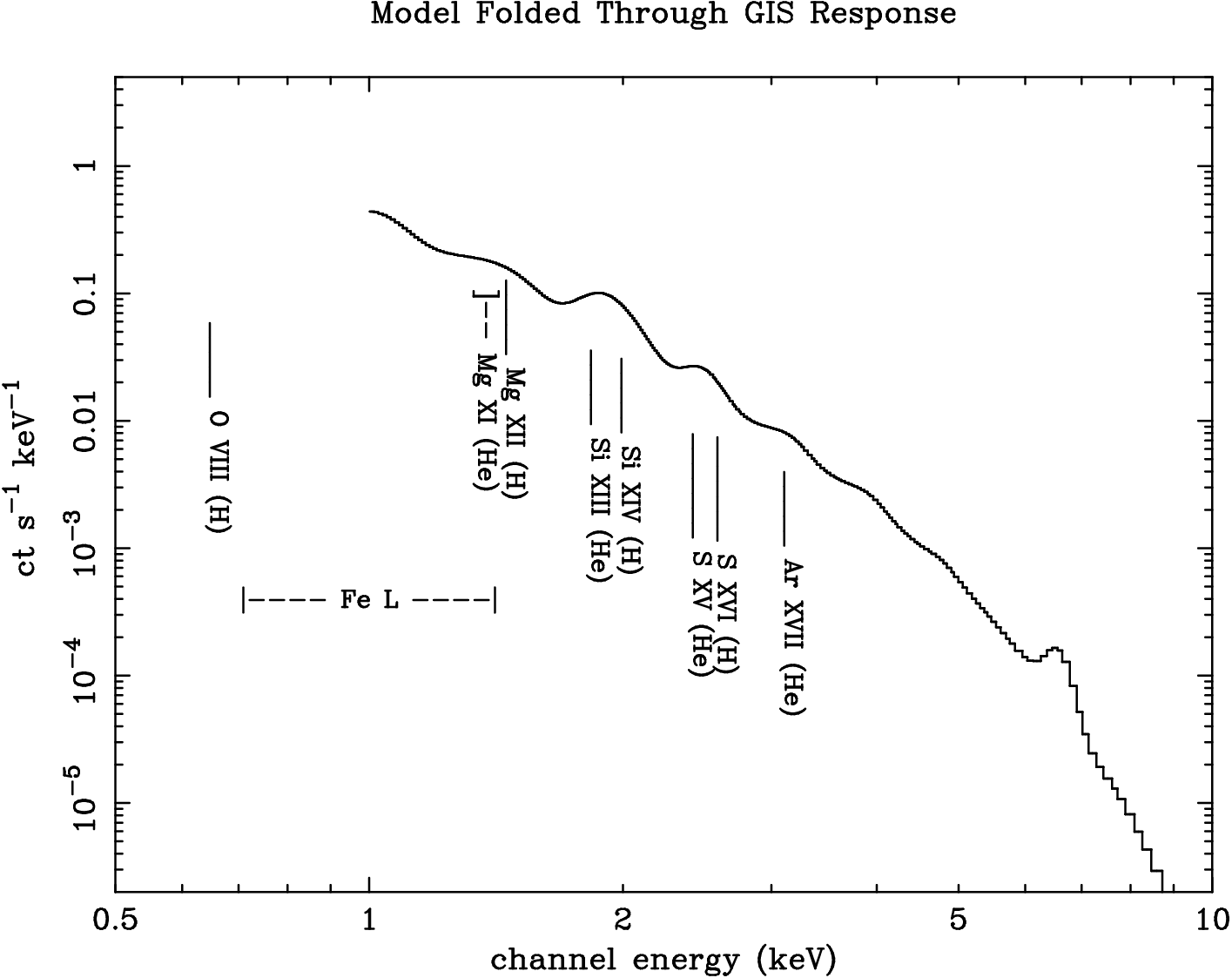,angle=-90,height=0.25\textheight}}
}
\parbox{0.49\textwidth}{
\centerline{\psfig{figure=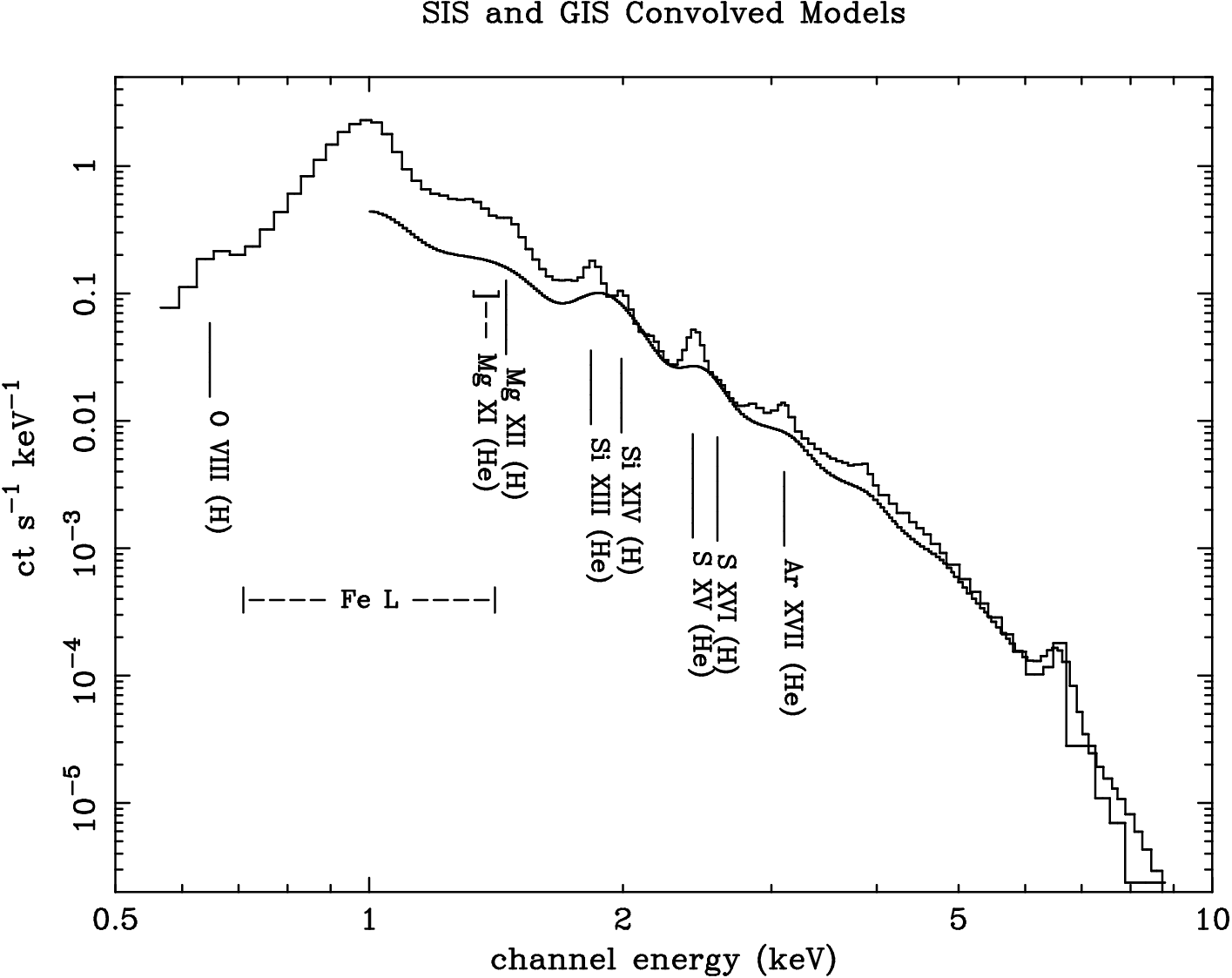,angle=-90,height=0.25\textheight}}
}
\caption{\label{fig.example} Comparison of the {\sl ASCA} SIS and GIS
for thermal plasmas appropriate for giant elliptical galaxies. The top
left panel shows a MEKAL model with $T=1$ keV and $Z=1Z_{\sun}$. The
emission measure is selected to be similar to NGC 4472 which
corresponds to 0.01 in units of $10^{-14}n_en_pV/4\pi D^2$ used in
XSPEC. The discreteness in the MEKAL model results from the binning of
the model in 500 energy bins over the range 0.4-10 keV. The bin sizes
for the SIS and GIS convolved models reflect those of the actual
response matrices we analyzed for NGC 4472.}

\end{figure*}

We begin our interpretation of the {\sl ASCA} data by fitting models
to the broad-band spectra of each galaxy.  For the SIS data we examine
energies 0.55-9 keV and for the GIS 1-9 keV. The cutoff at 9 keV is
chosen because at higher energies calibration errors are significant
\cite{gendreau}, and the background dominates the signal (thus
amplifying any error in the background level). The low energy cutoffs
are dictated by calibration uncertainties (see {\sl ASCA} GOF WWW
pages).

Since the X-ray emission of the ellipticals in our sample is dominated
by hot gas, we use coronal plasma models as our basic component. We
focus on the MEKAL code which is a modification of the original MEKA
code (Mewe, Gronenschild, \& van den Oord 1985; Kaastra \& Mewe 1993)
where the Fe L shell transitions crucial to the X-ray emission of
ellipticals have been re-calculated \cite{mekal}. For comparison, we
also use the Raymond-Smith code (RS) \cite{rs}. Although the RS code
does not include the updated Fe L calculations, nor does it include as
many lines as in MEKAL, we include it because it is frequently used in
related studies of ellipticals. For both models we take solar
(photospheric) abundances according to Anders \& Grevesse
\shortcite{ag} for consistency with most previous studies (Fe
abundance is $4.68\times 10^{-5}$ relative to H). (We also consider
versions of these codes, VMEKAL/VRAYMOND, that allow for the
abundances of each element to be different from solar.)

We account for absorption by our Galaxy using the photo-electric
absorption cross sections according to Baluci\'{n}ska-Church \&
McCammon \shortcite{phabs}. The absorber is modeled as a uniform
screen at zero redshift. To represent the integrated emission from
X-ray binaries we use a pure thermal bremsstrahlung model (BREM) which
is consistent with the functional form often used to describe the
spectrum of discrete sources in elliptical galaxies (e.g. Kim et
al. 1992; Matsumoto et al. 1997).

Finally, we consider a multiphase cooling flow model (CF)
\cite{rjcool}. This model assumes gas cools continuously at constant
pressure from some upper temperature, $T_{\rm max}$. The differential
emission measure is proportional to $\dot{M}/\Lambda(T)$, where
$\dot{M}$ is the mass deposition rate of gas cooling out of the flow,
and $\Lambda(T)$ is the cooling function of the gas (in our case, the
MEKAL plasma code). We remark that this is arguably the simplest model
of a cooling flow with mass drop-out. The advantage of this particular
model is that it is well studied, relatively easy to compute, and a
good fit to several ellipticals (BF).

All spectral fitting was performed with the software package {\sc
XSPEC} \cite{xspec}. We use the exact MEKAL models in {\sc XSPEC}
(i.e. model parameter ``switch'' set to 0), not those derived from
interpolating pre-computed tables.  For all fits we used the
${\chi^2}$ method implemented in its standard form in {\sc XSPEC}.

In order for the weights to be valid for the ${\chi^2}$ method we
followed the standard procedure and regrouped the PI bins for each
source spectrum so that each group had at least 20 counts. The
background templates, especially for the summed SIS and GIS data (see
end of section \ref{obs}), also generally have at least 20 counts when
their energy bins are grouped similarly to the source spectra.  The
background-subtracted count rate in a particular group can be small,
but the uncertainties in the source and background are correctly
propagated by {\sc XSPEC} to guarantee approximately gaussian
statistics for the statistical weights (K. Arnaud 1998, private
communication). Note that we did investigate using larger group sizes
(i.e. 50 and 100 counts) but found no qualitative differences in the
best-fitting models.

The normalizations of all model components are fitted separately for
each data set. For example, when jointly fitting multicomponent models
to the summed SIS and summed GIS data the normalizations of each model
component for the SIS and GIS are free parameters. This is done to
account for differences in the SIS and GIS fluxes due to small region
size differences and gain differences between the detectors. A
side-effect of this procedure is that in some instances the relative
normalizations of the model components for the SIS can differ
qualitatively from those of the GIS. We discuss these occurrences
below. (All of this also applies to joint fitting of SIS0 and SIS1
data and/or GIS2 and GIS3 data.)

The procedure we follow for spectral fitting is similar to that
described in BF. That is, we begin by fitting a single MEKAL model
modified by Galactic absorption \cite{stark} where the relative
abundances of the elements with respect to Fe are fixed at their solar
values; i.e. the free parameters are the temperature, metallicity, and
normalization. We then examine whether allowing $N_{\rm H}$ to be free
significantly improves the fit; note that any excess absorption so
detected is only an estimate of absorbing material intrinsic to the
galaxy in question (see BF).

We proceed to add other components, one at a time, until all free
parameters of consequence have been varied. We focus on the following
combinations of ``thermal+discrete'' models where each component is
modified by its own absorption: (1) 1T + BREM, (2) 2T, (3) 2T + BREM,
(4) CF+1T, (5) CF+1T+BREM. (Models having more components do not
significantly improve the fits to the {\sl ASCA} data.) For
multicomponent models with BREM we restricted the absorption on the
BREM component to be its Galactic value since the emission from
discrete sources is not expected to suffer from excess absorption. (At
any rate, we did not find much improvement in the fits when allowing
the BREM absorption to be a free parameter.) The cooling flow models
(4) and (5) account for gas that is not participating in the cooling
flow by adding a component of isothermal gas to the CF model where the
temperature of the isothermal component is tied to the maximum
temperature of the cooling flow component; e.g. our generic cooling
flow model (4) can be expressed as $\rm phabs\times CF + phabs\times
MEKAL$, where ``phabs'' represents the photo-electric absorption on
each component and $T_{\rm MEKAL} = T_{\rm CF}$.

Initially, the relative abundances with respect to Fe of any MEKAL/RS
components are fixed at their solar values; i.e. the Fe abundance is
the only abundance which is initially a free parameter. Once the
best-fit versions of models (1)-(5) have been found, we then
investigate allowing the relative abundances of other elements to be
free. We typically allow the abundances of the following
$\alpha$-process elements, one at a time, to be free parameters: Si,
S, Ar, Mg, Ne, O. However, we do examine whether allowing the
abundances of other elements can improve the fits of various models.
For multicomponent models the abundance of a particular element in one
component is tied to the corresponding value in the other
component. (Relaxing this restriction did not noticeably improve the
fits in any case.)

We emphasize that to achieve the global minimum for the multicomponent
models often requires some effort (see BF). Often the fits land in a
local minimum upon first adding another component to a model. To help
correct this problem, after a fit is completed we always reset a
subset of the free parameters and fit again several times until we are
satisfied that the minimum is stable. Stepping through the parameters
when determining confidence limits also was useful in assessing the
stability of the minimum. This is not the most rigorous method to find
the global minimum, but it is at present the most convenient way to do
it in {\sc XSPEC}.

\subsection{Relative importance of SIS and GIS}
\label{relative}

It is important to understand how much the SIS and GIS detectors
modify an incident galaxy spectrum in order to interpret the spectral
fitting results in the next section. For this discussion we consider a
MEKAL plasma with a temperature of 1 keV, solar abundances, and a
redshift and an emission measure similar to the galaxies in our
sample. This model is plotted in Figure \ref{fig.example}.

The most striking features of the plasma are the emission lines,
particularly the complex of Fe L shell lines around 1 keV. These Fe L
lines are generally the strongest and most temperature sensitive lines
for elliptical galaxies. Other prominent lines involve K$\alpha$
transitions of O, Mg, Si, S, and Ar. These lines individually tend not
to be as sensitive as Fe L to the plasma temperature, but in tandem
are very useful for probing the plasma emission measure
distribution. We refer the reader to Buote et al. \shortcite{b98} for
a discussion of the temperature sensitivity of these lines for models
of elliptical galaxies.

Much of the information provided by these lines is lost when folded
through the responses of the SIS and GIS detectors. However, the
degradation in the folded data is far from equal in the two cases. In
Figure \ref{fig.example} we plot the 1 keV plasma model folded through
the SIS and GIS responses. For the sake of visual clarity the folded
spectrum of each detector is shown separately. We also provide a
separate panel in Figure \ref{fig.example} showing both the SIS and
GIS spectra to facilitate visual comparison of the effective area as a
function of energy of the detectors.

The superior energy resolution of the SIS allows the He-like and
H-like K$\alpha$ line blends of Si and S to be clearly resolved
(though S XVI ly-$\alpha$ is weak). In contrast, these He-like and
H-like line complexes are unresolved by the GIS; i.e. only the blends
of the combined He-like and H-like K$\alpha$ lines are apparent.
Significant structure of the Fe L complex, though unresolved, is
readily apparent in the SIS unlike the GIS.

Given the importance of the Fe L lines and the other emission lines
around 2-3 keV, it is significant that the count rate of the SIS is
larger than the GIS, particularly near 1 keV where the ratio of count
rates is $\sim 5:1$. In fact, the count rate of the GIS only exceeds
the SIS for energies above $\sim 7$ keV where the count rate is very
low in both detectors.

Considering as well that the GIS does not have reliable data below
$\la 0.8$ keV where there is significant Fe L emission, it is clear
that for elliptical galaxies the SIS data possess much more
information than the GIS data and thus the SIS provides the most
important constraints on the temperatures and abundances of these
systems. These properties must be kept in mind when assessing the
quality of the spectral fits to which we now turn our attention.

\subsection{Quality of fits: isothermal vs multiphase models} 
\label{quality}

\begin{table*}
\begin{minipage}{180mm}
\caption{Quality of Spectral Fits}
\label{tab.quality}
\begin{tabular}{lcccccccc}
& & \multicolumn{2}{c}{1T+BREM} & 2T & \multicolumn{2}{c}{2T+BREM} &
CF+1T & CF+1T+BREM\\ 
& & Fix & Var & Fix & Fix & Var & \multicolumn{1}{c}{Fix} &
\multicolumn{1}{c}{Fix}\\ \\
\multicolumn{5}{l}{N1399:}\\ \\[-8pt]
MEKAL & $P$            & 1.9e-17     & 1.4e-8      & 2.7e-6      & 1.4e-2      & 3.1e-2      & 3.6e-9      & 2.9e-4\\
& ($\chi^2$/dof) & (651.6/373) & (539.1/368) & (509.3/372) & (431.3/369) & (414.8/363) & (553.2/373) & (473.0/372)\\

RS    & $P$            & 2.2e-16     & 9.1e-6      & 3.8e-14     & 9.4e-10     & 2.0e-3      & $\cdots$ & $\cdots$ \\
& ($\chi^2$/dof) & (640.1/373) & (496.0/368) & (613.9/372) & (556.3/369) & (446.4/364) & $\cdots$ & $\cdots$ \\ \\
\multicolumn{5}{l}{N4472:}\\ \\[-8pt]
MEKAL & $P$            & 5.3e-16     & 2.7e-9      & 5.5e-3      & 0.44        & 0.66        & 5.0e-4      & 2.8e-2\\
& ($\chi^2$/dof) & (624.3/364) & (537.9/359) & (435.2/363) & (363.6/360) & (342.5/354) & (459.3/364) & (413.9/361)\\
RS    & $P$            & 3.4e-6      & 9.4e-3      & 6.3e-5      & 1.2e-3      & 3.9e-2      & $\cdots$ & $\cdots$ \\
& ($\chi^2$/dof) & (498.3/364) & (424.9/359) & (475.6/363) & (446.9/360) & (404.4/356) & $\cdots$ & $\cdots$ \\ \\
\multicolumn{5}{l}{N4636:}\\ \\[-8pt]
MEKAL & $P$            & 1.5e-39     & 1.3e-4      & 3.2e-39     & 5.4e-9      & 1.0e-2      & 0.00       & 5.6e-20\\
& ($\chi^2$/dof) & (1346/727)  & (867.2/720) & (1342/726)  & (961.9/723) & (806.9/716) & (1786/727) & (1126/724)\\
RS    & $P$            & 0.00        & 1.1e-40     & 0.00        & 0.00        & 2.9e-10     & $\cdots$ & $\cdots$\\
& ($\chi^2$/dof) & (1826/727)  & (1349/721)  & (1822/726)  & (1586/723)  & (977.0/717) & $\cdots$ & $\cdots$\\ \\
\multicolumn{5}{l}{N5044:}\\ \\[-8pt]
MEKAL & $P$            & 2.0e-24     & 1.3e-8      & 5.6e-3      & 7.8e-3      & $\cdots$ & 1.2e-6      & 2.1e-5\\
& ($\chi^2$/dof) & (610.0/293) & (439.6/286) & (357.0/292) & (350.4/289) & $\cdots$ & (420.4/292) & (398.1/289)\\
RS & $P$            & 2.5e-9      & 1.9e-8      & 1.9e-11     & 6.1e-11     & 2.2e-6      & $\cdots$ & $\cdots$\\
& ($\chi^2$/dof) & (457.3/293) & (438.8/287) & (481.4/292) & (471.6/289) & (405.8/283) & $\cdots$ & $\cdots$\\
\end{tabular}

\medskip

The $\chi^2$ null hypothesis probability ($P$), the value of $\chi^2$,
and the number of degrees of freedom (dof) are listed for each model
fitted jointly to the SIS and GIS data with normalizations of each
data set free parameters. Models where the relative abundances of the
elements with respect to Fe are fixed at their solar values are
denoted by ``fix'' and those that allow variable relative abundances
are denoted by ``var''. ``MEKAL'' and ``RS'' indicate whether the 1T
and/or 2T thermal plasma components are computed using the MEKAL or
Raymond-Smith models respectively. See text for further explanation of
these fits.

\end{minipage}
\end{table*}

In this section we address our primary objective to determine whether
broad-band fits to the {\sl ASCA} spectra using models consisting of a
single temperature component of hot gas and a bremsstrahlung component
representing discrete sources (i.e. 1T+BREM) are of comparable quality
to fits consisting of at least two temperature components of hot gas
(i.e. 2T, 2T+BREM, CF+1T, and CF+1T+BREM). The quality of the fits and
the general properties of the best-fitting parameters are examined in
this section. We defer detailed discussion of the derived model
parameters to section \ref{detailed}.

Whereas BF emphasized joint fitting of the SIS0 and SIS1 data for each
available observation, we focus on analysis of the total summed SIS
data fitted jointly to the summed GIS data (see end of section
\ref{obs}). In addition to the gain in S/N offered by using the summed
data (which is also important for studying individual lines -- see
section \ref{lines}), the spectral analysis and presentation is
simplified greatly since only two spectra (i.e. the total SIS and GIS)
need to be simultaneously modeled for each galaxy. We discuss the
reliability of using the summed data sets in section \ref{calib}
below.

In Table \ref{tab.quality} we list the $\chi^2$ values and null
hypothesis probabilities $(P)$ for the best-fitting models of each
galaxy fitted jointly to the summed SIS and summed GIS data.  We plot
in Figures \ref{fig.n1399_1t}, \ref{fig.n1399_multit}, and
\ref{fig.n1399_ray} a subset of the best-fitting models listed in
Table \ref{tab.quality} for NGC 1399. The Fe abundance is a free
parameter in all of the fits. Results where relative abundances of the
elements with respect to Fe are either held fixed at their solar
values (fix) or allowed to vary (var) are presented for the the
1T+BREM and 2T+BREM models; the 2T model with variable relative
abundances is not displayed because the values of the derived
abundances and the improvement in $\chi^2$ are very similar to the
2T+BREM model in all cases.  Only fixed relative abundance results are
given for the cooling flow models because the current implementation
of the cooling flow model does not allow for variable relative
abundances owing to prohibitive computational expense.

\subsubsection{MEKAL models}
\label{mekal}

\begin{figure*}
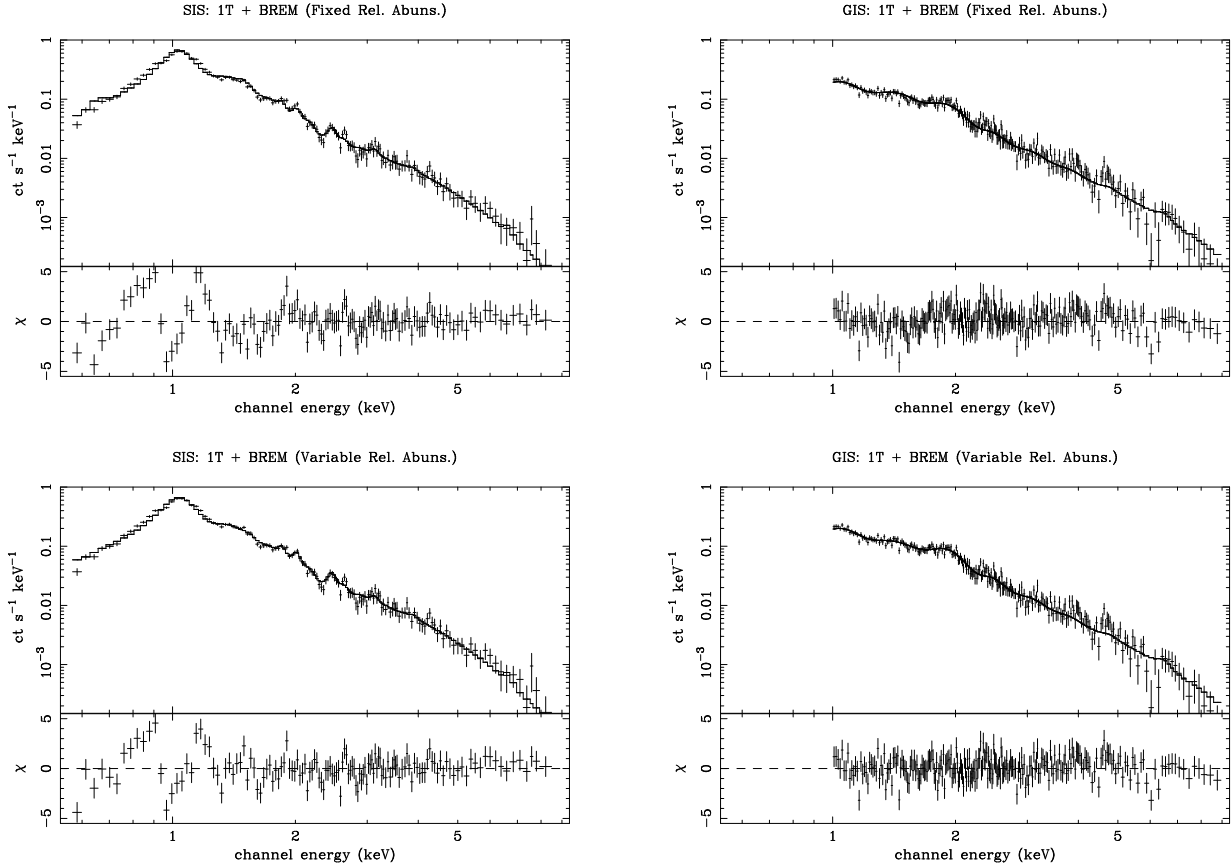

\parbox{0.49\textwidth}{
\centerline{\psfig{figure=fig2a.ps,angle=-90,height=0.23\textheight}}
}
\parbox{0.49\textwidth}{
\centerline{\psfig{figure=fig2b.ps,angle=-90,height=0.23\textheight}}
}

\vskip 0.4cm

\parbox{0.49\textwidth}{
\centerline{\psfig{figure=fig2c.ps,angle=-90,height=0.23\textheight}}
}
\parbox{0.49\textwidth}{
\centerline{\psfig{figure=fig2d.ps,angle=-90,height=0.23\textheight}}
}
\caption{\label{fig.n1399_1t} Best fitting 1T+BREM models for NGC 1399
for the cases where (1) the relative abundances with respect to Fe are
fixed at their solar values (top panels) and (2) the relative
abundances are allowed to vary (bottom panels). The models are fit
jointly to the SIS and GIS data, but the fits and residuals are shown
separately for clarity. The 1T component corresponds to a MEKAL model.
See text for further explanation of these models}

\end{figure*}

\begin{figure*}
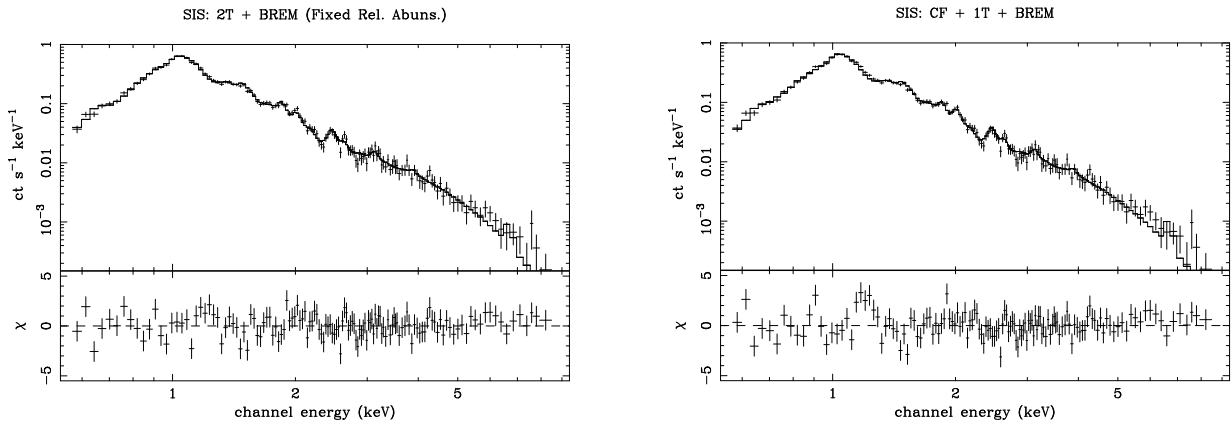

\parbox{0.49\textwidth}{
\centerline{\psfig{figure=fig3a.ps,angle=-90,height=0.23\textheight}}
}
\parbox{0.49\textwidth}{
\centerline{\psfig{figure=fig3b.ps,angle=-90,height=0.23\textheight}}
}
\caption{\label{fig.n1399_multit} As Figure \ref{fig.n1399_1t} except
displayed are the best-fitting 2T+BREM (left) and CF+1T+BREM (right)
models. Although each model is fitted jointly to the SIS and GIS data,
only the SIS data are shown because the relatively featureless GIS
residual patterns are essentially indistinguishable from those in
Figure \ref{fig.n1399_1t}.}

\end{figure*}

Let us consider the joint fits to the SIS and GIS data where the MEKAL
plasma code is used for the model components of hot gas, and let us
presently focus on the models with fixed relative abundances.
Examination of Table \ref{tab.quality} reveals that NGC 1399, NGC
4472, and NGC 5044 behave quite similarly when fitted by the models
shown. The 1T+BREM models are strongly excluded $(P\ll 1)$ for all of
the galaxies. From examination of the residuals in Figure
\ref{fig.n1399_1t} it is clear that the poor fits of the 1T+BREM
models are mostly the result of the SIS data in the Fe L energy region
$\sim 0.7$ keV - 1.4 keV.

If instead of the 1T+BREM model we consider a model with two plasma
temperature components (i.e. 2T model) we find that the fits for NGC
1399, NGC 4472, and NGC 5044 are substantially improved essentially as
a result of reducing the magnitude of the SIS residuals near 1 keV
present in the fits of the 1T+BREM models. The fits of NGC 1399 and
NGC 4472 are further improved when adding a BREM component (2T+BREM),
though the reduction in $\chi^2$ is not so large as those observed
between the 1T+BREM and 2T cases. The fit of the 2T+BREM model of NGC
4472 is formally very acceptable $(P=0.42)$ while those for NGC 1399
and NGC 5044 are formally marginal $(P\sim 0.01)$. However, as is
clear from Figure \ref{fig.n1399_multit}, these models provide quite
satisfactory fits considering their simplicity.  The cooling flow
models for NGC 1399, NGC 4472, and NGC 5044 behave almost exactly as
the two-temperature models except that they have slightly lower
quality fits as quantified by $P$.

NGC 4636, which has lower emission weighted temperature ($T\sim 0.7$
keV) than the others, has a qualitatively different behavior than the
other galaxies. Not only is the 1T+BREM model indicated to be highly
unacceptable, replacing the BREM with another MEKAL component does not
improve the fit significantly. However, the 2T+BREM model is a much
better fit than 1T+BREM, although it is still not formally
acceptable. Similarly, the CF+1T model behaves as a pure CF model and
has a larger $\chi^2$ than the 1T+BREM model. A noticeable improvement
in the fit is achieved when adding a BREM component, though again the
contribution of the 1T component is negligible; i.e. CF+1T+BREM
$\approx$ CF+BREM. Hence, it is clear that $P$ is substantially
improved for NGC 4636 when a bremsstrahlung component is added to the
multitemperature models (as is the case for NGC 1399 and NGC 4472 but
to a lesser extent).

When the relative abundances of the elements with respect to Fe are
allowed to vary the 1T+BREM models are improved significantly;
typically only the relative abundances of the $\alpha$-process
elements Si, S, Ar, Mg, Ne, and O were varied separately in these
fits, though in some cases other elements were varied as well (see
below). (Note, consistent with BF, we do not find improvement in the
fits when the $\alpha$-process elements were tied together and fitted
as one parameter.) For NGC 1399, NGC 4472, and NGC 5044 these 1T+BREM
models are still poor fits $(P\ll 1)$ and have much larger values of
$\chi^2$ than do the multitemperature models with fixed relative
abundances.  However, for NGC 4636 the 1T+BREM variable abundance model
is only marginally unacceptable $(P\sim 10^{-4})$ and provides a
better fit than the multitemperature models with fixed relative
abundances. Allowing for variable relative abundances in the two
temperature models only yields modest reductions in $\chi^2$ except
for NGC 4636 where the change is substantial and results in a
reasonably good fit.

The relative abundances with respect to Fe obtained from the fits of
the 1T+BREM models for some $\alpha$-process elements differ markedly
from their solar values.  For NGC 1399 and NGC 4472 we find that the
key improvements in $\chi^2$ for these models are the result of
reducing the abundances of O, Ne, and Mg such that $Z_{\rm
O,Ne}\approx 0$ and $Z_{\rm Mg}\ll Z_{\rm Fe}$. For NGC 5044, the
Nickel abundance is the most important. (The next important effect is
allowing the oxygen abundance to go to zero analogously to NGC 1399
and NGC 4472.)  It is clear that these abundances are primarily the
result of the 1T+BREM model trying to minimize the SIS residuals near
1 keV (see Figure \ref{fig.n1399_1t}) since over the {\sl ASCA} energy
band these elements have their strongest emission lines (O VIII, Ne X,
Mg XI and XII, Ni L shell) in the Fe-L energy region (with the
exception of O VII and Ni K shell).

Similar results hold for NGC 4636 except that varying the abundances
of each of the $\alpha$-process elements and Nickel significantly
improves $\chi^2$. Allowing the Na abundance to be a free parameter
gives the largest improvement of any element, with Mg showing the next
best improvement. (The energies of the strongest Na lines are at $\sim
1.1$ keV for the He-like ion and 1.24 keV for the the H-like ion.) The
resulting Na abundance, though uncertain, is many times solar; note
that NGC 1399 has a small $\chi^2$ improvement if Na is allowed to be
a free parameter and also implies a super-solar Na abundance. (These
abundances are discussed in more detail in section \ref{detailed}.)

In contrast, as stated above, for NGC 1399, NGC 4472, and NGC 5044 the
2T and 2T+BREM models with relative abundances with respect to Fe
fixed at their solar values are superior fits to all of the 1T+BREM
models.  We do not find significant improvement for these galaxies
when the relative abundances are varied, although for NGC 4472 the
best-fit O abundance is lower than Fe and for NGC 1399 the best-fit O
and Mg abundances are lower than Fe. However, these abundances are
consistent with the Fe abundances within their 90\% error
estimates. The abundances for the two-temperature fits of NGC 4636
behave similarly as for their 1T+BREM models.

\subsubsection{Comparison to Raymond-Smith}
\label{rs}

\begin{figure*}
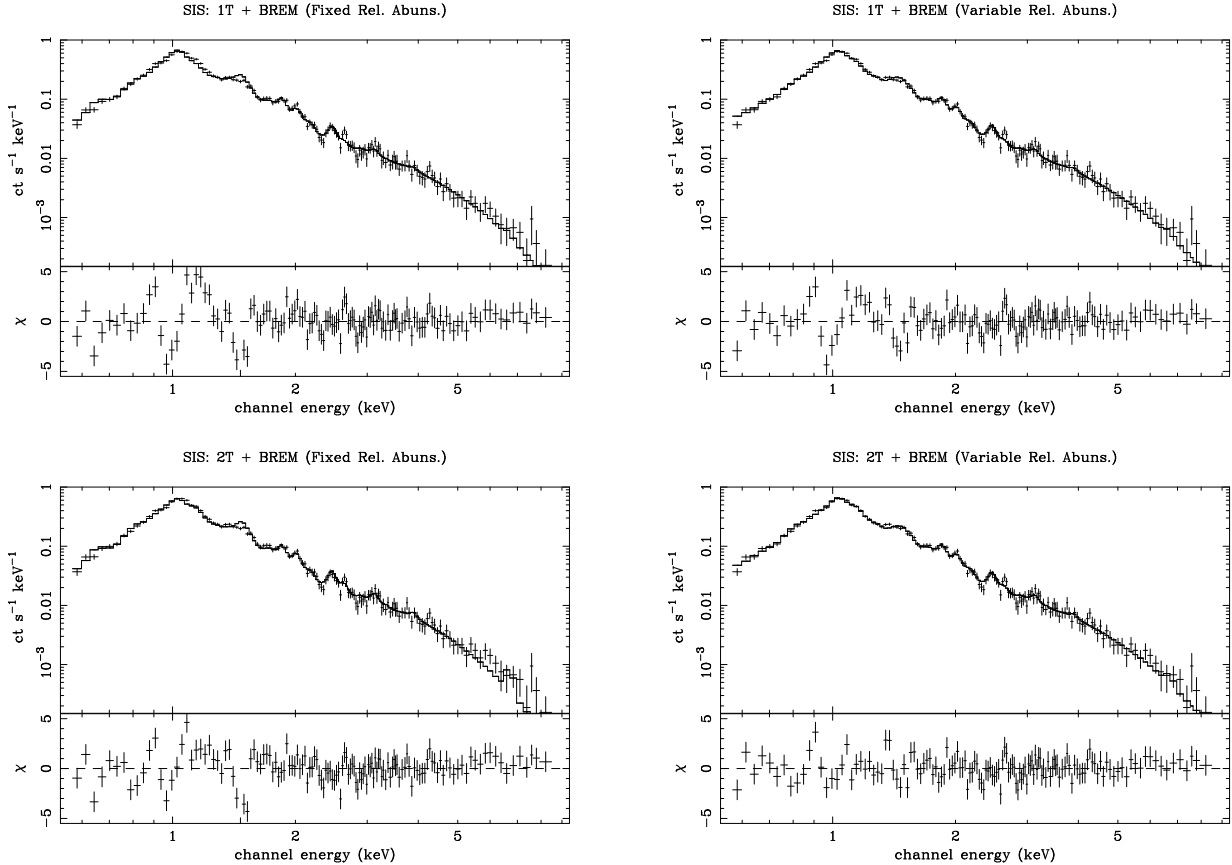

\parbox{0.49\textwidth}{
\centerline{\psfig{figure=fig4a.ps,angle=-90,height=0.23\textheight}}
}
\parbox{0.49\textwidth}{
\centerline{\psfig{figure=fig4b.ps,angle=-90,height=0.23\textheight}}
}

\vskip 0.4cm

\parbox{0.49\textwidth}{
\centerline{\psfig{figure=fig4c.ps,angle=-90,height=0.23\textheight}}
}
\parbox{0.49\textwidth}{
\centerline{\psfig{figure=fig4d.ps,angle=-90,height=0.23\textheight}}
}
\caption{\label{fig.n1399_ray} As Figure \ref{fig.n1399_multit} except
that Raymond-Smith (RS) code is used for the 1T and 2T model components.}

\end{figure*}

It is necessary to address the extent to which our results depend on
the accuracy of the plasma code. This is especially important since
the key constraints arise from the SIS data near 1 keV where emission
lines from the Fe L shell, which are not so accurately modeled as are
the K shell emission lines, dominate the emission. In particular, can
reasonable inaccuracies in the plasma code make a spectrum that is
actually 1T+BREM look like a 2T and/or 2T+BREM spectrum?

A simple way to test this is to perform the same fits of the previous
section and replace the MEKAL plasma model with that of Raymond-Smith
(1977, and updates; hereafter RS). Both the MEKAL and RS plasma codes
are identical in their treatment of the ionization balance as given by
Arnaud \& Raymond \shortcite{aray} for Fe and Arnaud \& Rothenflug
\shortcite{aroth} for the the other elements.  However, the RS code
has many fewer lines than does MEKAL, and, more importantly, it does
not incorporate the improved calculations of the Fe L shell lines by
Liedahl et al. \shortcite{mekal}. Although this is not intended as an
exhaustive test of all possible problems with the modeling of the Fe L
energy region, comparing the results of fitting MEKAL and RS models
does give a fair appraisal of the magnitude of the differences of
temperatures and abundances due to significantly different modeling of
the Fe L shell emission lines.

In analogy to the MEKAL fits of the previous section, we list the null
hypothesis probabilities and $\chi^2$ values for corresponding models
in Table \ref{tab.quality}. Similarly, the RS fits of 1T+BREM and
2T+BREM models for NGC 1399 are displayed in Figure
\ref{fig.n1399_ray}. Inspection of Table \ref{tab.quality} reveals
that, as would be expected, it is again the SIS data in the Fe L shell
energy region (see Figure \ref{fig.n1399_ray}) which figure most
prominently in the model constraints.

In contrast to the fits with the MEKAL code, the 1T+BREM models with
variable relative abundances fit better than two-temperature models that
have fixed relative abundances. When allowing for variable relative
abundances the two-temperature models do fit better than the corresponding
isothermal models similar to the models with fixed relative
abundances. The relative abundances for these two-temperature models are
similar to the isothermal models.

Eliminating the Mg abundance is important for both the isothermal and
two-temperature RS models. In Figure \ref{fig.n1399_ray} the prominent
absorption feature in the residuals of the SIS data near 1.5 keV
(i.e. near Mg XII ly-$\alpha$) for the models with fixed relative
abundances can be partially compensated for by lowering the Mg
abundance to zero. However, as indicated by the residuals of the
corresponding variable relative abundance models this procedure does
not remove all of the residuals near 1.5 keV.

The best MEKAL model fits (i.e. 2T+BREM) are much better than any fit
using RS; this result agrees with the previous study of BF who
analyzed the {\sl ASCA} data of these galaxies using 1T and 2T
MEKAL/RS models. Moreover, the relative abundances of several elements
with respect to Fe are required to be very different from their solar
values for the best 1T+BREM and multitemperature RS fits whereas the
multitemperature fits with MEKAL models (except for NGC 4636) provide
acceptable fits for the relative abundances fixed at their solar
values.

Problems with older plasma codes like RS near 1.5 keV arising from
inaccurate 4-2 Fe L transitions were first suggested from analysis of
{\sl ASCA} data of the Centaurus cluster by Fabian et
al. \shortcite{acf_fel}. These authors were able to obtain
substantially better fits when the energies near 1.5 keV were
excluded. Indeed, for NGC 1399 and NGC 4472 which have the highest
temperatures in our sample and thus have the most important
contribution from the Fe L lines near 1.5 keV, we find significant
improvements in the RS fits if we exclude the energies 1.35-1.55
keV. In particular, the two-temperature models with fixed relative
abundances now are better fits than all the 1T+BREM models; e.g., for
NGC 1399 we obtain ($\chi^2$/dof/$P$ = 422.2/342/2.0e-3) for the
1T+BREM variable abundance model and (402.1/341/1.3e-2) for the
2T+BREM fixed relative abundance model. Note, however, that the
residuals near 1 keV are still present (as in Figure
\ref{fig.n1399_ray}) and thus even when the energies 1.35-1.55 keV are
excluded the resulting fits are not as good as with the MEKAL plasma
code.

\subsubsection{Overall assessment}
\label{ass}

The spectra of the galaxies in our sample generally are better
described by models with two thermal plasma components regardless of
whether the MEKAL or RS plasma code is employed in the fits; note that
it is the SIS data that provide the key constraints. The temperatures
derived for the two-temperature MEKAL and RS models are similar and
imply multiple phases of hot gas, although the precise values obtained
from the MEKAL and RS models differ to some degree as expected (see
section \ref{detailed}). Given the differences in the modeling of the
Fe L shell transitions between these codes, it is evident that such
differences are insufficient to cause a spectrum that is actually
1T+BREM to masquerade as a 2T or 2T+BREM system. 

The 2T and 2T+BREM models using the MEKAL code with the relative
abundances fixed at the solar values are superior fits to all 1T+BREM
MEKAL models. Moreover, these two-temperature MEKAL models provide
substantially better fits than the 1T+BREM, 2T, and 2T+BREM models
which employ the RS code, even when allowing the relative abundances
with respect to Fe to vary in the RS models. The key differences are
in the Fe L energy region ($\sim 0.7-1.4$ keV) where the best-fitting
two-temperature MEKAL models have negligible residuals (Figure
\ref{fig.n1399_multit}) in contrast to the substantial residuals
obtained for all best-fit RS models investigated (Figure
\ref{fig.n1399_ray}); i.e. the residuals arising from the known errors
in the Fe L emission near 1.5 keV of the RS code cannot be removed by
adding another temperature component.

Hence, these {\sl ASCA} data (particularly the SIS) demonstrate that
the MEKAL plasma code (as expected) is significantly more accurate
than the RS code for the important energies $\sim 0.7-1.4$ keV, and
the small residuals for the best-fitting two-temperature MEKAL models
imply that remaining inaccuracies in the Fe L energy region for the
MEKAL code are insufficient to change qualitatively the results for
data of the present quality.  Consequently, the results obtained for
the models using the MEKAL code should be given precedence over the
corresponding results using the RS code.

We thus conclude from the broad-band spectral analysis that (except
for possibly NGC 4636) isothermal (1T+BREM) models cannot provide
acceptable fits to the {\sl ASCA} data within the $r\sim 5\arcmin$
apertures of these galaxies.

\subsection{Detailed properties of models} 
\label{detailed}

\subsubsection{Calibration and systematic errors}
\label{calib}

\begin{figure*}
\centerline{\psfig{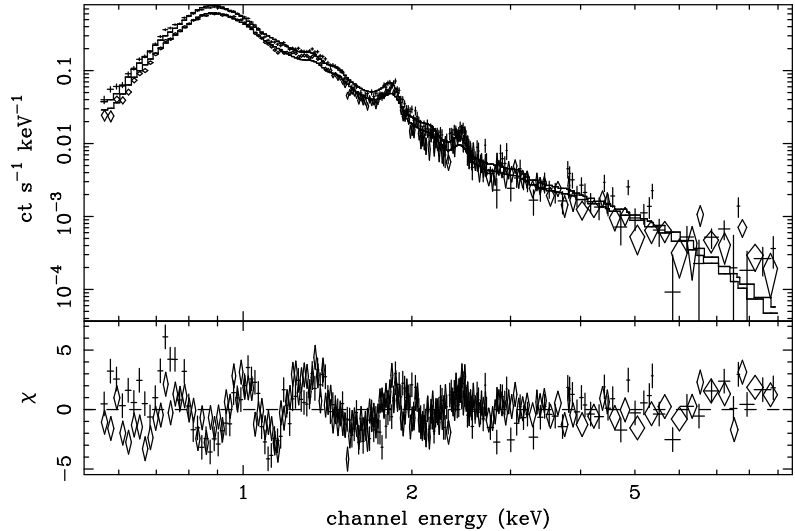}}
\caption{\label{fig.n4636_s0s1} The SIS0 (crosses) and SIS1 (diamonds)
data of NGC 4636 plotted along with the best-fitting 1T+BREM model
(relative abundances with respect to Fe fixed at solar) that is fitted
jointly to the two data sets. Note in particular the residuals of the
model below $\sim 0.8$ keV where the SIS0 deviations systematically
lie above those of the SIS1 in contrast to the excellent agreement
seen at other energies. This is evidence of a relative error in the
calibration of the SIS0 and SIS1 response similar to that described
for some other sources on the {\sl ASCA} GOF WWW pages. See text in
section \ref{calib} for details.}
\end{figure*}

We have examined the reliability of analyzing the summed SIS and
summed GIS data for the galaxies in our sample and find that the
best-fitting parameters we derived from fitting the summed data sets
generally differ by $<10\%$ from those derived from simultaneous
fitting of the available SIS0, SIS1, GIS2, and GIS3 data sets;
usually, the results agree to within a few percent.

For example, consider a 2T MEKAL model (no BREM) fitted to the NGC
1399 SIS data. When fitting the summed SIS data we obtain best-fitting
values of $Z=1.04Z_{\sun}$ for the metallicity and $N_{\rm H}^c =
3.1\times 10^{21}$ cm$^{-2}$ and $T_c=0.76$ (keV) for the colder
component and $N_{\rm H}^h = 0.4\times 10^{21}$ cm$^{-2}$ and
$T_h=1.72$ (keV) for the hotter component. If instead we
simultaneously fit the SIS0 and SIS1 data for the two observation
sequences listed in Table \ref{tab.obs} (i.e. a total of four data
sets) we obtain best-fitting values of $Z=1.07Z_{\sun}$ for the
metallicity and $N_{\rm H}^c = 3.1\times 10^{21}$ cm$^{-2}$ and
$T_c=0.76$ (keV) for the colder component and $N_{\rm H}^h = 0.4\times
10^{21}$ cm$^{-2}$ and $T_h=1.72$ (keV) for the hotter
component. Hence, the potential inaccuracies involved in combining the
individual SIS0/SIS1 and GIS2/GIS3 data appear to be unimportant for
the data of the elliptical galaxies in our present study.

We also find this level of agreement with our previous study (BF)
which included {\sl ASCA} data of these galaxies where the
simultaneous fitting of the available SIS0 and SIS1 data was
emphasized. In BF we tied together the values of $N_{\rm H}$ for the
colder and hotter components of two-temperature models. However, we
have examined the actual data sets that were reduced and analyzed by
BF and have obtained results consistent with those stated above when
untying the $N_{\rm H}$ values. That is, the differences in
calibration and data reduction between our study and that of BF do not
result in significant differences in the best-fit column densities,
temperatures, and abundances.

We do find some evidence for errors in the calibration of the photon
energies for the SIS and GIS data. For the galaxies in our sample the
fits are generally improved when the redshifts of the SIS and GIS data
are allowed to be free parameters. For the example above of the 2T
model fitted to the summed SIS and summed GIS data of NGC 1399, the
value of $\chi^2$ is lowered by 35 (372 dof) when the redshifts are
allowed to be free parameters. In this case the best-fitting redshift
for the SIS is zero and for the GIS $\sim 0.03$ which implies energy
shifts of $\sim 10$ eV at 2 keV for the SIS and $\sim 50$ eV for the
GIS. (The magnitude of these shifts are not overly sensitive to the
model.)  These shifts are due to known calibration problems. (See the
{\sl ASCA} GOF WWW pages.) In particular the relatively large shift
for the GIS is the result of the energy scale of the GIS RMF being
slightly wrong below the Xe-L edge (4.8 keV). (Note that the
magnitudes of these shifts are similar if we instead simultaneously
fit the SIS0, SIS1, GIS2, and GIS3 data.)

Hence, to account for these (small) calibration errors in the energy
scale of the RMFs we always allow the redshifts to be free parameters
in our fits; i.e. the model components fitted to the SIS data have one
redshift, and those components fitted to the GIS data have a different
redshift. Once the best-fitting model is found we freeze the redshifts
at their best-fit values for determining error bars on other
interesting parameters. This is done to speed up the computations and
has no discernible effect on the sizes of the error bars obtained for
the other parameters.

For NGC 4636 we find important differences between the SIS0 and SIS1
data at low energies that are not apparent in the data of the other
galaxies. In Figure \ref{fig.n4636_s0s1} we plot the SIS0 and SIS1
spectral data of NGC 4636 along with the best-fitting 1T+BREM model
obtained from joint fitting of the SIS0 and SIS1 data (relative
abundances fixed at solar; BREM component modified by Galactic
absorption). Above $\sim 0.8$ keV the residuals of the fit are similar
for the SIS0 and SIS1 data as would be expected. However, there is a
noticeable discrepancy below $\sim 0.8$ keV where the residuals of the
SIS0 fit lie above those of the SIS1.

These differences at low energies translate to slightly different
column densities inferred by the SIS0 and SIS1. If the SIS0 and SIS1
are fitted separately with the 1T+BREM model, we obtain $N_{\rm H} =
(1.1,1.5)\times 10^{21}$ cm$^{-2}$ for the (SIS0,SIS1) respectively
with statistical uncertainties of $\sim 10\%$ percent. We are unable
to affect these low energy differences by changing the event selection
criteria of the data. Most importantly, no effect is seen when
selecting data with minimal contamination from solar X-rays (i.e. mkf
parameter ``sunshine=0'').

Differences in $N_{\rm H}$ of $\sim 0.2\times 10^{21}$ cm$^{-2}$
between the SIS0 and SIS1 are expected due to remaining calibration
errors in the low energy response of the SIS (see Orr et al. 1998 and
the {\sl ASCA} GOF WWW pages).  Although the differences in $N_{\rm
H}$ for NGC 4636 are about a factor of 2 larger, the discrepancy is
reduced to expected levels when the relative abundances are allowed to
be free parameters. We have verified that the PV-phase data analyzed
by BF also shows these discrepancies, but the effect is of lower
significance because of the lower S/N and, presumably, because these
calibration errors are smaller for the data taken during the PV-phase
({\sl ASCA} GOF WWW pages).

However, it should be emphasized that the Fe abundance is also
affected by these differences in $N_{\rm H}$; i.e.  $Z_{\rm Fe}\approx
0.3Z_{\sun}$ for the model fitted to the SIS0 data and $Z_{\rm
Fe}\approx 0.5Z_{\sun}$ for the SIS1 data. If the relative abundances
are varied, this discrepancy for Fe remains but the ratios of other
elements to Fe (e.g. Si/Fe) are the same for the SIS0 and SIS1 within
statistical errors. The results for these parameters listed below
using the summed SIS data thus reflect an average of the results from
the SIS0 and SIS1 data.

\subsubsection{Isothermal models} 
\label{1t}

\begin{table*}
\begin{minipage}{180mm}
\caption{Parameters of Isothermal and Two-Temperature Models}
\label{tab.params}
\begin{tabular}{rcccccccc}
& \multicolumn{2}{c}{NGC 1399} & \multicolumn{2}{c}{NGC 4472} &
\multicolumn{2}{c}{NGC 5044} & \multicolumn{2}{c}{NGC 4636}\\
& MEKAL & RS & MEKAL & RS & MEKAL & RS & MEKAL & RS\\ \\ \\
\multicolumn{1}{l}{Isothermal Model}\\ \\[-5pt]
\multicolumn{1}{c}{1T:}\\ \\[-7pt]
$N_{\rm H}$ ($10^{21}$ cm$^{-2}$) & $0.45_{-0.14}^{+0.13}$ &
$0.66_{-0.16}^{+0.19}$ & $0.37_{-0.15}^{+0.17}$ &
$0.72_{-0.26}^{+0.28}$ & $1.74_{-0.42}^{+0.46}$ &
$0.74_{-0.18}^{+0.22}$ & $0.24_{-0.19}^{+0.19}$ & $0.43_{-0.12}^{+0.13}$\\ 
$T$ (keV) & $1.25_{-0.02}^{+0.02}$ & $1.11_{-0.01}^{+0.01}$ &
$1.04_{-0.02}^{+0.02}$ & $1.02_{-0.02}^{+0.02}$ &
$0.70_{-0.02}^{+0.02}$ & $0.95_{-0.02}^{+0.02}$ &
$0.66_{-0.01}^{+0.01}$ & $0.66_{-0.01}^{+0.01}$\\
$EM_{\rm sis}$ & $10.37_{-0.75}^{+0.66}$ & $10.41_{-1.54}^{+0.94}$ &
$10.56_{-1.01}^{+1.05}$ & $10.29_{-1.47}^{+0.97}$ &
$23.79_{-14.16}^{+9.72}$ & $25.48_{-3.06}^{+3.45}$
& $2.93_{-1.18}^{+1.11}$ &  $8.88_{-0.60}^{+0.63}$\\
$EM_{\rm gis}$ & $15.80_{-1.12}^{+1.02}$ & $15.88_{-2.34}^{+1.42}$ &
$15.36_{-1.50}^{+1.52}$ & $15.54_{-2.27}^{+1.42}$ &
$33.64_{-20.12}^{+14.36}$ & $38.43_{-5.02}^{+5.69}$ &
$3.79_{-1.52}^{+1.43}$ & $11.17_{-0.73}^{+0.76}$\\
Fe & $0.35_{-0.03}^{+0.03}$ & $0.43_{-0.05}^{+0.08}$ &
$0.28_{-0.03}^{+0.03}$ & $0.41_{-0.05}^{+0.08}$ &
$0.22_{-0.07}^{+0.35}$ & $0.31_{-0.03}^{+0.04}$ &
$0.80_{-0.21}^{+0.48}$ & $0.16_{-0.02}^{+0.02}$\\
O  & $0.00_{-0.00}^{+0.05}$ & $0.00_{-0.00}^{+0.15}$ &
$0.00_{-0.00}^{+0.06}$ & $0.09_{-0.09}^{+0.23}$ &
$0.24_{-0.18}^{+0.47}$ & $0.00_{-0.00}^{+0.12}$ &
$0.46_{-0.16}^{+0.29}$ & $0.14_{-0.05}^{+0.06}$\\
Ne & $0.00_{-0.00}^{+0.07}$ & $0.03_{-0.03}^{+0.26}$ &
$0.00_{-0.00}^{+0.07}$ & $0.00_{-0.00}^{+0.16}$ &
$1.18_{-0.36}^{+1.78}$ & $0.00_{-0.00}^{+0.05}$ &
$2.16_{-0.62}^{+1.44}$ & $0.00_{-0.00}^{+0.03}$\\
Mg & $0.01_{-0.01}^{+0.13}$ & $0.00_{-0.00}^{+0.02}$ &
$0.10_{-0.10}^{+0.10}$ & $0.00_{-0.00}^{+0.10}$ &
$0.36_{-0.14}^{+0.61}$ & $0.15_{-0.06}^{+0.07}$ &
$1.83_{-0.54}^{+1.28}$ & $0.30_{-0.04}^{+0.04}$\\
Si & $0.45_{-0.07}^{+0.08}$ & $0.43_{-0.07}^{+0.08}$ &
$0.40_{-0.07}^{+0.07}$ & $0.40_{-0.06}^{+0.09}$ &
$0.50_{-0.15}^{+0.72}$ & $0.30_{-0.05}^{+0.07}$ &
$1.69_{-0.47}^{+1.13}$ & $0.50_{-0.04}^{+0.05}$\\
S  & $0.40_{-0.10}^{+0.11}$ & $0.44_{-0.11}^{+0.12}$ &
$0.48_{-0.12}^{+0.11}$ & $0.49_{-0.12}^{+0.14}$ &
$0.65_{-0.22}^{+0.86}$ & $0.28_{-0.08}^{+0.09}$ &
$2.42_{-0.32}^{+0.67}$ & $0.83_{-0.11}^{+0.12}$\\
Na & $\cdots$ & $\cdots$ & $\cdots$ & $\cdots$ & $\cdots$ & $\cdots$ &
$21_{-7}^{+16}$ & $\cdots$\\
Ni & $\cdots$ & $\cdots$ & $\cdots$ & $\cdots$ &
$3.17_{-1.06}^{+5.10}$ & $0.53_{-0.15}^{+0.18}$ &
$0.00_{-0.00}^{+0.26}$ & $1.96_{-0.17}^{+0.19}$\\
\multicolumn{1}{c}{BREM:}\\ \\[-7pt]
$T$ (keV) & $>14$  & $>6$ & $>7$ & $>6$ & $1.93_{-0.49}^{+0.97}$ &
$3.58_{-1.28}^{+3.31}$ & $5.22_{-1.29}^{+2.58}$ & $>17$\\
$EM_{\rm sis}$ & $0.11_{-0.03}^{+0.03}$ & $0.12_{-0.03}^{+0.08}$ &
$0.07_{-0.03}^{+0.05}$ & $0.08_{-0.04}^{+0.04}$ &
$0.59_{-0.30}^{+0.62}$ & $0.07_{-0.04}^{+0.07}$ &
$0.07_{-0.01}^{+0.02}$ & $0.08_{-0.01}^{+0.01}$\\
$EM_{\rm gis}$ & $0.20_{-0.05}^{+0.05}$ & $0.18_{-0.04}^{+0.12}$ &
$0.15_{-0.05}^{+0.07}$ & $0.16_{-0.07}^{+0.05}$ &
$1.12_{-0.51}^{+0.96}$ & $0.21_{-0.10}^{+0.18}$ &
$0.09_{-0.02}^{+0.02}$ & $0.08_{-0.01}^{+0.01}$\\ \\ \\
\multicolumn{1}{l}{Two-Temperature Model}\\ \\[-5pt]
\multicolumn{1}{c}{2T:}\\ 
\multicolumn{1}{c}{(colder+hotter)}\\ \\[-7pt]
$N_{\rm H}^{\rm c}$ ($10^{21}$ cm$^{-2}$) & $4.91_{-0.89}^{+0.64}$ &
$5.40_{-4.36}^{+1.85}$ & $2.86_{-0.93}^{+0.62}$ &
$5.93_{-1.53}^{+3.69}$ & $2.47_{-0.61}^{+0.51}$ &
$7.79_{-2.61}^{+2.21}$ & $0.02_{-0.01}^{+0.01}$ & $0.81_{-0.39}^{+0.41}$\\
$N_{\rm H}^{\rm h}$ ($10^{21}$ cm$^{-2}$) & $0.07_{-0.07}^{+0.32}$ &
$0.34_{-0.34}^{+0.29}$ & $0.04_{-0.04}^{+0.70}$ & $0.16_{-0.16}^{+0.51}$ &
$0.63_{-0.36}^{+0.42}$ & $0.51_{-0.38}^{+0.29}$ &
$0.24_{-0.24}^{+0.41}$ & $4.78_{-0.20}^{+0.44}$\\
$T_{\rm c}$ (keV) & $0.69_{-0.03}^{+0.03}$ & $0.71_{-0.07}^{+0.27}$ &
$0.72_{-0.03}^{+0.03}$ & $0.71_{-0.27}^{+0.25}$ &
$0.70_{-0.03}^{+0.02}$ & $0.56_{-0.10}^{+0.12}$ &
$0.52_{-0.06}^{+0.05}$ & $0.24_{-0.01}^{+0.01}$\\
$T_{\rm h}$ (keV) & $1.54_{-0.07}^{+0.08}$ & $1.27_{-0.06}^{+0.30}$ &
$1.39_{-0.09}^{+0.11}$ & $1.06_{-0.03}^{+0.02}$ &
$1.23_{-0.06}^{+0.07}$ & $0.97_{-0.02}^{+0.02}$ &
$0.76_{-0.03}^{+0.03}$ & $0.71_{-0.01}^{+0.01}$\\
$EM_{\rm sis}^{\rm c}$ & $2.49_{-0.71}^{+0.82}$ &
$3.67_{-1.64}^{+2.97}$ & $1.67_{-0.79}^{+0.77}$ & 
$4.22_{-1.45}^{+1.60}$ & $12.92_{-2.85}^{+3.08}$ &
$15.32_{-8.77}^{+26.37}$ & $1.75_{-0.45}^{+0.37}$ & $4.29_{-1.31}^{+2.60}$\\
$EM_{\rm gis}^{\rm c}$ & $2.92_{-1.38}^{+1.46}$ &
$5.14_{-4.27}^{+5.12}$ & $3.15_{-1.15}^{+1.64}$ &
$4.79_{-1.23}^{+6.64}$ & $10.05_{-8.07}^{+6.11}$ &
$11.71_{-11.71}^{+39.57}$ & $0.00_{-0.00}^{+0.55}$ & $5.34_{-1.57}^{+3.20}$\\
$EM_{\rm sis}^{\rm h}$ & $2.32_{-0.70}^{+0.81}$ &
$6.65_{-2.09}^{+1.21}$ & $1.30_{-0.69}^{+1.00}$ &
$4.79_{-1.69}^{+0.98}$ & $9.63_{-1.62}^{+1.88}$ &
$19.22_{-6.70}^{+4.25}$ & $2.37_{-0.55}^{+0.60}$ & $4.33_{-1.03}^{+1.22}$\\
$EM_{\rm gis}^{\rm h}$ & $4.39_{-1.40}^{+1.63}$ &
$10.78_{-3.33}^{+2.32}$ & $1.54_{-0.72}^{+1.25}$ &
$6.92_{-2.62}^{+1.69}$ & $21.56_{-4.45}^{+5.72}$ &
$30.67_{-11.04}^{+8.73}$ & $4.69_{-1.29}^{+1.33}$ & $5.81_{-1.29}^{+1.61}$\\
Fe  & $1.62_{-0.61}^{+0.74}$ & $0.61_{-0.10}^{+0.25}$ &
$2.00_{-0.97}^{+2.40}$ & $0.61_{-0.13}^{+0.30}$ &
$0.57_{-0.08}^{+0.10}$ & $0.41_{-0.06}^{+0.18}$ &
$0.73_{-0.16}^{+0.28}$ & $0.99_{-0.29}^{+0.45}$\\
O   & $\cdots$ & $0.10_{-0.10}^{+0.37}$ & $\cdots$ &
$0.25_{-0.25}^{+0.47}$& $\cdots$ & $0.11_{-0.11}^{+0.20}$ &
$0.39_{-0.14}^{+0.23}$ & $0.14_{-0.05}^{+0.04}$\\
Ne  & $\cdots$ & $0.76_{-0.44}^{+0.53}$ & $\cdots$ &
$0.28_{-0.26}^{+0.44}$& $\cdots$ & $0.00_{-0.00}^{+0.15}$ &
$0.92_{-0.28}^{+0.26}$ & $2.42_{-0.74}^{+1.11}$\\
Mg  & $\cdots$ & $0.00_{-0.00}^{+0.16}$ & $\cdots$ &
$0.18_{-0.11}^{+0.19}$& $\cdots$ & $0.21_{-0.07}^{+0.14}$ &
$1.40_{-0.35}^{+0.57}$ & $1.25_{-0.36}^{+0.51}$\\
Si  & $\cdots$ & $0.55_{-0.11}^{+0.13}$ & $\cdots$ &
$0.52_{-0.12}^{+0.21}$& $\cdots$ & $0.36_{-0.07}^{+0.10}$ &
$1.24_{-0.29}^{+0.46}$ & $1.20_{-0.30}^{+0.40}$\\
S   & $\cdots$ & $0.55_{-0.13}^{+0.18}$ & $\cdots$ &
$0.68_{-0.18}^{+0.29}$& $\cdots$ & $0.36_{-0.10}^{+0.11}$ &
$1.64_{-0.36}^{+0.46}$ & $1.46_{-0.19}^{+0.40}$\\
Na  & $\cdots$ & $\cdots$ & $\cdots$ & $\cdots$ & $\cdots$ & $\cdots$
& $15_{-4}^{+5}$ & $\cdots$\\
Ni  & $\cdots$ & $\cdots$ & $\cdots$ & $\cdots$ & $\cdots$ &
$0.83_{-0.24}^{+0.34}$ & $0.00_{-0.00}^{+0.15}$ & $4.92_{-1.30}^{+0.99}$\\
\multicolumn{1}{c}{BREM:}\\ \\[-7pt]
$T$ (keV) & $5(>3)$ & $>5$ & $3.23_{-0.77}^{+1.35}$ &
$5.46_{-1.75}^{+3.21}$ & $\cdots$ & $2.65_{-0.70}^{+1.29}$ & $9>5$ & $>12$\\
$EM_{\rm sis}$  & $0.16_{-0.07}^{+0.23}$ & $0.10_{-0.03}^{+0.08}$ &
$0.18_{-0.05}^{+0.06}$ & $1.59_{-0.56}^{+0.33}$ & $\cdots$ &
$0.15_{-0.07}^{+0.11}$ & $0.06_{-0.01}^{+0.02}$ & $0.05_{-0.01}^{+0.03}$\\
$EM_{\rm gis}$ & $0.21_{-0.11}^{+0.32}$ & $0.16_{-0.04}^{+0.12}$ &
$0.32_{-0.15}^{+0.16}$ & $2.29_{-0.87}^{+0.56}$ & $\cdots$ &
$0.38_{-0.16}^{+0.24}$ & $0.06_{-0.01}^{+0.02}$ & $0.05_{-0.01}^{+0.03}$\\

\end{tabular}

\medskip

Best-fitting values and 90\% confidence limits on one interesting
parameter $(\Delta\chi^2=2.71)$ are listed for selected isothermal
and two-temperature models jointly fit to the SIS and GIS data. We list
results for each model using both the MEKAL and Raymond-Smith (RS)
plasma codes. (See Table \ref{tab.quality} for the $\chi^2$ values of
these models.) The isothermal models are 1T+BREM models with
variable relative abundances. Only the abundances listed with error
estimates are free parameters; the remaining abundances (not listed)
and those with ``$\cdots$'' are tied to Fe. All abundances are quoted
in units of their (photospheric) solar values \cite{ag}.  The emission
measures $(EM)$ are quoted in units of $10^{-17}n_en_pV/4\pi D^2$
similar to what is done in {\sc XSPEC}. (We have converted the BREM
emission measures to this standard as well.)  The BREM component is
also modified by photo-electric absorption fixed at the Galactic value
(see Table \ref{tab.prop}). For the two-temperature models the abundances
for the colder and hotter temperature components are tied together in
the fits.

\end{minipage}
\end{table*}

\begin{figure*}
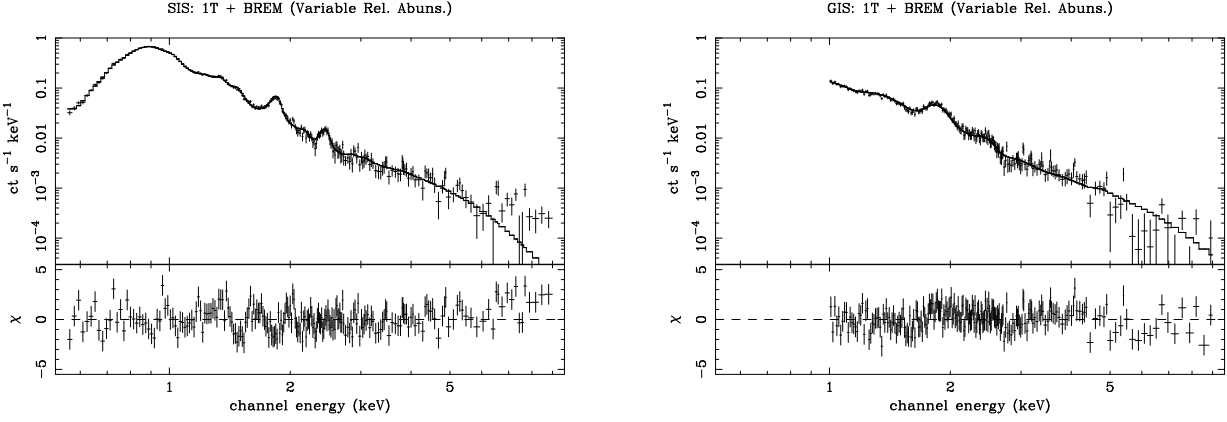

\parbox{0.49\textwidth}{
\centerline{\psfig{figure=fig6a.ps,angle=-90,height=0.23\textheight}}
}
\parbox{0.49\textwidth}{
\centerline{\psfig{figure=fig6b.ps,angle=-90,height=0.23\textheight}}
}
\caption{\label{fig.n4636_1t} Best-fitting 1T+BREM (MEKAL) model with
variable relative abundances for NGC 4636. The models are fitted
jointly to the SIS and GIS data, and for display purposes only the
energy bins above $\sim 5$ keV have been re-binned such that S/N $> 3$
in each bin. The parameters of these plots are listed in Tables
\ref{tab.params} and \ref{tab.cf}.}

\end{figure*}

\begin{figure*}
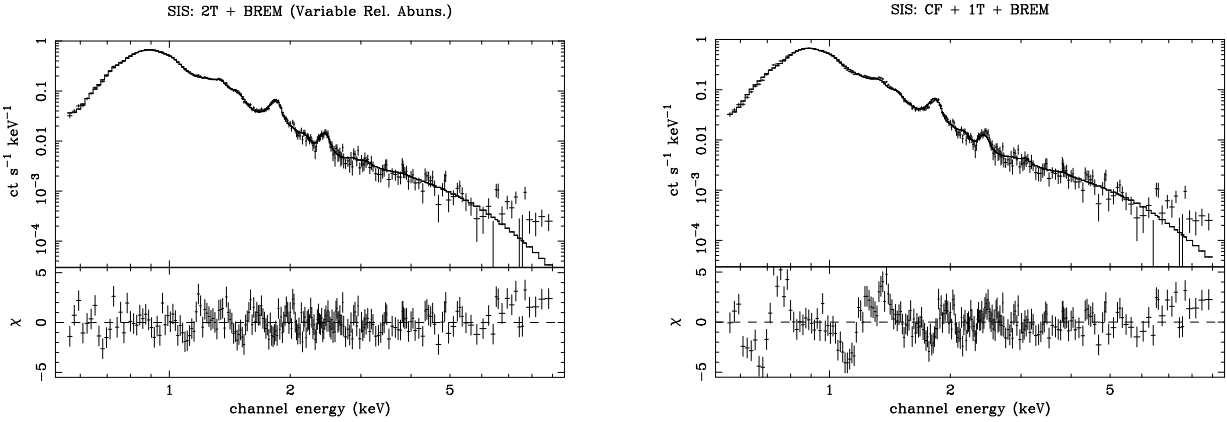

\parbox{0.49\textwidth}{
\centerline{\psfig{figure=fig7a.ps,angle=-90,height=0.23\textheight}}
}
\parbox{0.49\textwidth}{
\centerline{\psfig{figure=fig7b.ps,angle=-90,height=0.23\textheight}} 
}
\caption{\label{fig.n4636_multit} As Figure \ref{fig.n4636_1t} except
the 2T+BREM and CF+1T+BREM models are displayed. Although these models
were fitted jointly to the SIS and GIS data, only the SIS data are
shown.}

\end{figure*}

Recall that our isothermal model (1T+BREM) actually has two
components where the first component is an isothermal coronal plasma
(MEKAL or RS) modified by variable photo-electric absorption, and the
second component is thermal bremsstrahlung modified by absorption
fixed at the appropriate Galactic value. We shall focus on the models
with variable relative abundances since they provided superior fits
for each galaxy (see Table \ref{tab.quality} in section
\ref{quality}). However, in most cases the results we obtain for the
temperatures (plasma and bremsstrahlung), column densities, and Fe
abundances of these variable abundance models are quite similar to the
case where the relative abundances are fixed at their solar values.

Our procedure for varying the relative abundances with respect to Fe
begins with the best-fitting model with relative abundances fixed at
the solar values. Then we proceed to untie the abundances of the
$\alpha$-process elements Si, S, Mg, Ne, and O; a new best-fit model
is found after each element is freed. After fitting these parameters
we examined whether untying any other elements improved the fits.

In Table \ref{tab.params} we list the best-fitting parameters and 90\%
confidence limits determined for these 1T+BREM models.  We again refer
the reader to Figures \ref{fig.n1399_1t} and \ref{fig.n1399_ray}
for plots of the 1T+BREM models for NGC 1399 and to Figure
\ref{fig.n4636_1t} for NGC 4636. Related single-temperature fits to the
SIS data of NGC 4472 and NGC 5044 are presented in Figure 1 of
BF. (The 1T MEKAL models presented in Figure 1 of BF have relative
abundances fixed at solar and do not include a BREM
component. However, the pattern of residuals is very similar to the
1T+BREM variable relative abundance models.)

The galaxies NGC 1399, NGC 4472, and NGC 5044 have similar spectral
properties. Each has $T\sim 1$ keV for the plasma component and,
except for NGC 5044, has a bremsstrahlung temperature at least as high
as 6 keV. These high bremsstrahlung temperatures are consistent with
emission from discrete sources in these systems (e.g. Canizares et
al. 1987; Kim et al. 1992).  The hot plasma components dominate these
systems with the bremsstrahlung components contributing only $\sim
0.1\%$ to the total emission measures.

In the case of NGC 5044, the bremsstrahlung temperature is $T\approx
2$ keV which is inconsistent with emission from discrete sources and
rather indicates a contribution from another plasma component in NGC
5044. For the 1T+BREM MEKAL model of NGC 5044 the larger relative
contribution $(\sim 1\%)$ of this bremsstrahlung component to the
total emission measure also supports a plasma origin since NGC 5044,
which has a larger $L_{\rm x}/L_{\rm B}$ ratio than the other
galaxies, is expected to have a smaller relative contribution to its
X-ray emission by discrete sources (e.g. Kim et al. 1992; BF).

Excess absorption above the Galactic value is also indicated for these
three galaxies. For NGC 1399 and NGC 4472 the modest excess $N_{\rm
H}$ implied by the MEKAL model is approximately 0.2-$0.3\times
10^{21}$ cm$^{-2}$ which is probably significant within the
uncertainties in the low energy calibration of the SIS as discussed in
section \ref{calib}. (The columns inferred using the RS model are
within a factor of 2 of these values.) However, for NGC 5044 the
inferred (MEKAL) $N_{\rm H}$ exceeds the Galactic value by over
$1\times 10^{21}$ cm$^{-2}$. These values of $N_{\rm H}$ depend on the
model, but excess absorption in the brightest ellipticals appears to
be common (BF).

These results for the temperatures and column densities generally
agree with previous studies that have fit similar 1T and/or 1T+BREM
models to the SIS and GIS data of these galaxies (or 2T models to only
the GIS data); e.g. Awaki et al. 1994; Matsushita et al. 1994;
Fukazawa et al. 1996; Arimoto et al. 1997; Matsumoto et al. 1997;
BF. Similar agreement is found for the Fe abundances.  We find that
$Z_{\rm Fe}\sim 0.3 (<0.5)Z_{\sun}$ for NGC 1399, NGC 4472, and NGC
5044 for these models, and we again remark that these values are
essentially unchanged for models where the relative abundances of the
elements are fixed at their solar values.

The Si and S abundances are consistent with Fe for NGC 1399 and the RS
models of NGC 4472 and NGC 5044, but they exceed Fe for the MEKAL
models of NGC 4472 and NGC 5044. The Si and S abundances obtained for
the RS models are consistent with previous published results for NGC
4472 \cite{awaki} and NGC 5044 \cite{fuk} considering the small
differences in calibration, extraction regions, and background between
those studies and ours.

In contrast to previous studies, we find that varying each of the O,
Ne, and Mg abundances has an important contribution to lowering
$\chi^2$ and each abundance is, for the most part, very different from
Fe. The O, Ne, and Mg abundances are approximately zero for NGC 1399
and NGC 4472. This situation is similar for NGC 5044 for the RS case,
but when the plasma component is a MEKAL model the O and Mg abundances
are essentially consistent with the Fe abundance. The Ne abundance,
however, is about 6 times larger than Fe.

Because these models are poor fits to the SIS data around 1 keV for
NGC 1399, NGC 4472, and NGC 5044, these abundances unlikely have
physical meaning. However, as we discussed in section \ref{quality}
errors in the plasma codes are not the key reason for the poor
fits. We can illustrate this further by considering the Mg abundance.

As we mentioned at the end of section \ref{rs}, a serious error in the
RS (and old MEKA) plasma code near the strong Mg transitions ($\sim
1.5$ keV) was noted first by Fabian et al. \shortcite{acf_fel}. The
MEKAL code includes new computations of the 4-2 transitions of Fe to
correct these errors. However, we find that the Mg abundance is zero
for both the RS and MEKAL isothermal models of NGC 1399 and NGC
4472. If indeed a zero Mg abundance is unphysical, then a different
model for the temperature structure is probably required rather than
more precise atomic physics calculations.

For NGC 5044 the situation is complicated because we find that the
fits are substantially improved when Ni is allowed to be a free
parameter for the MEKAL case though only modestly improved for
RS. Although the implied large Ni abundance $(Z_{\rm Ni}\approx
3Z_{\sun})$ could indicate other inaccuracies in the MEKAL plasma
code, the $\sim 2$ keV bremsstrahlung temperature already shows that a
model of an isothermal plasma with discrete sources is
inadequate. Thus, like NGC 1399 and NGC 4472, the abundances and
temperatures for NGC 5044 themselves show the inadequacy of the
1T+BREM model as already indicated by the poor quality of the fits
described in section \ref{quality}.

NGC 4636 has a lower plasma temperature ($\approx 0.66$ keV) than the
other galaxies in our sample which is similar for both the MEKAL and
RS models. However, of all the galaxies in our sample NGC 4636 shows
the most dramatic difference in the quality of the fits between models
using the MEKAL code and those using the RS code (see Table
\ref{tab.quality}). For the 1T+BREM model with variable relative
abundances we obtained a decent fit $(P\sim 10^{-4})$ using the MEKAL
code whereas a totally unacceptable fit $(P \sim 10^{-40})$ was
obtained for the RS code. Residuals are still prominent for the MEKAL
case in both the SIS and GIS data (see Figure \ref{fig.n4636_1t}), and
we shall consider these below.

The inferred column densities for the 1T+BREM model of NGC 4636 imply
a slight excess above the Galactic value that is consistent with the
Galactic value within the errors in the low-energy calibration of the
SIS. The $\sim 5$ keV bremsstrahlung temperature of the MEKAL model
can be reasonably attributed to discrete sources (as is the case for
the 1T+BREM RS model). These results for the temperatures of the
plasma and bremsstrahlung components and for the column densities are
are not substantially different if the relative abundances are fixed
at their solar values in the fits; e.g. plasma temperatures increase
to $\sim 0.75$ keV -- see section \ref{lines}.

The abundances differ substantially for the MEKAL and RS 1T+BREM
models of NGC 4636. The abundances of most of the elements determined
using the MEKAL model greatly exceed those determined using RS. The Fe
abundance is $\approx 0.8$ solar for the MEKAL model versus $\approx
0.2$ solar for the RS model. (Note that the MEKAL Fe abundance is
lowered to $\approx 0.6$ solar if the relative abundances are fixed at
their solar values.) However, the relative abundances of Mg, Si, and S
are not too dissimilar for the MEKAL and RS models; i.e. Mg/Fe, Si/Fe
$\sim 2$ and S/Fe $\sim 3$ for MEKAL while Mg/Fe $\sim 2$, Si/Fe $\sim
3$ and S/Fe $\sim 5$ for RS. The O abundance is consistent with the Fe
abundance for RS but is about half that of Fe for MEKAL.

In contrast, the Ne, Na, and Ni abundances have striking qualitative
differences for the MEKAL and RS 1T+BREM models for NGC 4636. First,
Ne is zero for RS while about twice solar for MEKAL. Second, the fits
of both models are significantly improved when allowing the Ni
abundance to be a free parameter, but the resulting value is very
different in both cases: zero Ni abundance for MEKAL and $\sim
2Z_{\sun}$ for RS. Finally, the MEKAL fit is also substantially
improved if the Na abundance is allowed to be a free parameter with a
best fit value of $\sim 21Z_{\sun}$; RS does not have any Na lines in
its code.

Our results for NGC 4636 are not inconsistent with those of Matsushita
et al. \shortcite{matsushita} who analyzed the same deep {\sl ASCA}
observation. These authors fit 1T+BREM models jointly to the SIS and
GIS data where: (1) the bremsstrahlung temperature was fixed at 10
keV; (2) the abundances of Fe and Ni were tied together and fitted as
a free parameter; (3) the abundances of O, Ne, Mg, Si, and S (and the
rest of the elements) were tied together and fitted as one free
parameter. They were unable to obtain an acceptable fit with any
plasma model (e.g. MEKAL and RS) to the Fe L energy region using this
method. This is consistent with our results since we required separate
fitting of several of the abundances they tied together to obtain an
acceptable fit.

Matsushita et al.'s solution was instead to add a systematic error to
the data in the Fe L region. (This is tantamount to excluding the Fe L
data -- a procedure we do not favor because (1) the Fe L lines are the
most important for constraining the temperatures and Fe abundances of
ellipticals, and (2) we are able to obtain good fits for the other
galaxies with the Fe L included.) Upon excluding the Fe L region they
obtained acceptable fits for the MEKAL model (and other models) with
Fe and Ni abundances of $\sim 0.9Z_{\sun}$ and an abundance of $\sim
1.1Z_{\sun}$ for the rest. These abundances (aside from Ni) are
consistent with our results in Table \ref{tab.params} if the
$\alpha$-process abundances are tied together in the fits. The
principal difference between our results and those of Matsushita et
al. is that we have narrowed down the locations of candidate
problematic emission lines in the Fe L regions: i.e. energies near the
strongest O, Na, and Ni lines. Below in section \ref{fel} we discuss
the implications of these abundances for the plasma codes.

\subsubsection{Two-temperature models} 
\label{2t}

Next we consider the two-temperature models consisting of two thermal plasma
components (each modified by their own photo-electric absorption) with
or without a thermal bremsstrahlung component (modified by Galactic
absorption) to account for emission from discrete sources. The
abundances of the two plasma components are tied together as we did
not obtain significant improvement in the fits when doing
otherwise. We initially determined the best-fitting 2T or 2T+BREM
models with the relative abundances fixed at their solar values; for
the 2T+BREM case the BREM component is added after the best 2T model
is obtained. Then, analogously to the isothermal models, the
abundances of the $\alpha$-process elements (and others) are untied
sequentially until the best fit was achieved.

Recall from section \ref{quality} (Table \ref{tab.quality}) that the
2T MEKAL models with relative abundances fixed at their solar values
are substantially better fits than isothermal (1T+BREM) models for NGC
1399, NGC 4472, and NGC 5044. The temperatures for the ``colder''
components are typically $T_{\rm c}\sim 0.7$ keV and for the
``hotter'' components $T_{\rm h}\sim 1.2$-1.6 keV for the MEKAL
models. These values indicate different temperatures in the hot gas
and are inconsistent with the temperatures expected for discrete
sources in agreement with the results of BF for these galaxies.

If a bremsstrahlung component is added to these galaxies there is some
small improvement in the fits for NGC 1399 and NGC 4472 though
essentially no improvement for NGC 5044 (Table \ref{tab.quality}). We
list the 2T+BREM model parameters for NGC 1399 and NGC 4472 in Table
\ref{tab.params}. For NGC 5044 we list the 2T model parameters for the
MEKAL plasma code because the bremsstrahlung component is not clearly
required by the fits; however, for illustration, we do give the
2T+BREM parameters for the RS code.

The values of $T_{\rm c}$ are mostly unaffected by the addition of the
bremsstrahlung component, though the values of $T_{\rm h}$ decrease by
$\sim 0.2$ keV for NGC 1399 and NGC 4472. For NGC 1399 the temperature
of the bremsstrahlung component is consistent with discrete sources,
but the $\sim 3$ keV temperature (MEKAL) for NGC 4472 is perhaps too
low and may indicate additional temperature structure in the hot gas.

The column densities for the colder and hotter components of NGC 1399,
NGC 4472, and NGC 5044 are very similar. Excess absorption above the
Galactic value is strongly indicated for the colder components in
these galaxies with $N_{\rm H}$ ranging from about 2-$5\times 10^{21}$
cm$^{-2}$. In contrast, the absorption of the hotter components is
small and consistent with the Galactic value for each of these
galaxies. 

This distinction in column densities between the colder and hotter
components for elliptical galaxies has not been reported
previously. (In BF we tied together the columns of the colder and
hotter components, though upon re-examining the data analyzed by BF we
have verified the differences between the two components for these
galaxies.)  This pattern of excess intrinsic absorption on the colder
gas which is presumably more centrally concentrated than the hotter
gas (having only Galactic absorption) is similar to what is found in
the cores of some galaxy clusters with strong cooling flows
(e.g. Fabian 1994).

The emission measures of the colder and hotter components of NGC 1399,
NGC 4472, and NGC 5044 are approximately equal such that $0.5\la \rm
EM^c/EM^h \la 1$. Although for NGC 1399 and NGC 5044 $EM^c/EM^h$ is
larger for the SIS as would be expected because it is more sensitive
to energies $\sim 0.5-1$ keV, this is not the case for NGC 4472.  That
is, there is no systematic difference in $EM^c/EM^h$ between the SIS
and GIS for these galaxies.

Probably the most significant differences between the parameters of
the isothermal and two-temperature models are the Fe abundances for the
MEKAL plasma models of NGC 1399, NGC 4472, and NGC 5044. The Fe
abundances are approximately 1.6 solar and 2 solar for the 2T+BREM
MEKAL models for NGC 1399 and NGC 4472 compared to approximately 0.3
solar for the isothermal models. For the 2T models the Fe abundances
of NGC 1399 and NGC 4472 are $\sim 1.1$ solar and have smaller error
bars in excellent agreement with the results in Table 5 of BF. That
is, adding the bremsstrahlung component significantly loosens the
constraints on the upper limit of the Fe abundance for the two-temperature
MEKAL model. When determining the upper limits for the Fe abundances
obtained from the 2T+BREM models, the bremsstrahlung temperatures were
fixed to the best-fit values; i.e. for NGC 1399 and NGC 4472.

The Fe abundance for the 2T MEKAL model for NGC 5044 ($\sim 0.6$
solar) is also larger than that determined from the isothermal model
($\sim 0.2$ solar) also in good agreement with Table 5 of BF. These
large differences in the Fe abundances between the isothermal and
two-temperature models highlight the importance of finding the correct
emission measure distribution as a function of temperature for
interpreting the abundances derived from spectral fitting, a point
that has been stressed in some previous X-ray studies of ellipticals
(Buote \& Canizares 1994; Trinchieri et al. 1994; BF) and by a recent
study of poor galaxy groups \cite{b99}.

As already mentioned in section \ref{quality}, allowing the relative
abundances to be free parameters offers no significant improvements in
the fits of 2T or 2T+BREM models to NGC 1399, NGC 4472, and NGC 5044
using the MEKAL plasma code. In contrast, the abundances of the
two-temperature RS models behave very similarly to their isothermal
counterparts except that the error bars are larger for the two-temperature
case as would be expected.  (Note if the energies 1.35 - 1.55 keV are
excluded from the RS fits to NGC 1399 and NGC 4472, the 2T+BREM models
with fixed relative abundances have larger Fe abundances: $\sim
1.0Z_{\sun}$ for NGC 1399 and $\sim 0.8Z_{\sun}$ for NGC 4472; cf. end
of section \ref{rs}.)

For NGC 4636 the 2T+BREM MEKAL model with variable relative abundances
offers some improvement over the variable abundance 1T+BREM model as
discussed in section \ref{quality}. Only the SIS data actually require
the additional temperature component (Table \ref{tab.params}): $T_{\rm
c}\sim 0.5$ keV and $T_{\rm h}\sim 0.75$ keV. The absorbing columns on
both components are consistent with zero and the abundances are very
similar to those obtained for the 1T+BREM case.

Although the 2T+BREM and 1T+BREM models with variable relative
abundances give marginally acceptable fits to the data in terms of
$\chi^2$ it is clear from the residuals of the fits (Figure
\ref{fig.n4636_multit}) that the energies below 2 keV are still not
adequately modeled. Moreover, the highly significant super-solar Na
abundance and zero Ni abundance for these models places added doubt on
the correctness of these models for NGC 4636.

The 2T+BREM RS model, in contrast, offers a substantial improvement in
the fit over the 1T+BREM models for NGC 4636, though the fit is still
not as good as that with the MEKAL model. The temperature of the
colder component, $T_{\rm c}\sim 0.25$ keV, is half that derived for
the MEKAL model. Moreover, substantial excess absorption is required
for the hotter component (larger than that for the colder component)
unlike for the MEKAL model for NGC 4636 and the other galaxies in the
sample. The Fe abundance for the 2T+BREM RS model is about 1 solar,
more than five times the value obtained for the 1T+BREM
models. Although most of the other abundances are also larger for the
2T+BREM case, the O abundance remains small (0.14 solar) while the Ni
abundance is even larger than before (about 4 solar). Again, like the
case of the MEKAL model the 2T+BREM RS model has significant residuals
below 2 keV and abundances (especially Ni) that cast doubt on the
model.

Finally, we have computed the 0.5-10 keV luminosities of the BREM
components for the the 2T+BREM models of NGC 1399, NGC 4472, and NGC
4636. Using the values of $L_B$ listed in Table 1 of BF we obtain
best-fitting results for $L_{\rm x}/L_B$ of 0.006 for NGC 1399, 0.001
for NGC 4472, and 0.002 for NGC 4636. These ratios are consistent
within their 90\% errors indicating that the BREM components are
consistent with the expectation that the X-ray emission from discrete
sources should be proportional to $L_B$.

\subsubsection{Multiphase cooling flows}
\label{cf}

\begin{table*}
\caption{Parameters of Cooling Flow Models}
\label{tab.cf}
\begin{tabular}{lccccc}
& \multicolumn{1}{c}{NGC 1399} & \multicolumn{1}{c}{NGC 4472} &
\multicolumn{1}{c}{NGC 5044} & \multicolumn{1}{c}{NGC 4636}\\ \\
\multicolumn{1}{c}{CF:}\\ \\[-5pt]
$N_{\rm H}$ ($10^{21}$ cm$^{-2}$) & $2.96_{-0.26}^{+0.27}$ &
$2.86_{-0.26}^{+0.29}$ & $3.20_{-0.26}^{+0.27}$ & $1.44_{-0.24}^{+0.25}$\\
$T$ (keV) & $1.62_{-0.08}^{+0.10}$ & $1.30_{-0.06}^{+0.11}$ &
$1.15_{-0.05}^{+0.07}$ & $0.73_{-0.01}^{+0.01}$\\
$Z (Z_{\sun})$ & $1.14_{-0.22}^{+0.43}$ & $1.41_{-0.43}^{+0.59}$ &
$0.79_{-0.22}^{+0.88}$ & $1.04_{-0.18}^{+0.35}$\\
$\dot{M}_{\rm sis}$ ($M_{\sun}$ yr$^{-1}$) & $1.62_{-0.18}^{+0.20}$ &
$2.54_{-0.37}^{+0.34}$ & $41.32_{-5.42}^{+4.59}$ & $2.65_{-0.40}^{+0.45}$\\
$\dot{M}_{\rm gis}$ ($M_{\sun}$ yr$^{-1}$) & $3.90_{-0.67}^{+0.61}$ &
$4.56_{-0.96}^{+0.81}$ & $27.09_{-23.76}^{+20.80}$ & $1.03_{-0.93}^{+1.34}$\\ \\
\multicolumn{1}{c}{1T:}\\ \\[-5pt]
$N_{\rm H}$ ($10^{21}$ cm$^{-2}$) & $0.02_{-0.01}^{+0.01}$ &
$0.00_{-0.00}^{+0.88}$  & $0.67_{-0.64}^{+0.65}$
& $2.72_{-0.29}^{+0.29}$\\
$Z (Z_{\sun})$ & $\cdots$ & $\cdots$ & $0.44_{-0.15}^{+0.19}$ & $\cdots$\\
$EM_{\rm sis}$ & $2.20_{-0.65}^{+0.59}$ & $0.85_{-0.45}^{+0.55}$ &
$6.33_{-2.55}^{+2.97}$ & $3.25_{-0.76}^{+0.61}$\\
$EM_{\rm gis}$ & $0.00_{-0.00}^{+1.49}$ & $0.00_{-0.00}^{+0.42}$ &
$21.25_{-8.23}^{+8.53}$ & $5.88_{-1.41}^{+1.23}$\\ \\
\multicolumn{1}{c}{BREM:}\\ \\[-5pt]
$T$ (keV) & $>4$ & $3.71_{-0.98}^{+6.44}$ & $\cdots$ & $>8$\\
$EM_{\rm sis}$ & $0.06_{-0.03}^{+0.07}$ & $0.14_{-0.08}^{+0.06}$ &
$\cdots$ & $0.06_{-0.01}^{+0.03}$\\
$EM_{\rm gis}$ & $0.16_{-0.04}^{+0.11}$ & $0.28_{-0.12}^{+0.11}$ &
$\cdots$ & $0.05_{-0.01}^{+0.03}$\\

\end{tabular}

\medskip
\raggedright

Best-fitting and 90\% confidence limits on one interesting parameter
$(\Delta\chi^2=2.71)$ are listed for selected cooling flow models;
$\chi^2$ values for these models are listed in Table
\ref{tab.quality}. The temperatures and abundances of the CF and 1T
components are tied together in the fits except for the abundances for
NGC 5044; the relative abundances with respect to Fe for all the
cooling flow models are fixed at solar. The mass deposition rates
assume distances of 18 Mpc for NGC 1399, NGC 4472, and NGC 4636 and 39
Mpc for NGC 5044 ($H_0=70$ km s$^{-1}$ Mpc$^{-1}$). The bremsstrahlung
component is modified by Galactic absorption (Table \ref{tab.prop})
and the emission measures are quoted in units as explained in Table
\ref{tab.params}.

\end{table*}

We have learned in section \ref{quality} that for NGC 1399, NGC 4472,
and NGC 5044 the multiphase cooling flows have $\chi^2$ values
approximately midway between those of the 1T+BREM variable relative
abundance models and the 2T+BREM models having relative abundances
with respect to Fe fixed at their solar values. If only models with
solar relative abundances are compared (as is appropriate for the
cooling flow models since they are restricted to solar relative
abundances) then the cooling flow models have $\chi^2$ values that are
substantially smaller than the 1T+BREM models and similar (though
slightly worse) than the 2T+BREM models. For NGC 4636 this also
applies if the comparison is restricted to models with solar relative
abundances.

The fits of the cooling flow models are quite similar to the
two-temperature models of NGC 1399, NGC 4472, and NGC 5044. The
pattern of residuals in the SIS data seen for the CF+1T+BREM model for
NGC 1399 in Figure \ref{fig.n1399_multit} are very similar to those of
NGC 4472 and to a lesser extent those of NGC 5044. For NGC 4636
(Figure \ref{fig.n4636_multit}) the residuals are larger and more
complex.

The properties of the cooling flow models are similar in many
important respects to the two-temperature models. In Table
\ref{tab.cf} we list the derived parameters for CF+1T+BREM models of
NGC 1399, NGC 4472, and NGC 4636 and for the CF+1T model of NGC
5044. (The CF+1T model is focused on for NGC 5044 since a
bremsstrahlung component is not required as also found for the
isothermal and two-temperature models.)  The temperatures of the
isothermal components (which also equal the maximum temperature of the
cooling flow components) are very similar to $T_{\rm h}$ for the
two-temperature models (Table \ref{tab.params}). The emission-weighted
temperatures of the cooling flow components are also approximately
equal to $T_{\rm c}$ (using $\langle T\rangle \approx 0.6T$ -- Buote
et al. 1998).

Moreover for NGC 1399, NGC 4472, and NGC 5044, analogously to the
two-temperature models, the ``hotter'' component of the cooling flow
models (i.e. 1T component) has small absorption consistent with zero
whereas the colder component (i.e. CF component) requires significant
absorption in excess of the Galactic value.  Excess absorption
somewhat larger than the cooling-flow component is indicated for the
1T component of NGC 4636 similar to the two-temperature MEKAL model.

The metal abundances for the cooling flows are also reasonably
consistent with the two-temperature models. For NGC 5044 we are able to
place interesting constraints on the metal abundances of both the
cooling flow and isothermal components. We find that the abundance of
the cooling flow is about 0.8 solar which is twice that of the
isothermal component.  The principal reason for this constraint is the
lack of an additional bremsstrahlung component; i.e. adding a
bremsstrahlung component weakens this constraint, but the best-fit
abundances still indicate a higher abundance for the cooling flow
component. This also applies to NGC 1399 and NGC 4472: if the
bremsstrahlung component is omitted, the abundance of the cooling flow
component tends to exceed the isothermal component, though at a lower
significance level than found for NGC 5044.

The mass deposition rates of a few solar masses per year are typical
for ellipticals (e.g. BF). However, the large $\dot{M}_{\rm sis}\sim
40$ $M_{\sun}$ yr$^{-1}$ for NGC 5044 is even larger than for typical
galaxy groups \cite{b99} and is in fact more similar to that of a
galaxy cluster (e.g. Fabian 1994).
 
\subsubsection{Remaining problems with Fe L in the plasma codes} 
\label{fel}

Although, as we discussed in section \ref{ass}, the MEKAL code is
superior to the RS code for analysis of {\sl ASCA} data of the
elliptical galaxies in our sample, it is likely that problems still
exist in the MEKAL code. Examination of the (admittedly small) SIS
residuals for the two-temperature fits of NGC 1399 (Figure
\ref{fig.n1399_multit}) reveals systematic excesses near 1.1 keV and
0.9 keV (and elsewhere to a lesser extent). Comparing these residuals
to those of the cooling flow model for NGC 1399 shows that the
residuals at 0.9 keV and 1.1 keV exist for the cooling flow model but
are more pronounced. Related behavior is observed for NGC 4472 and NGC
5044. Taken by themselves these small residual patterns reasonably
could be attributed to the model of the emission measure distribution
as a function of temperature (i.e. 2T+BREM, CF+1T+BREM, etc.) rather
than the plasma code.

The case of NGC 4636 lends support to the plasma code explanation. In
order to obtain formally acceptable fits to the SIS data of NGC 4636
we require that the Na abundance be 15-20 solar and the Ni abundance
be zero: both of these values greatly differ from the derived Fe
abundance of 0.8 solar. These abundances of Na and Ni are inconsistent
with standard models for the enrichment of the hot gas in ellipticals
(e.g. David et al. 1991; Ciotti et al. 1991).

If indeed the super-solar Na abundance is unphysical, then an error in
the MEKAL code near 1.1 keV is implied since that is location of the
strong K$\alpha$ lines of the He-like Na ion. This is precisely the
location of the weak residuals discussed above for the other galaxies.
If the Ni abundance is also unphysical, then additional errors likely
exist in either the Fe L shell lines or in the Ni L shell emission of
the MEKAL code. Since much more attention has been devoted to
correcting the strong Fe L shell emission lines (e.g. Liedahl et
al. 1995; Wargelin et al. 1998; Brown et al. 1998), it is possible
that the neglected Ni L lines require some adjustments as well.

However, as we show in section \ref{rosasca}, temperature gradients
may partially explain the Na abundance inferred for NGC 4636.

\section{Analysis of individual line blends with {\sl ASCA}}
\label{lines}

We have shown that the important constraints from the broad-band
spectral fitting are largely determined from the data near 1 keV which
are dominated by emission lines from the Fe L shell. Although good
fits for these galaxies are obtained for multiphase models using the
MEKAL plasma code, it is instructive to examine whether the {\sl ASCA}
data outside the Fe L region of the spectrum also favor multiphase
models over isothermal models. In particular, the emission lines from
the K$\alpha$ transitions of the He-like and H-like ions of Si and S
provide constraints on the temperature complementary to the Fe L
emission lines. These emission lines have the advantage that they are
well separated in energy from the strong Fe L shell lines and their
properties are relatively simple and thus are known to much higher
accuracy than the Fe L lines .

\subsection{Deduced line properties from {\sl ASCA} data} 
\label{prop}

\begin{figure*}
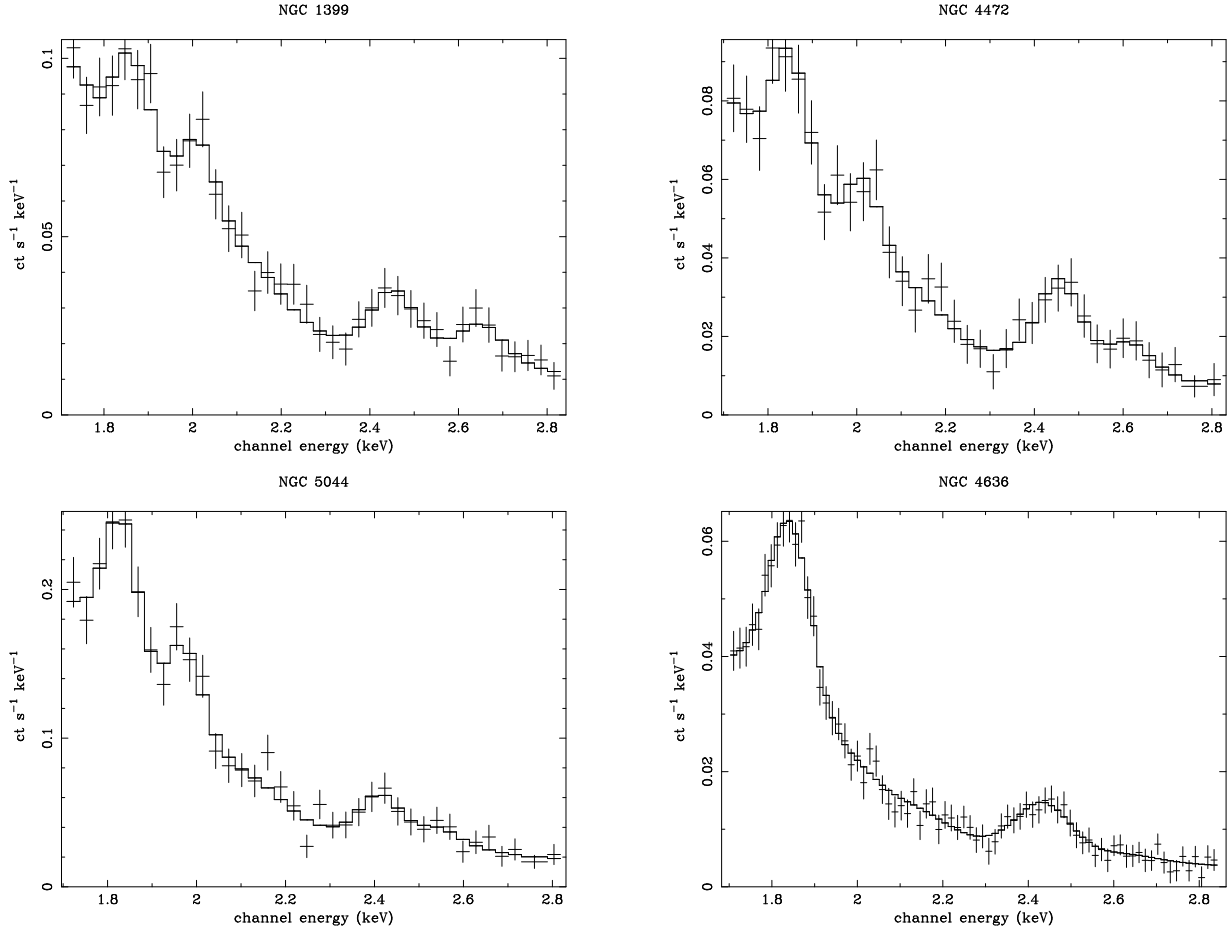

\parbox{0.49\textwidth}{
\centerline{\psfig{figure=fig8a.ps,angle=-90,height=0.25\textheight}}
}
\parbox{0.49\textwidth}{
\centerline{\psfig{figure=fig8b.ps,angle=-90,height=0.25\textheight}}
}
\vskip 0.25cm
\parbox{0.49\textwidth}{
\centerline{\psfig{figure=fig8c.ps,angle=-90,height=0.25\textheight}}
}
\parbox{0.49\textwidth}{
\centerline{\psfig{figure=fig8d.ps,angle=-90,height=0.25\textheight}}
}
\caption{\label{fig.lines} {\sl ASCA} SIS data between energies 1.7
and 2.85 keV for NGC 1399, NGC 4472, NGC 5044, and NGC 4636. Also
shown for each galaxy is the best-fitting model consisting of a
thermal bremsstrahlung component to represent the continuum emission
and four gaussians of zero intrinsic width to model respectively the
major K$\alpha$ emission line complexes of Si XIII, Si XIV, S XV, and
S XVI. See Table \ref{tab.lines} for the parameters of these emission
lines.}
\end{figure*}

\begin{table*}

\caption{Emission Lines Measured from {\sl ASCA} Data}
\label{tab.lines}
\begin{tabular}{lrcrrr}
& $E$ & $T$ & EW & 
\multicolumn{2}{c}{Flux}\\ 
& (keV) & (keV) & (eV) & ($10^{-5}$ ph cm$^{-2}$ s$^{-1}$) & ($10^{-13}$ erg cm$^{-2}$ s$^{-1}$)\\ 

\\NGC 1399:\\[+3pt]

Mg XI    (He)  & $1.38_{-0.03}^{+0.05}$ & $0.99_{-0.31}^{+0.58}$& $ 14_{-11}^{+13}$ & $1.82_{-1.43}^{+1.58}$ & $0.40_{-0.32}^{+0.35}$\\
Mg XII   (H)   & $1.49_{-0.05}^{+0.03}$ & $0.99_{-0.31}^{+0.58}$& $ 18_{-15}^{+20}$ & $1.96_{-1.61}^{+1.84}$ & $0.47_{-0.38}^{+0.44}$\\
Si XIII  (He)  & $1.87_{-0.01}^{+0.02}$ & $1.09_{-0.24}^{+0.38}$& $ 96_{-29}^{+32}$ & $5.05_{-1.37}^{+1.38}$ & $1.51_{-0.41}^{+0.41}$\\
Si XIV   (H)   & $2.01_{-0.01}^{+0.02}$ & $1.09_{-0.24}^{+0.38}$& $ 84_{-26}^{+28}$ & $3.52_{-1.03}^{+1.02}$ & $1.13_{-0.33}^{+0.33}$\\
S XV     (He)  & $2.45_{-0.02}^{+0.02}$ & $1.09_{-0.24}^{+0.38}$& $140_{-53}^{+64}$ & $3.02_{-1.00}^{+1.06}$ & $1.18_{-0.39}^{+0.42}$\\
S XVI    (H)   & $2.65_{-0.03}^{+0.03}$ & $1.09_{-0.24}^{+0.38}$& $108_{-65}^{+74}$ & $1.74_{-0.97}^{+0.88}$ & $0.74_{-0.41}^{+0.37}$\\
Ar XVII  (He)  & $3.13_{-0.07}^{+0.07}$ & $2.54_{-0.85}^{+2.23}$& $ 77_{-60}^{+72}$ & $0.87_{-0.67}^{+0.69}$ & $0.44_{-0.33}^{+0.35}$\\

\\NGC 4472:\\[+3pt]

Mg XII   (H)   & $1.47_{-0.02}^{+0.02}$ & $0.49_{-0.09}^{+0.14}$& $ 32_{-19}^{+ 20}$ & $3.02_{-1.72}^{+1.53}$ & $0.71_{-0.40}^{+0.36}$\\
Si XIII  (He)  & $1.85_{-0.01}^{+0.01}$ & $0.91_{-0.21}^{+0.37}$& $136_{-42}^{+ 48}$ & $5.67_{-1.49}^{+1.48}$ & $1.68_{-0.44}^{+0.44}$\\
Si XIV   (H)   & $2.01_{-0.02}^{+0.02}$ & $0.91_{-0.21}^{+0.37}$& $ 90_{-35}^{+ 39}$ & $2.82_{-1.01}^{+1.01}$ & $0.91_{-0.32}^{+0.32}$\\
S XV     (He)  & $2.46_{-0.02}^{+0.02}$ & $0.91_{-0.21}^{+0.37}$& $244_{-84}^{+100}$ & $3.62_{-1.04}^{+1.04}$ & $1.42_{-0.41}^{+0.41}$\\
S XVI    (H)   & $2.62_{-0.06}^{+0.05}$ & $0.91_{-0.21}^{+0.37}$& $100_{-78}^{+100}$ & $1.14_{-0.86}^{+0.86}$ & $0.48_{-0.36}^{+0.36}$\\

\\NGC 4636:\\[+3pt]

Mg XI    (He)  & $1.34_{-0.01}^{+0.01}$ & $0.32_{-0.03}^{+0.03}$& $ 48_{- 9}^{+10}$ & $4.11_{-0.67}^{+0.67}$ & $0.88_{-0.14}^{+0.14}$\\
Mg XII   (H)   & $1.47_{-0.02}^{+0.01}$ & $0.32_{-0.03}^{+0.03}$& $ 32_{-14}^{+17}$ & $1.58_{-0.67}^{+0.69}$ & $0.37_{-0.16}^{+0.16}$\\
Si XIII  (He)  & $1.84_{-0.00}^{+0.00}$ & $0.98_{-0.18}^{+0.26}$& $264_{-34}^{+37}$ & $5.41_{-0.46}^{+0.47}$ & $1.60_{-0.14}^{+0.14}$\\
Si XIV   (H)   & $1.98_{-0.03}^{+0.03}$ & $0.98_{-0.18}^{+0.26}$& $ 45_{-20}^{+21}$ & $0.74_{-0.31}^{+0.31}$ & $0.24_{-0.10}^{+0.10}$\\
S XV     (He)  & $2.43_{-0.01}^{+0.02}$ & $0.98_{-0.18}^{+0.26}$& $228_{-53}^{+61}$ & $1.78_{-0.33}^{+0.33}$ & $0.69_{-0.13}^{+0.13}$\\
S XVI    (H)   & $2.64$                 & $0.98_{-0.18}^{+0.26}$& $ 25_{-25}^{+57}$ & $0.14_{-0.14}^{+0.28}$ & $0.06_{-0.06}^{+0.12}$\\

\\NGC 5044:\\[+3pt]

Mg XI    (He)  & $1.35_{-0.03}^{+0.03}$ & $0.44_{-0.05}^{+0.15}$& $ 20_{-14}^{+11}$ & $ 6.35_{-4.33}^{+3.18}$ & $1.37_{-0.94}^{+0.69}$\\
Mg XII   (H)   & $1.49_{-0.03}^{+0.02}$ & $0.44_{-0.05}^{+0.10}$& $ 36_{-25}^{+16}$ & $ 7.14_{-4.78}^{+2.58}$ & $1.70_{-1.14}^{+0.62}$\\
Si XIII  (He)  & $1.83_{-0.01}^{+0.01}$ & $0.97_{-0.19}^{+0.29}$& $115_{-29}^{+34}$ & $10.43_{-2.11}^{+2.12}$ & $3.07_{-0.62}^{+0.62}$\\
Si XIV   (H)   & $1.98_{-0.01}^{+0.01}$ & $0.97_{-0.19}^{+0.29}$& $108_{-28}^{+32}$ & $ 7.70_{-1.83}^{+1.83}$ & $2.44_{-0.58}^{+0.58}$\\
S XV     (He)  & $2.42_{-0.02}^{+0.02}$ & $0.97_{-0.19}^{+0.29}$& $126_{-46}^{+50}$ & $ 4.59_{-1.55}^{+1.55}$ & $1.78_{-0.60}^{+0.60}$\\
S XVI    (H)   & $2.55_{-0.09}^{+0.09}$ & $0.97_{-0.19}^{+0.29}$& $ 54_{-49}^{+56}$ & $ 1.50_{-1.34}^{+1.33}$ & $0.61_{-0.55}^{+0.54}$\\

\end{tabular}

\medskip

\raggedright

Properties of emission lines derived from fitting simple
continuum+lines models to the summed {\sl ASCA} SIS data.  The
continuum is represented by a thermal bremsstrahlung component and the
lines by gaussians of zero intrinsic width. Lines that have the same
continuum temperature, $T$, are cases where multiple gaussians are
joined by one bremsstrahlung component. See text for additional
details regarding the model fitting. Quoted errors are 90 per cent
confidence on one interesting parameter. Line energies without error
bars were fixed at their best-fitting values.

\end{table*}

\begin{table*}
\caption{Line Ratios}
\label{tab.ratios}
\begin{tabular}{rcccc}
& NGC 1399 & NGC 4472 & NGC 4636 & NGC 5044\\
Si XIV/Si XIII & $0.75_{-0.33}^{+0.58}$ & $0.54_{-0.27}^{+0.45}$ &
$0.15_{-0.07}^{+0.08}$ & $0.79_{-0.29}^{+0.44}$\\
S XVI/SXV      & $0.62_{-0.42}^{+0.78}$ & $0.34_{-0.27}^{+0.49}$ &
$0.09_{-0.09}^{+0.23}$ & $0.35_{-0.32}^{+0.64}$\\

\end{tabular}

\medskip

\raggedright

Ratios of K$\alpha$ emission lines for the H-like and He-like ions of
Si and S using the fluxes from Table \ref{tab.lines}. Quoted
errors represent ratios of 90 per cent confidence lower limits to 90
per cent confidence upper limits and vice versa. Thus, the quoted
error ranges reflect larger than a 90 per cent confidence interval and
typically correspond to $\sim 2\sigma$ confidence.

\end{table*}

For our study of the Si and S line blends we must pay special
attention to the calibration problem in the energy response near 2.2
keV owing to the optical constants in the XRT
\cite{gendreau}. Although the ARF files generated with the standard
software (see section \ref{obs}) correct for much of this problem,
there are low-level residuals extending throughout the energies of the
key Si and S lines. For compactness of presentation, the results we
present for the joint SIS0+SIS1 fits for NGC 1399 (8003800) and NGC
4472 (60029000) refer to one observation sequence in each case.

We focus our analysis of individual line blends on the SIS data
because, unlike the GIS, the SIS is able to resolve the He-like and
H-like blends of Si and S (see section \ref{relative}). In no instance
did we find that constraints on the line properties improved
noticeably when incorporating the GIS data into the fits.

Our procedure for obtaining the properties of the line blends
commences by fitting a thermal bremsstrahlung model to represent the
continuum emission. We desire a measurement of the local continuum
near the line blends of interest so that the assumption of a single
temperature is reasonable even if the spectrum actually consists of
multiple temperature components. However, because of S/N
considerations and the finite energy resolution of the SIS we allowed
the continuum component to extend over multiple line blends that are
nearby in energy.

The line blends of interest were then modeled as gaussians of zero
intrinsic width on top of the continuum component. For analysis of the
Si and S line blends we fitted the bremsstrahlung model over the open
interval (1.7,2.85) keV with gaussian components initially at 1.85 keV
(Si XIII), 2.0 keV (Si XIV), 2.45 keV (S XV), and 2.62 keV (S
XVI). The energies of the lines were free parameters in the fits. When
determining the error bars on the equivalent widths and fluxes for a
particular line the energies of the other lines were fixed at their
best-fit values. This was done to insure that the other lines did not
mix during the error search.

For completeness, when available we also derived properties for the Mg
and Ar line blends. The continuum for the Mg XI (He) and Mg XII (H)
lines was defined over the open interval (1.23,1.6) keV. For the Ar
XVII (He) blend we used the open interval (2.85,5) keV.  The
properties of the all detected lines are listed in Tables
\ref{tab.lines} and \ref{tab.ratios}.  In Figure \ref{fig.lines} we
plot the SIS data and best-fitting model over the spectral region used
to determine the properties of the Si and S lines for each
galaxy. Since the continuum+lines model fit well for each system we do
not show the residuals in the figure.

The K$\alpha$ line blends of the He-like and H-like ions of Si and S
are detected for each galaxy except the S XVI blend of NGC 4636. Both
the Mg XI and XII K$\alpha$ line blends are detected for NGC 4636 and
NGC 5044 in each SIS; these blends are detected only in the combined
SIS data for NGC 1399 and only for Mg XII for NGC 4472. We also detect
weak Ar XVII (He) emission in NGC 1399 in both the SIS0 and SIS1.

Generally the line energies computed for the SIS0 and SIS1 agree
within their estimated 90\% confidence limits. The line equivalent
widths (EWs) and fluxes mostly agree within these limits with some
notable exceptions: S XIV for NGC 4636 and both S complexes for NGC
1399. The ratios of the H-like/He-like K$\alpha$ fluxes for both Si
and S (Table \ref{tab.ratios}) also agree within the estimated
$2\sigma$ limits for the SIS0 and SIS1. 

Therefore, although the marginal agreement (or rather disagreement) at
the 90\% confidence level for the fluxes computed for the SIS0 and
SIS1 may point to underlying systematic errors as discussed above,
such errors are mostly blurred by the S/N of these data. As done
previously with the broad-band spectral analysis, we shall henceforth
focus on analysis of the summed SIS data which essentially averages
over the systematic differences in the SIS0 and SIS1 data. (The
possible underlying systematic differences between the SIS0 and SIS1
must be regarded as a caveat to any analysis of the emission lines of
these galaxies with the {\sl ASCA} SIS, including the analysis
presented in the following section.)

\subsection{Comparison to isothermal and multiphase models}  
\label{models}

We compare the equivalent widths and ratios of the Si and S line
blends computed using the continuum+gaussian approximation with those
predicted by the models obtained from the broad-band spectral fitting.
Although our focus is on these lines that are far from the Fe L energy
region, we comment on the Mg lines near the end of this
section. Examination of Figure \ref{fig.lines} and Tables
\ref{tab.lines} and \ref{tab.ratios} reveals substantial Si XIV
ly-$\alpha$ emission for NGC 1399, NGC 4472, and NGC 5044 with respect
to NGC 4636. Similarly, there is significant S XVI ly-$\alpha$
emission for NGC 1399 and NGC 4472 (and marginally for NGC 5044)
whereas only upper limits are obtained for NGC 4636.

The fact that the Si XIV/XIII and S XVI/XV ratios are comparable for
NGC 1399, NGC 4472, and NGC 5044 but are clearly smaller for NGC 4636
indicates that the plasma temperature of NGC 4636 is significantly
smaller than the others -- a statement that is independent of the Si
and S abundances\footnote{This insensitivity to the abundance is exact
for an isothermal plasma, but it should be noted that for multiphase
plasmas like cooling flows there can be a small metallicity dependence
for such line ratios \cite{b98}.}. This demonstrates at least a
``zeroth-order'' level of consistency between the temperatures derived
from the K$\alpha$ lines of Si and S with those determined largely by
the Fe L lines from the broad-band spectral fitting analysis in
section \ref{broad}.

Let us now examine in detail how well the models obtained from the
broad-band spectral fitting reproduce the observed Si and S line
ratios and equivalent widths. Our method to extract the properties of
the Si and S line complexes using a bremsstrahlung component and four
zero-width gaussians is a convenient approximation appropriate for the
energy resolution of the SIS data. To account for any biases in this
procedure we compute the line properties from the models using the
same procedure. That is, for each best-fitting isothermal,
two-temperature, and cooling flow model listed in Table \ref{tab.params} we
simulated a SIS observation (using {\sc xspec}) with an exposure time
sufficiently long so that the statistical errors in our derived
parameters for each model are generally $<10\%$. We found that
exposure times of $10^7$s - $10^8$s were satisfactory for this
purpose.

\begin{figure*}
\parbox{0.49\textwidth}{
\centerline{\psfig{figure=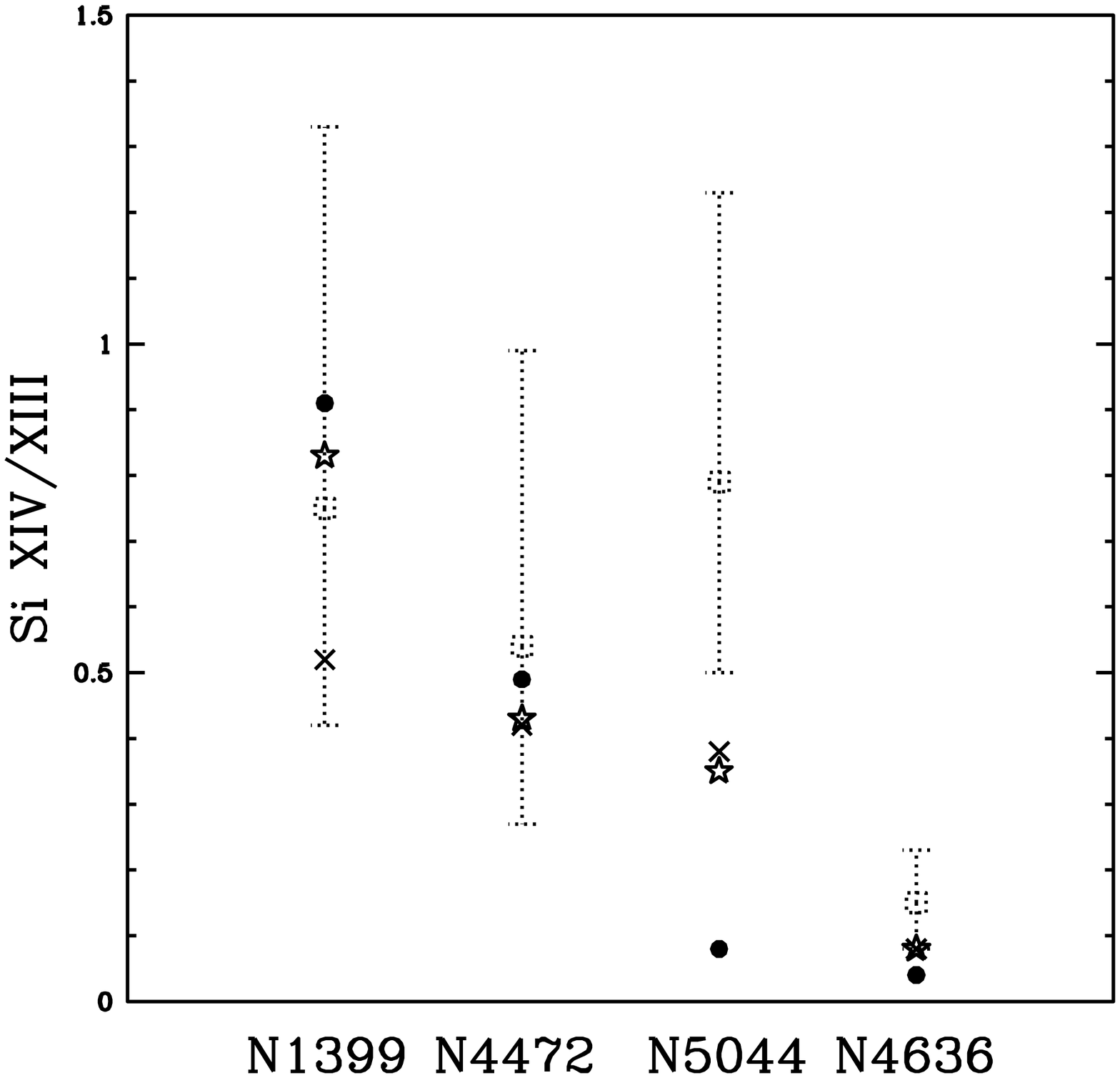,angle=0,height=0.25\textheight}}
}
\parbox{0.49\textwidth}{
\centerline{\psfig{figure=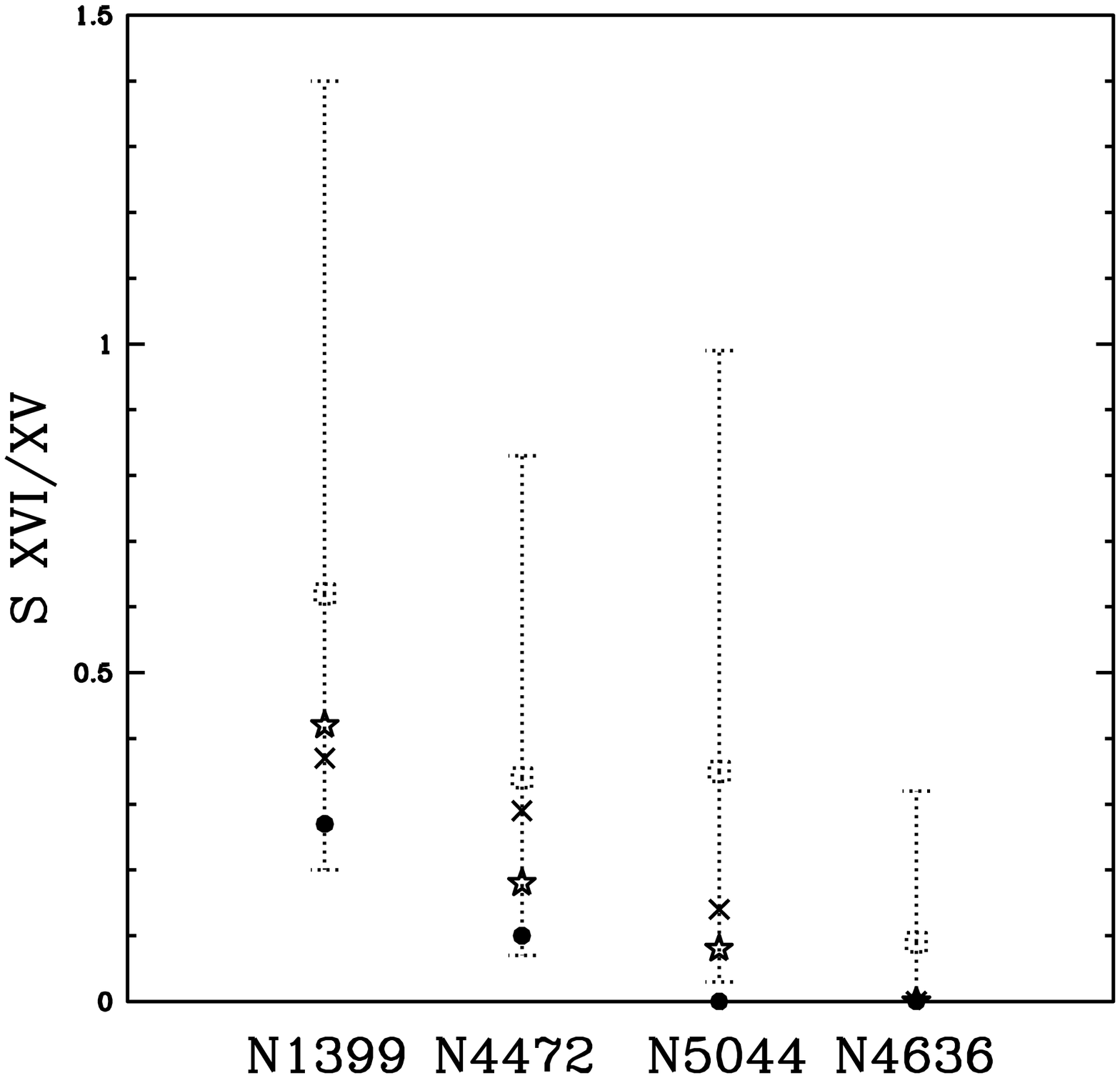,angle=0,height=0.25\textheight}}
}
\caption{\label{fig.ratios} K$\alpha$ line ratios of H-like to He-like
ions of Si (left) and S (right) for the {\sl ASCA} SIS data (dotted)
with $\sim 2\sigma$ error bars (Table \ref{tab.ratios}) plotted along
with ratios computed from the best-fit (MEKAL) models of Tables
\ref{tab.params} and \ref{tab.cf}: isothermal (filled circle);
two-temperature (crosses); multiphase cooling flow (stars).}
\end{figure*}

\begin{figure*}
\parbox{0.49\textwidth}{
\centerline{\psfig{figure=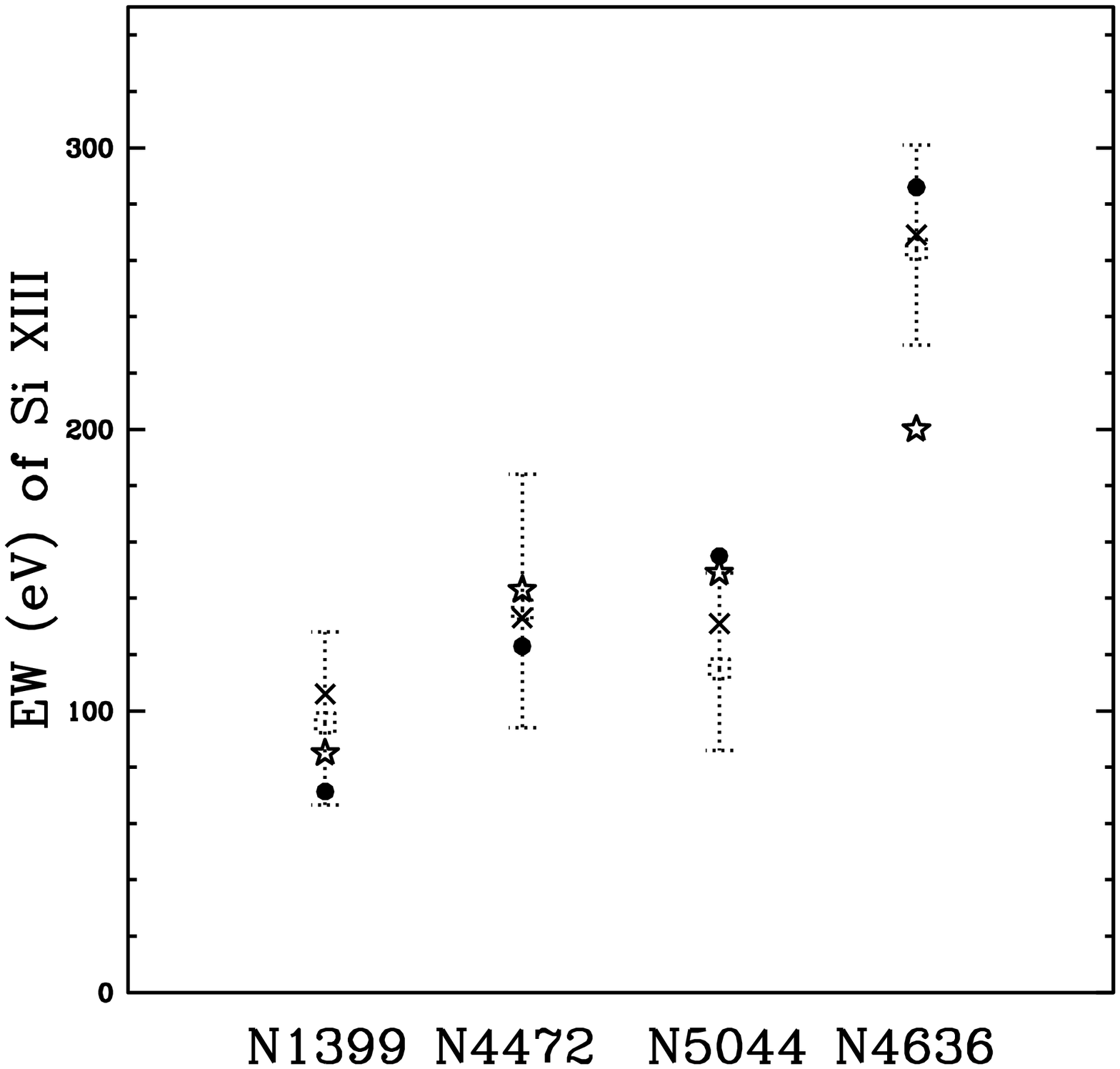,angle=0,height=0.25\textheight}}
}
\parbox{0.49\textwidth}{
\centerline{\psfig{figure=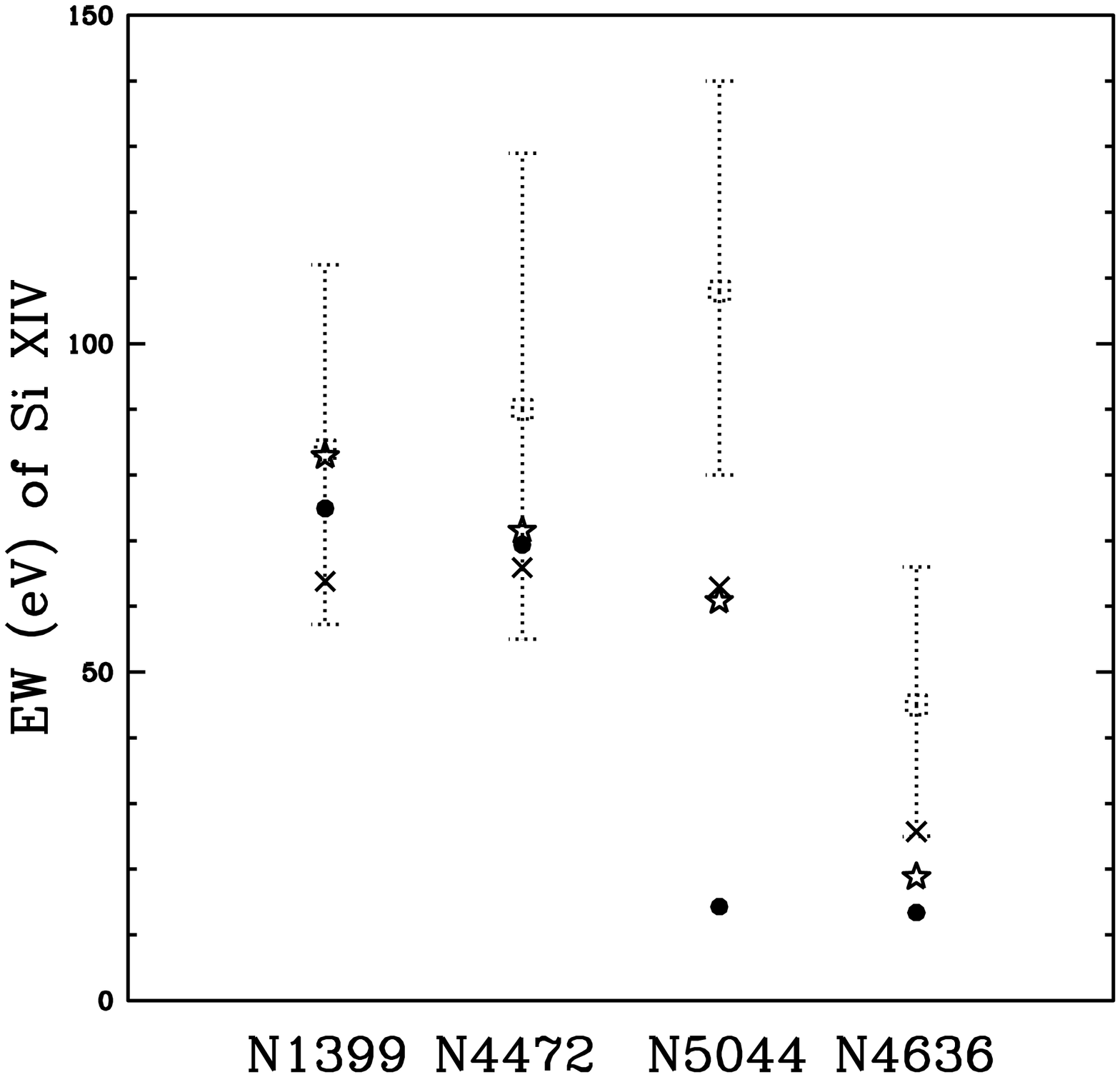,angle=0,height=0.25\textheight}}
}
\vskip 0.25cm
\parbox{0.49\textwidth}{
\centerline{\psfig{figure=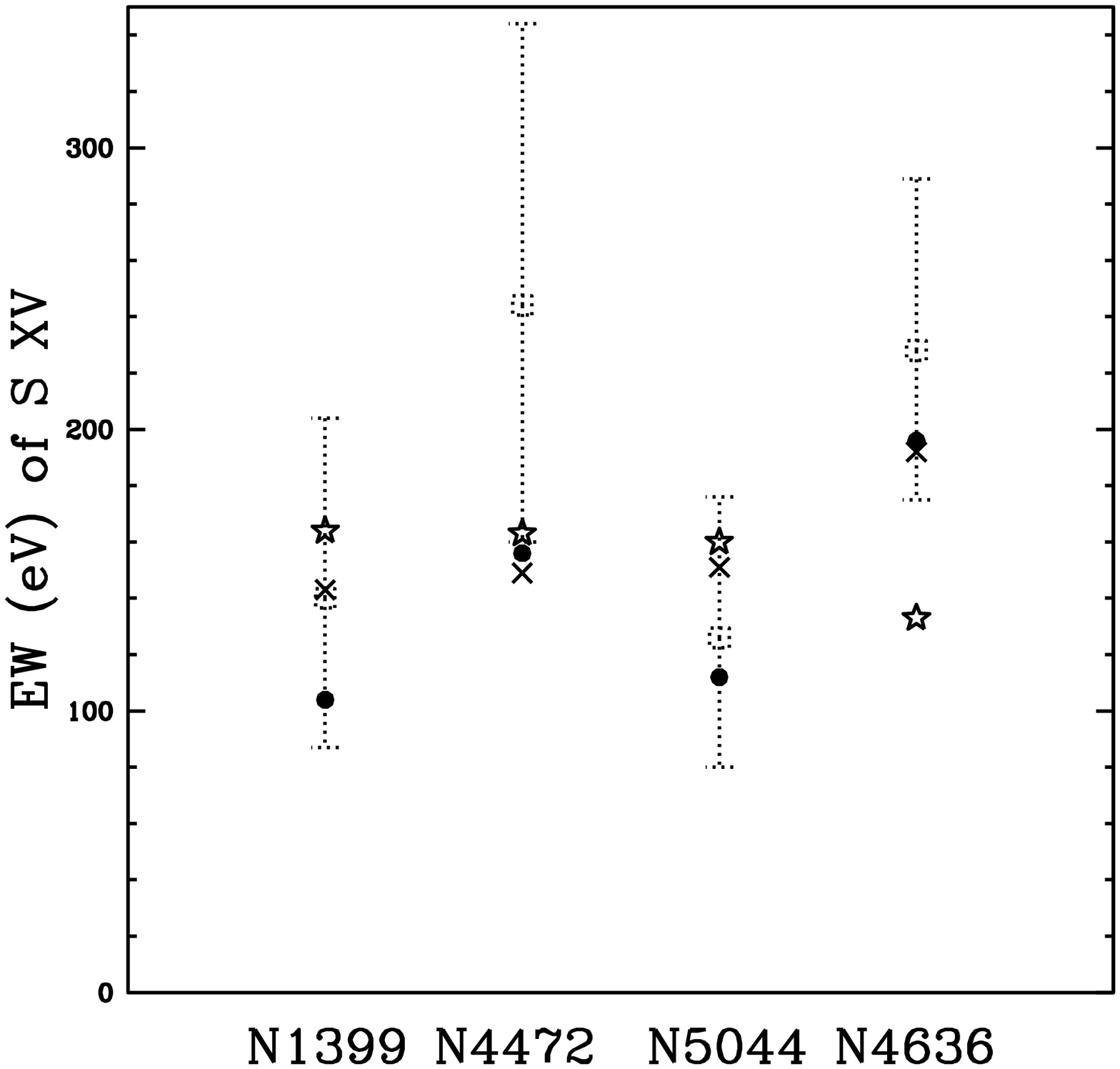,angle=0,height=0.25\textheight}}
}
\parbox{0.49\textwidth}{
\centerline{\psfig{figure=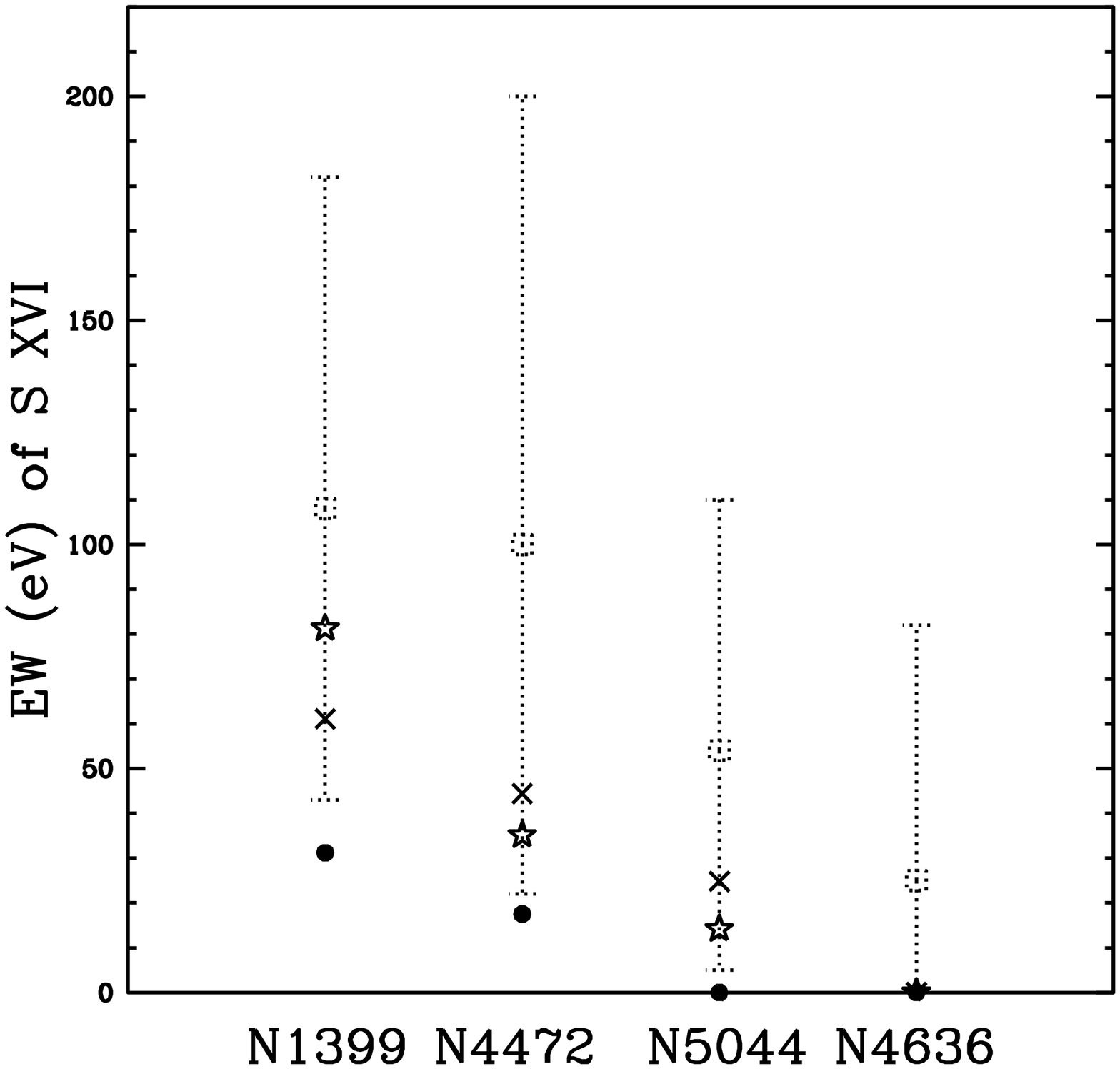,angle=0,height=0.25\textheight}}
}
\caption{\label{fig.ews} As Figure \ref{fig.ratios} except that the
equivalent widths of the lines for the data and models are shown and
the error bars represent 90\% confidence levels
($\Delta\chi^2=2.71$).}
\end{figure*}

In Figure \ref{fig.ratios} we plot the K$\alpha$ line ratios Si
XIV/XIII and S XVI/XV of the models and data. The equivalent widths of
the line complexes are displayed in Figure \ref{fig.ews} for the
models and data. Note that the error bars on the ratios in Figure
\ref{fig.ratios} for the data are $\sim 2\sigma$ while the error bars
for the equivalent widths in Figure \ref{fig.ews} are 90\% confidence
limits.

Inspection of Figure \ref{fig.ratios} reveals that the Si and S ratios
of each of the three models for NGC 1399 are consistent within the
$2\sigma$ limits. Although all the models are consistent, the cooling
flow model best reproduces these ratios. The isothermal model agrees
with the Si ratio better than the two-temperature model and vice versa
for the S ratio. The equivalent widths (Figure \ref{fig.ews}) for all
of the models are also consistent with the data of NGC 1399 except for
the S XVI line blend where the isothermal model under-predicts the
equivalent width. In fact, except for the Si XIV line, the isothermal
model tends to under-predict the equivalent widths of the lines with
respect to the multiphase models.

To facilitate our comparison of equivalent widths we define the
quantity,
\begin{equation}
\Delta_{\rm EW} \equiv \sum_i {\left( {\rm EW}^{data}_i - {\rm
EW}^{model}_i \right)^2 \over \sigma^2}, \label{eqn.ew} 
\end{equation}
where $\sigma$ is the one-sided 68\% confidence level of the
equivalent widths estimated from the data and the summation runs over
the two Si and two S line blends. The purpose of $\Delta_{\rm EW}$ is
to provide a number to indicate the relative agreement of the various
models analogously to $\chi^2$.

For NGC 1399 we have $\Delta_{\rm EW} = (1.59,0.75,0.30)$ for
respectively the best-fitting isothermal, two-temperature, and cooling
flow models. That is, the isothermal model clearly shows the most
deviations and the cooling flow model agrees better than the
two-temperature model. Since each model has comparable
emission-weighted temperature, the smaller equivalent widths of the
isothermal model are due primarily to the smaller Si and S abundances
of the model (see Table \ref{tab.params}).

Loewenstein \& Mushotzky \shortcite{lmushy} have asserted that the Si
ratio computed for {\sl ASCA} SIS data of NGC 1399 is consistent with
a isothermal model but is inconsistent with the two-temperature model
presented in BF. Our two-temperature models agree well with BF and, as we
have just shown, also predict Si ratios that agree with the data. It
is true that for the Si ratio the isothermal model gives better
agreement than the two-temperature model. However, we have shown this to be
misleading since when the S ratio and the equivalent widths of the Si
and S line blends are also considered together with the Si ratio we
find that the two-temperature and cooling flow models agree much better with
the data than does the isothermal model.

The Si and S ratios for each model also agree with the data of NGC
4472. The agreement is good for all models for the Si ratio, but for
the S ratio the two-temperature model fares much better than the
isothermal case which is only marginally in agreement at the
$2\sigma$ lower limit. With the exception of the Si XIII blend where
the agreement is good for all models, the models tend to under-predict
the equivalent widths for the Si and S lines though the only actual
discrepancy is for the isothermal prediction of S XVI. In terms of
$\Delta_{\rm EW}$ we have (1.47,1.28,1.07) for respectively the
best-fit isothermal, two-temperature, and cooling flow models. Hence, all
models are comparable, though the multiphase models, especially the
cooling flow, is the best fit to the Si and S equivalent widths
similar to NGC 1399.

For NGC 5044 we find that no model is able to reproduce the Si ratio
within the $2\sigma$ error. The multiphase models, though, give Si
ratios in much better agreement with the data than do the isothermal
models. The model ratios also tend to be low for S, though the
multiphase models are consistent within the errors while the
isothermal ratio is inconsistent. These results for the Si and S
ratios suggest that higher temperature gas is needed in the
models. The need for additional temperature components is not
inconsistent with the broad-band fits since, e.g., even the best
cooling flow models are only marginally acceptable in terms of their
null hypothesis probabilities (see Table \ref{tab.quality}).

The equivalent widths of Si and S further support the need for higher
plasma temperature components in NGC 5044. All models tend to
over-predict the He-like Si XIII and S XV emission while
under-predicting the H-like Si XIV and S XVI emission, though the most
significant disagreement lies with the isothermal models:
$\Delta_{\rm EW} = (8.38,2.41,3.07)$ for respectively the best-fit
isothermal, two-temperature, and cooling flow models.

The Si ratio of NGC 4636 is marginally acceptable for the multiphase
models and inconsistent for the isothermal model within the
$2\sigma$ errors. If only the relative abundances of the
$\alpha$-process elements are varied for the two-temperature model (i.e. Na
and Ni abundance tied to Fe) then the predicted ratio for the
two-temperature model rises to 0.15. This excellent agreement for the
two-temperature model is not shared by the corresponding isothermal model
with variable relative abundances for only the $\alpha$-process
elements. All the models predict a value of zero for the S ratio which
is consistent with the data.

Only the two-temperature model predicts equivalent widths for each Si and S
blend that are at least marginally consistent with the data. The
largest deviation is found for Si XIV where the isothermal model
produces too small an equivalent width. However, the cooling flow
model generally has the most deviations: $\Delta_{\rm EW} =
(2.38,1.16,5.48)$ for respectively the best-fit isothermal,
two-temperature, and cooling flow models. Similar to the results from the
broad-band analysis, the relatively poor performance of the cooling
flow for NGC 4636 is probably in large part due to the restriction
that the model have relative abundances fixed at their solar values.

The equivalent widths of Si and S and especially the Si ratio give
strong support for multiphase gas in NGC 4636 (or, rather, gas hotter
than predicted by the isothermal models). Although the broad-band
fits slightly favor the two-temperature model (Table
\ref{tab.quality}), the difference in $\chi^2$ between the
isothermal and two-temperature models is not compelling: i.e. the Si and S
lines give stronger evidence for multiphase gas in NGC 4636 than do
the broad-band fits.

Recently Loewenstein \shortcite{lconf} has argued that the Si XIV/XIII
ratio of NGC 4636 is consistent with isothermal gas. From inspection
of Figure 4 of Loewenstein we find he derives a value of $\sim 0.15$
for the Si ratio from the data in excellent agreement with our value
for the summed SIS data in Table \ref{tab.ratios}. He also asserts
that this ratio is consistent with isothermal gas with a temperature
$\sim 0.75$ keV. As we have shown in Table \ref{tab.params}
isothermal models (both MEKAL and RS) with variable relative
abundances have temperatures $\sim 0.66$ keV which, consistent with
Loewenstein's figure, cannot produce the Si ratios; note that the
isothermal models (both MEKAL and RS) with relative abundances fixed
at their solar values do have temperatures near 0.75 keV but their
fits are substantially worse and have very poor null hypothesis
probabilities (see Table \ref{tab.quality}). Thus, the Si ratio of NGC
4636 cannot be produced by isothermal models which also best match
data over the rest of the {\sl ASCA} spectrum.

We mention that for all of the galaxies the equivalent widths and Mg
XII/XI ratios (where available) predicted by the models are mostly
consistent with the data within the $2\sigma$ errors . The most notable
discrepancy is for NGC 1399 where the isothermal model substantially
under-predicts the equivalent width of Mg XII while the multi-phase
models substantially over-predict the equivalent width.

Hence, for all of the galaxies we find that the line equivalent widths
and H-like/He-like ratios for Si and S generally favor multiphase
models over isothermal.  For NGC 1399 and NGC 4472 the isothermal
models tend to under-predict the equivalent widths which is due to the
smaller abundances of those models. The better performance of the
multiphase models of NGC 4636 and NGC 5044 is primarily the result of
the need for a higher temperature component than present in the
isothermal models.

\section{Radial temperature gradients from {\sl ROSAT}}
\label{rosat}

The evidence we have obtained for multiphase gas with approximately
solar abundances rests on analysis of the aggregate {\sl ASCA} spectra
within a radius of $\sim 5\arcmin$ for each galaxy. The data of these
galaxies available from the {\sl ROSAT} satellite \cite{rosat} do not
provide useful ``aggregate'' spectral constraints over those obtained
from the {\sl ASCA} data because of the limited energy resolution
($\Delta E\sim 500$ eV at 1 keV) and bandwidth (0.1-2.4 keV) of the
{\sl ROSAT} Position Sensitive Proportional Counter (PSPC) with
respect to {\sl ASCA}. However, the superior spatial resolution of the
PSPC (PSF $\sim 30\arcsec$ FWHM) allows the radial variation of the
spectra to be probed within the $\sim 5\arcmin$ apertures.

Although the PSPC cannot distinguish very well between complex
spectral models of elliptical galaxies, it can provide relatively
tight constraints on a model consisting of a single temperature
component (e.g. Buote \& Canizares 1994; Trinchieri et al. 1994).
This is significant if the aggregate (multiphase) spectra within the
$\sim 5\arcmin$ {\sl ASCA} apertures can be approximated as a series
of isothermal components; i.e. one temperature for each radial bin
within the aperture. This is a reasonable approximation if in each
radial bin one temperature component dominates the emissivity. 

Previous analyses of the PSPC data of these galaxies using
isothermal models in radial bins generally have found significant
temperature gradients (Forman et al. 1993; David et al. 1994;
Trinchieri et al. 1994; Rangarajan et al. 1995; Irwin \& Sarazin 1996;
Jones et al. 1997). The temperature profiles are qualitatively similar
for all of these galaxies: starting from a minimum at the center the
temperature rises sharply out to a radius similar to our {\sl ASCA}
apertures and then gently declines at larger radii. The abundances
also rise and then fall, typically having approximately solar
abundances within the radii of our {\sl ASCA} apertures (except for
the innermost radial bin) and smaller abundances at larger
radii. Excess absorption is not found from analysis of PSPC data of
these ellipticals (e.g. section 4 of BF) except by Rangarajan et
al. \shortcite{vijay} for the central radial bin of NGC 1399.

This evidence for multitemperature structure with approximately solar
abundances within $\sim 5\arcmin$ radii of NGC 1399, NGC 4472, NGC
4636, and NGC 5044 appears to corroborate our single-aperture {\sl
ASCA} spectral analysis. We do not expect the spectral models derived
from the PSPC data to agree exactly with our {\sl ASCA} models because
a single-temperature model for each radial bin may not be an accurate
representation. It is our goal in this section to assess the extent to
which the temperature profiles derived from the PSPC data agree with
the isothermal and multiphase models obtained from our {\sl ASCA}
spectral analysis.

\subsection{Observations and data reduction}
\label{rosobs}

\begin{table*}
\begin{minipage}{110mm}
\caption{{\sl ROSAT} PSPC Observations}
\label{tab.rosobs}
\begin{tabular}{lcccc}
& NGC 1399 & NGC 4472 & NGC 4636 & NGC 5044\\ \\[-7pt]
Sequence & rp600043n00 & rp600248n00 & rp600016n00 & rp800020n00\\
Exposure (ks) & 23.4 & 25.4 & 10.8 & 24.3\\

\end{tabular}

\medskip

The exposures include time filtering. 

\end{minipage}
\end{table*}

We obtained {\sl ROSAT} PSPC observations from the HEASARC archive. The
observation sequences are listed in Table \ref{tab.rosobs}. The data
were reduced using the standard {\sc FTOOLS} (v4.1) software according
to the procedures described in the OGIP Memo OGIP/94-010 (``ROSAT data
analysis using xselect and ftools'') and the WWW pages of the {\sl
ROSAT} Guest Observer Facility (GOF) (See
http://heasarc.gsfc.nasa.gov/docs/rosat.) 

The events files of each observation were cleaned of after-pulse
signals by removing any events following within 0.35 ms of a
precursor. We corrected the Pulse Invariant (PI) bins of each data set
for spatial and long-term temporal gain variations using the most
up-to-date calibration files. For the {\sc ftool} {\sc pcecor}, which
corrects for the variation in the linearity of the PSPC response, we
used the in-flight calibration data for the correction. We then
removed large fluctuations in the light curves indicative of scattered
light from the Bright Earth, Sun, or SAA. The resulting filtered
exposure times are listed in Table \ref{tab.rosobs}.

We generated images in the 0.5-2 keV band from the reduced PCPC data
of each galaxy. Within the radii defined by the apertures used in our
{\sl ASCA} analysis (see section \ref{obs}) the surface brightness
distributions appear mostly symmetrically distributed about the center
of each system with the possible exception of NGC 4472 which has a
weak enhancement in the surface brightness approximately 2.5
arcminutes SE of the center. (We also confirm the N-S asymmetry in the
surface brightness at distances larger than our {\sl ASCA} radii for
NGC 1399 as noted by Jones et al. 1997.)

For each galaxy we extracted spectra in a series of radial bins
centered by eye on each galaxy center. The bin sizes were chosen to
agree with those of Jones et al. \shortcite{jones} for NGC 1399, Irwin
\& Sarazin \shortcite{is} for NGC 4472, Trinchieri et
al. \shortcite{trin} for NGC 4636, and David et
al. \shortcite{david94} for NGC 5044. The largest radial bin used for
each galaxy has inner radius smaller than the radius of the
corresponding aperture used in our {\sl ASCA} analysis. 

We regrouped the PI bins of each spectrum so that each group had at
least 20 counts as done for the {\sl ASCA} data (see section
\ref{obs}). We generated ARF files for the spectrum of each radial bin
using the {\sc FTOOL pcarf} being careful to correctly assign the
deltx and delty keywords as described on the {\sl ROSAT} GOF WWW
pages. The RMF files were taken from the HEASARC database and used for
each galaxy according to when the galaxy was observed:
pspcb\_gain1\_256.rmf for NGC 1399 and NGC 5044; pspcb\_gain2\_256.rmf
for NGC 4472 and NGC 4636. Similar to the {\sl ASCA} data analysis,
the response matrix for each PSPC spectrum is the product of the RMF
and ARF files.

Finally, we obtained background spectra from source-free regions at
large off-axis radii similar to those used in the previous {\sl ROSAT}
studies. To correct for exposure differences between source and
background spectra we multiplied the background spectra by a constant
factor representing the ratio of effective exposures at the
position of the given source annulus to that of the background. 

\subsection{Spectral analysis of PSPC data}
\label{rosspec}

\begin{table*}
\caption{{\sl ROSAT} PSPC Spectral Properties of Central and Outermost
Radial Bins} 
\label{tab.center}
\begin{tabular}{ccccccccc}
& \multicolumn{2}{c}{$N_{\rm H}$} & \multicolumn{2}{c}{$T$} &
\multicolumn{2}{c}{$Z$} & \multicolumn{2}{c}{Goodness-of-Fit} \\ 
Radius & \multicolumn{2}{c}{($10^{21}$ cm$^{-2}$)} &
\multicolumn{2}{c}{(keV)} & \multicolumn{2}{c}{($Z_{\sun}$)} & 
\multicolumn{2}{c}{($P/\chi^2/\rm dof$)} \\
(arcmin) & 0.2-2 keV & 0.5-2 keV & 0.2-2 keV & 0.5-2 keV & 0.2-2 keV &
0.5-2 keV & 0.2-2 keV & 0.5-2 keV\\
NGC 1399:\\
0-1 & $0.15_{-0.03}^{+0.03}$ & $1.05_{-0.51}^{+0.66}$ &
$0.91_{-0.03}^{+0.03}$ & $0.81_{-0.07}^{+0.06}$ &
$0.51_{-0.10}^{+0.14}$ & $0.30_{-0.07}^{+0.10}$ &  
0.11/145.8/126 & 0.38/104.9/101\\
4-6 & $0.11_{-0.03}^{+0.03}$ & $0.20_{-0.20}^{+0.58}$ &
$1.49_{-0.10}^{+0.13}$ & $1.46_{-0.20}^{+0.18}$ &
$0.81_{-0.23}^{+0.37}$ & $0.70_{-0.42}^{+0.63}$ & 
0.61/133.9/139 & 0.84/95.3/110\\
NGC 4472:\\
0-1 & $0.07_{-0.03}^{+0.03}$ & $0.58_{-0.58}^{+0.70}$ &
$0.87_{-0.02}^{+0.02}$ & $0.83_{-0.05}^{+0.05}$ &
$1.12_{-0.24}^{+0.41}$ & $1.62_{-0.64}^{+2.37}$ & 
0.12/146.3/127 & 0.23/111.0/101\\
4-7 & $0.27_{-0.14}^{+0.34}$ & $1.20_{-0.72}^{+0.94}$ &
$1.34_{-0.06}^{+0.07}$ & $1.20_{-0.14}^{+0.13}$ &
$2.57_{-1.30}^{+4.16}$ & $0.98_{-0.49}^{+1.38}$ &
0.53/137.9/140 & 0.63/104.5/110\\
NGC 4636:\\ 
0-1 & $0.15_{-0.06}^{+0.05}$ & $0.92_{-0.82}^{+1.04}$ &
$0.64_{-0.02}^{+0.02}$ & $0.58_{-0.08}^{+0.06}$ &
$0.56_{-0.17}^{+0.34}$ & $1.0(>0.5)$ &
0.11/119.7/102 & 0.05/100.9/79\\
4-6 & $0.10_{-0.10}^{+0.18}$ & $1.04_{-1.04}^{+2.48}$ &
$0.96_{-0.06}^{+0.06}$ & $0.88_{-0.19}^{+0.13}$ & $1.5(>0.6)$ &
$1.7(>0.5)$ & 0.77/67.8/77 & 0.82/43.6/53\\
NGC 5044:\\
0-1 & $0.40_{-0.04}^{+0.04}$ & $1.54_{-0.48}^{+0.65}$ &
$0.81_{-0.01}^{+0.01}$ & $0.71_{-0.05}^{+0.04}$ &
$0.72_{-0.12}^{+0.17}$ & $0.69_{-0.16}^{+0.39}$ & 
0.01/193.7/152 & 0.05/149.9/123\\
5-6 & $0.47_{-0.13}^{+0.17}$ & $0.32_{-0.32}^{+0.55}$ &
$1.24_{-0.07}^{+0.08}$ & $1.27_{-0.12}^{+0.11}$ &
$0.87_{-0.31}^{+0.61}$ & $0.84_{-0.42}^{+0.86}$ &
0.80/103.9/117 & 0.90/79.0/96\\

\end{tabular}

\medskip
\raggedright

Results of fitting a single absorbed MEKAL model to the PSPC spectra
over the energy ranges 0.2-2 keV and 0.5-2 keV for the central radial
bin and outermost radial bin (which overlaps the {\sl ASCA} SIS
aperture) for each galaxy. Quoted errors are 90\% confidence on one
interesting parameter ($\Delta\chi^2=2.71$).

\end{table*}

\begin{figure*}
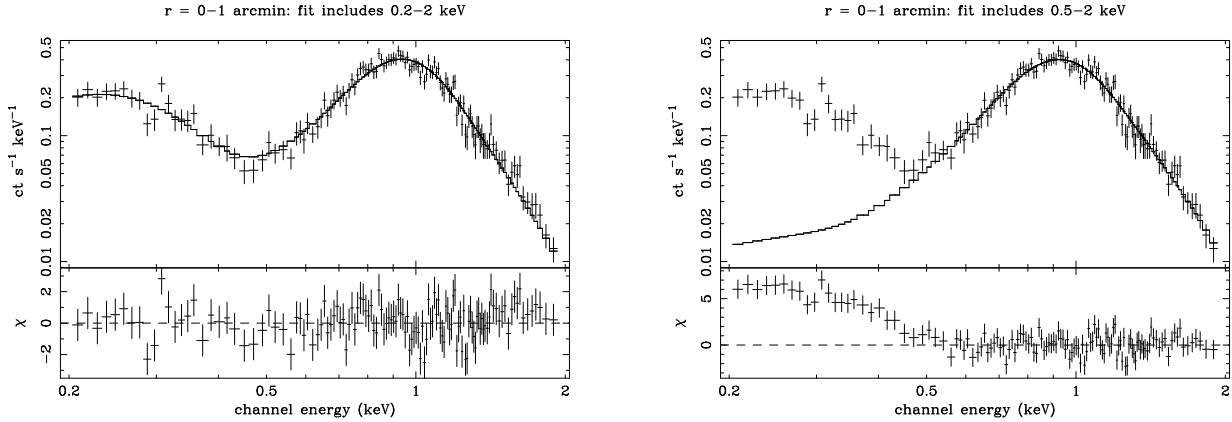

\parbox{0.49\textwidth}{
\centerline{\psfig{figure=fig11a.ps,angle=-90,height=0.23\textheight}}
}
\parbox{0.49\textwidth}{
\centerline{\psfig{figure=fig11b.ps,angle=-90,height=0.23\textheight}}
}
\caption{\label{fig.pspc} Absorbed MEKAL models fit to the {\sl ROSAT}
PSPC data of the central radial bin ($r=0\arcmin$-$1\arcmin$) and the
outer radial bin ($r=4\arcmin$-$6\arcmin$) of NGC 1399. The left panel
shows the result of fitting the entire 0.2-2 keV energy range while
the right panel shows the result of fitting only the 0.5-2 keV energy
range. (The best-fitting model parameters are listed in Table
\ref{tab.center}.)}

\end{figure*}

For each galaxy we fit a single absorbed MEKAL model to the 0.2-2 keV
PSPC spectrum of each radial bin. In Table \ref{tab.center} we list
the results for the central radial bin and outermost radial bin for
each galaxy. The column densities and temperatures agree with previous
studies within our 90\% confidence limits. As expected, the plasma
code differences are most pronounced for the abundances which do show
significant disagreement in some cases; e.g. Irwin \& Sarazin obtain
$Z=0.4Z_{\sun}$ in the $0\arcmin$ - $1\arcmin$ bin which is $\sim 1/3$
the value we obtained from the MEKAL fit.

In all we are able to reproduce to reasonable accuracy the rising
temperature gradients found in previous studies within the radii
defined by our {\sl ASCA} SIS apertures considering that we use the
MEKAL plasma codes whereas previous studies used RS. We also find
abundance gradients similar to previous studies. It should be
mentioned that the abundances rise and then fall for NGC 1399 and NGC
5044 within the inner and outer apertures listed in Table
\ref{tab.center}; i.e. the abundance reaches a maximum of $Z\sim
1.5Z_{\sun}$ for $r=2\arcmin$-$4\arcmin$ for NGC 1399, and for NGC
5044 the maximum of $Z\sim 1Z_{\sun}$ is achieved at
$r=2\arcmin$-$3\arcmin$.

Let us define the temperature of the central bin as $T_{\rm in}$ and
the temperature of the outer bin as $T_{\rm out}$. The values of
$T_{\rm in}$ and $T_{\rm out}$ listed for the 0.2-2 keV fits in Table
\ref{tab.center} qualitatively agree with respectively the values of
$T_{\rm c}$ and $T_{\rm h}$ that we derived for the two-temperature models
fitted to the {\sl ASCA} spectra (Table \ref{tab.params}).  The values
of $T_{\rm in}$ tend to exceed $T_{\rm c}$ and actually agree better
with the emission-weighted temperatures of the cooling flow models for
NGC 1399, NGC 4472, and NGC 5044 (see section \ref{cf}).  The
radially-averaged abundances of the PSPC fits agree, within their
large errors, with the 1-2 solar abundances of the two-temperature and
cooling flow models obtained from the {\sl ASCA} data.

The column densities in the central radial bins determined from 0.2-2
keV fits to the PSPC spectra do not exceed the Galactic columns in
stark contrast to the excess absorption demanded by the colder
components of the multiphase {\sl ASCA} models. This well-known
discrepancy in $N_{\rm H}$ between {\sl ROSAT} and {\sl ASCA} studies
has been discussed in the context of ellipticals by BF (see their
section 4). The origin of this discrepancy is still controversial as
it may be due to a calibration error in one of the instruments, or it
may be an artifact of the models used to fit the data.

We have re-examined this issue using these {\sl ROSAT} PSPC data.  In
every case the isothermal model has the worst fit in the central bin
for each galaxy and has a formally marginal $\chi^2$ null hypothesis
probability $\sim 0.1$ ($\sim 0.01$ for NGC 5044). Moreover, in the
central bin $N_{\rm H}$ actually tends to be {\em less} than the
Galactic value, especially for NGC 1399 and NGC 5044. That is, a
isothermal MEKAL plasma modified by Galactic absorption does not
produce enough X-ray emission at the lowest energies in the 0.2-2 keV
PSPC band.

If this excess soft X-ray emission is due to other temperature
components, then we would expect that the parameters derived from
fitting an isothermal model should depend on the energy range over
which the model is fitted. When restricting the fits to the energy
range 0.5-2 keV we indeed find significant differences in the
isothermal model parameters of the central bins for NGC 1399 and NGC
5044 (see Table \ref{tab.center}). The most substantial change is for
the column densities: NGC 1399 and NGC 5044 require excess absorption
that agrees much better with the multiphase models derived from the
{\sl ASCA} data. (Using somewhat different arguments, Rangarajan et
al. 1995 also found excess absorption at the center of NGC 1399 in the
PSPC data.) The slightly smaller temperatures also agree better with
the two-temperature {\sl ASCA} models. Finally, the quality of the
fits is also improved most for these galaxies when fitting only to
0.5-2 keV. NGC 4636 and NGC 4472 appear to follow similar trends but
their changes for $N_{\rm H}$ and $T_{\rm in}$ are not as significant.

These effects occur only for the central radial bin of each
galaxy. Although some hints of these model differences due to fitting
over different energy ranges exist in the second radial bins, the
differences diminish rapidly with increasing radius. In particular,
for the outer bins listed in Table \ref{tab.center} the column
densities, temperatures, and abundances do not change within their
90\% confidence limits for NGC 1399 and NGC 5044 when restricting the
fits to 0.5-2 keV. When fitted over the restricted energy range of
0.5-2 keV the column densities for the outer bins of NGC 4472 and NGC
4636 become very uncertain. Although the column density for NGC 4636
is consistent with no change, some excess $N_{\rm H}$ is implied
within the 90\% confidence limit for NGC 4472.

The spectra of the central and outer bins are therefore qualitatively
different for each galaxy (see Figure \ref{fig.pspc}). Focusing our
attention on NGC 1399 and NGC 5044 we see that the outer bin is very
consistent with isothermal gas with Galactic absorption (regardless
of the energy range fitted). For the central bin of these systems the
marginal quality of the fits over 0.2-2 keV coupled with the
significant differences in model parameters when confining the fits to
0.5-2 keV implies that the isothermal model is inadequate at the
centers. As the panels in Figure \ref{fig.pspc} vividly demonstrate,
the isothermal model with excess absorption which best describes the
0.5-2 keV spectrum under-predicts the emission at lower energies. The
excess emission at lower energies is consistent with the existence of
additional lower temperature components.

Hence, the PSPC spectra (in particular for NGC 1399 and NGC 5044)
indicate significant excess absorption for the central radial bin
which decreases rapidly with increasing radius. The central bins also
have the coldest temperatures and thus excess absorption is tied to
the coldest temperature components. These results concur with the
multiphase models obtained from the ``single-aperture'' analysis of
the {\sl ASCA} data in sections \ref{2t} and \ref{cf}. Moreover, the
temperatures of the central and outer bins of the PSPC data agree
reasonably well with those of the {\sl ASCA} multiphase models. Though
the abundances derived from the PSPC data are uncertain, they are
approximately solar and are consistent with the multiphase {\sl ASCA}
models.

\subsection{Simulated {\sl ASCA} observations of {\sl ROSAT} PSPC
models} 
\label{rosasca}

\begin{table*}
\caption{Radial Variation of Isothermal Models of the  {\sl ROSAT} PSPC
data of NGC 1399} 
\label{tab.n1399}
\begin{tabular}{ccccccccc}
Radius & $N_{\rm H}$ & $T$ & $Z$ & $EM$\\
(arcmin) & ($10^{21}$ cm$^{-2}$) & (keV) & ($Z_{\sun}$) &
$(10^{-17}n_en_pV/4\pi D^2)$\\
0-1 & 1.05 & 0.81 & 0.30 & 3.65\\
1-2 & 0.22 & 1.35 & 1.10 & 0.83\\
2-4 & 0.34 & 1.44 & 1.11 & 1.90\\
4-6 & 0.20 & 1.46 & 0.70 & 2.04\\
\end{tabular}

\medskip
\raggedright

Best-fitting absorbed MEKAL model for the PSPC data of each radial bin
of NGC 1399. Only the energies 0.5-2 keV are included in these fits
which is most important for the central bin as discussed in section
\ref{rosspec}.  (See Table \ref{tab.center} for errors on the
parameters of the central and outer bins.)

\end{table*}

\begin{figure*}
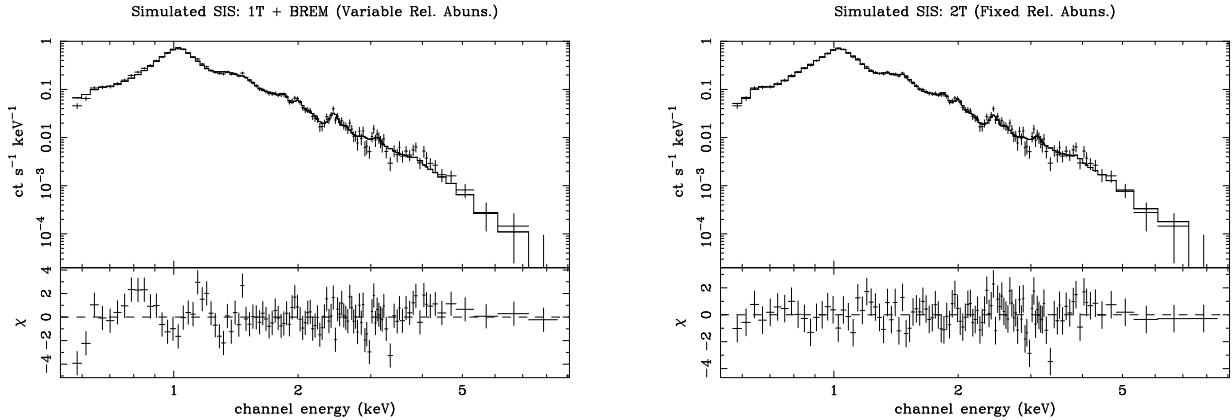

\parbox{0.49\textwidth}{
\centerline{\psfig{figure=fig12a.ps,angle=-90,height=0.23\textheight}}
}
\parbox{0.49\textwidth}{
\centerline{\psfig{figure=fig12b.ps,angle=-90,height=0.23\textheight}}
}
\caption{\label{fig.sim} Simulated {\sl ASCA} SIS data of NGC 1399
constructed by summing the radially varying isothermal models
obtained from the {\sl ROSAT} PSPC data. The parameters of the
composite PSPC models are listed Table \ref{tab.n1399} except that
normalizations are rescaled to match the real {\sl ASCA} data. The
best-fitting 1T+BREM model with variable relative abundances (left)
and two-temperature model with relative abundances fixed at their
solar values (right) are shown. Both models use the MEKAL plasma
code.}

\end{figure*}

We have shown that there is good overall agreement between the
multiphase models obtained for the {\sl ASCA} data and the radial
temperature profiles obtained by representing the {\sl ROSAT} PSPC
spectra as a series of isothermal models as a function of radius
within the radii defined by the {\sl ASCA} apertures.  These
temperature gradients implied by the PSPC, however, appear to be very
inconsistent with the isothermal {\sl ASCA} models for NGC 1399 and
NGC 5044. Owing to the larger uncertainties for NGC 4472 and NGC 4636,
the PSPC temperature gradients do not clearly favor any of the {\sl
ASCA} models over another.

We would like to better assess the extent to which the temperature
gradients inferred from the PSPC data favor the multiphase {\sl ASCA}
models. Although joint fitting of the PSPC and {\sl ASCA} data, as
mentioned above, is not useful, an approach that takes advantage of
the spatial resolution of the PSPC is to construct a model for each
galaxy that is the sum of the isothermal models within the {\sl
ASCA} aperture. This composite model can then be fitted directly to
the {\sl ASCA} data. The only free parameters are the total
normalization and the redshifts (because of the calibrations issues
discussed in section \ref{calib}).

By summing the best-fitting isothermal models over the radial bins
within the {\sl ASCA} aperture, we constructed composite PSPC models
for each system; in Table \ref{tab.n1399} we list the parameters of
NGC 1399 for illustration.  We use the PSPC models derived from
fitting the restricted energy range 0.5-2 keV since the isothermal
approximation is better suited for the central bin of each galaxy for
this energy range as discussed above. Upon fitting these models to the
{\sl ASCA} SIS data we obtain good qualitative fits over the whole SIS
energy range for NGC 1399 and NGC 5044. For each of these galaxies
there is a substantial residual near 1 keV which prevents a good
formal fit; this residual can be largely removed in each case by
changing the individual temperatures of the composite PSPC model
within their 90\% errors.

Agreement can not be so easily achieved with NGC 4472 and NGC 4636
because, in addition to residuals near 1 keV similar to NGC 1399 and
NGC 5044, the PSPC models are clearly deficient in emission above
$\sim 3$ keV. This is not surprising since the limited bandwidth of
the PSPC does not allow strong constraints on components (from plasma
or discrete sources) which have temperatures $\ga 3$ keV. Upon adding
a bremsstrahlung model to the PSPC model of NGC 4636 the fit improves
substantially. Interestingly, the residuals of the fit of this
PSPC+BREM model for NGC 4636 are almost identical in shape to those of
the cooling flow model fit to the {\sl ASCA} data in Figure
\ref{fig.n4636_multit}.

Perhaps a more interesting comparison of the {\sl ROSAT} and {\sl
ASCA} models is achieved via simulation. That is, for each of the
composite isothermal PSPC models we simulate an {\sl ASCA} SIS
observation appropriate for each galaxy; i.e. the same exposure and
response as the total summed SIS data (see section \ref{obs}). (Prior
to each simulation the PSPC models are re-scaled to best match the
flux of the SIS data though the relative normalizations of the
individual components are preserved.)  We then analyzed these
simulated data analogously to the real {\sl ASCA} data as done in
section \ref{broad}. Our goal is to examine whether fitting these
simulated {\sl ASCA} data can reproduce in detail the rich behavior
found when fitting the real {\sl ASCA} data.

In Figure \ref{fig.sim} we plot the simulated SIS data for NGC 1399
with the best-fitting isothermal and two-temperature models. The
isothermal model consists of an absorbed MEKAL model with variable
relative abundances and a bremsstrahlung component modified by
Galactic absorption. This 1T+BREM model delivers a marginally
unacceptable fit to the simulated SIS data: $P=10^{-3}$,
$\chi^2=181.1$ for 127 dof. The residuals near 1 keV are very similar
to those in Figure \ref{fig.n1399_1t} for the real SIS data, though of
somewhat smaller magnitude. (Note that the 1T+BREM model with relative
abundances fixed at their solar values is a considerably worse fit
$(P=2\times 10^{-5})$ and has larger residuals near 1 keV in better
agreement with Figure \ref{fig.n1399_1t}.)  The best-fitting
temperature, $T=1.20$ keV, and Fe abundance, $Z_{\rm
Fe}=0.39Z_{\sun}$, are very similar to those determined from the real
{\sl ASCA} data (Table \ref{tab.params}); the column density, $N_{\rm
H}=0.13 \times 10^{21}$ cm$^{-2}$ is somewhat smaller. Moreover, the
Ne abundance is zero for this 1T+BREM fit and the O and Mg abundances
are $\sim 1/2$ that of Fe.

The two-temperature (Figure \ref{fig.sim}) and multiphase cooling flow
models are excellent fits to the simulated SIS data of NGC 1399 and
also have parameters in good overall agreement with those determined
from the real {\sl ASCA} data (Tables \ref{tab.params} and
\ref{tab.cf}): $P=(0.59,0.48)$, $\chi^2=(126.6,132.3)$ for dof =
(131,132) for the 2T and CF+1T model respectively. (Both models have
relative abundances with respect to Fe fixed at their solar values and
note that a bremsstrahlung components is not included since it does
not improve the fits in either case.)  The values of $T_{\rm c}=0.73$
keV and $T_{\rm h}=1.36$ keV for the 2T model and $T=1.36$ keV for the
cooling-flow model agree reasonably well with the real {\sl ASCA}
fits.

We also reproduce the Galactic columns on the hotter components and
excess absorption on the colder components, though the excess is still
a factor of two below the values listed in Table \ref{tab.params} and
\ref{tab.cf} for NGC 1399. Both the cooling-flow and two-temperature models
give $Z=0.73Z_{\sun}$ which is about half that obtained from the
actual {\sl ASCA} data. However, this is consistent with the PSPC
model listed in Table \ref{tab.n1399} but should not be considered
``real'' in particular because the low abundance of the inner bin is
probably an artifact of fitting a isothermal model to a spectrum
composed of multiple temperature components.

The fits to the simulated SIS data of NGC 5044 behave analogously to
NGC 1399 and essentially reproduce the isothermal and multiphase
results of fitting the real {\sl ASCA} data. The 1T+BREM models,
though, deserve special mention since we find that fitting the
simulated SIS data requires a Ne abundance of $\sim 0.9$ solar and a
Ni abundance of $\sim 1$ solar. The Ne and Ni abundances are very
different from Fe (0.3 solar) and allowing them to be free parameters
significantly improved the fits as also found when fitting the real
{\sl ASCA} data. Thus, these peculiar abundances for the 1T+BREM
models do not have to be explained by problems with the MEKAL plasma
code.

As expected, the results for NGC 4472 are similar to NGC 1399 and NGC
5044 except that the derived temperatures are lower than obtained from
the real {\sl ASCA} data which can be attributed to the deficiency in
emission above $\sim 3$ keV mentioned above.  On the other hand, we
find that fitting the simulated SIS data for NGC 4636 gives results
very similar to those found for NGC 1399 and NGC 5044 except that the
1T model with variable relative abundances is an acceptable fit;
unlike the real {\sl ASCA} data, we do not require a bremsstrahlung
component for the PSPC models consistent with the deficiency of
emission above 3 keV mentioned above. For the most part the parameters
derived for the 1T model agree very well with those listed in Table
\ref{tab.params} for the 1T+BREM fits to the real {\sl ASCA}
data. Even super-solar Na abundance is required, though the best-fit
value of 5.9 solar is less than that listed in Table
\ref{tab.params}. 

In contrast to the broad-band fits of the real {\sl ASCA} data of NGC
4636, the two-temperature models are much better fits to the simulated {\sl
ASCA} SIS data than the isothermal models and they are not improved
significantly when allowing for variable relative abundances. The
best-fitting temperatures of the 2T model actually agree well with
those found for the real data in Tables \ref{tab.params} and
\ref{tab.cf}; i.e. $T_{\rm c}=0.56$ keV and $T_{\rm h}=0.81$
keV. Although the best-fitting abundance $(Z=1.16Z_{\sun})$ is larger
than that inferred from the real data, there is considerable leverage
in the abundance errors in the PSPC models to resolve this
discrepancy. Similarly, excess absorption is implied for both
components (similar amounts on each component) of the 2T model, much
of which can be reduced within the errors of 0.5-2 keV PSPC fits (see
Table \ref{tab.center}). Overall the fits of the multiphase models to
the simulated {\sl ASCA} data of NGC 4636 behave quite similarly to
the fits of the real (and simulated) {\sl ASCA} data of the other
galaxies. It very likely that that the emission above $\sim 3$ keV
lacking in the {\sl ROSAT} PSPC isothermal models of NGC 4636
accounts for the differences in the multiphase fits to the real and
simulated SIS data.

In conclusion, the temperature gradients constructed by radially
varying isothermal models of the {\sl ROSAT} PSPC data are very
consistent with the {\sl ASCA} data of NGC 1399, NGC 4472, and NGC
5044 and strongly favor multiphase models of the hot gas in these
systems.  Our analysis also shows that many of the peculiar abundances
derived for the isothermal models of the real {\sl ASCA} data (Table
\ref{tab.params}) of all the galaxies can be produced by forcing an
isothermal model to fit the multiphase spectrum generated by the
composite PSPC models.  Although the PSPC models suggest the spectral
properties of NGC 4636 are quite similar to the other galaxies, strong
conclusions cannot be drawn from the radially varying isothermal
models of the PSPC data of NGC 4636 and NGC 4472 since they fail to
produce the {\sl ASCA} emission above $\sim 3$ keV.

\section{Discussion}
\label{disc}

\subsection{Evidence for a ``Standard Model''?}  
\label{standard}

As remarked in the Introduction most previous studies of ellipticals
based on {\sl ASCA} data within a single aperture find that the
spectra of these objects are well represented by a component of
isothermal hot gas and (if needed) a high temperature bremsstrahlung
component to account for possible emission from discrete sources. The
evidence for this ``Standard Model'' is reviewed by Loewenstein \&
Mushotzky \shortcite{lmushy} and Loewenstein \shortcite{lconf}. These
review articles go further and assert that the K$\alpha$ emission line
ratios Si XIV/Si XIII favor the isothermal models obtained from
broad-band spectral fitting and are at best marginally consistent with
the multiphase models described in BF.

The accumulated body of evidence from our X-ray analysis of these
galaxies strongly favors multiphase models for the hot gas over
isothermal models. First, in every case the broad-band spectral fits
of the {\sl ASCA} data within a single aperture rule out the
isothermal models having relative abundances with respect to Fe fixed
at the solar values. Even if the relative abundances are allowed to
vary the isothermal models remain poor fits for NGC 1399, NGC 4472,
and NGC 5044 especially when the MEKAL model (with its superior
modeling of the Fe L shell transitions with respect to RS) is used.
The multiphase models with relative abundances fixed at their solar
values provide superior fits to these galaxies.  (The two-temperature
model for NGC 4636 is also a slightly better fit though it requires
variable relative abundances.)

We have shown in section \ref{lines} that isothermal models
generally under-predict the equivalent widths of the Si and S lines of
NGC 1399, NGC 4472, NGC 5044, and NGC 4636 while much better agreement
with the {\sl ASCA} data is provided by multiphase models. The
multiphase models, as a whole, also better describe the measured Si
XIV/Si XIII and S XVI/XV ratios.  Even for NGC 4636, which displays
the smallest improvement for the two-temperature model over the isothermal
model for the broad-band spectral fits (section \ref{broad}), the line
ratios and equivalent widths of Si and S clearly favor the two-temperature
model. Hence, whereas the previously cited review articles found
support for isothermal models using only the Si ratio, we find that
consideration of both the ratios and equivalent widths of the Si and S
line blends strongly favors the multiphase models.

Our final evidence is obtained by exploiting the superior spatial
resolution of the {\sl ROSAT} PSPC. The PSPC data are generally
consistent with a series of isothermal models which vary with radius
within the apertures used for analysis of the {\sl ASCA} data. Using
different approaches (sections \ref{rosspec} and \ref{rosasca}) we
find that the temperature gradients inferred from the {\sl ROSAT} PSPC
data (section \ref{rosat}) are inconsistent with the isothermal
models derived from fitting the {\sl ASCA} data within a
single-aperture but are very consistent with the multiphase models,
especially for NGC 1399 and NGC 5044.

In light of this evidence we would argue that the Standard Model for
these brightest ellipticals should be revised as follows: (1) the
spectra within the {\sl ASCA} apertures consist of at least two
temperature components for the hot gas; (2) the temperature declines
towards the center as indicated by the temperature profiles deduced
from the {\sl ROSAT} PSPC; (3) the Fe abundances are approximately 1-2
solar and the relative abundances of the elements with respect to Fe
are also consistent with solar; (4) excess absorption above the
Galactic value is indicated for the colder temperature components
which declines rapidly with increasing radius until the absorption is
consistent with the Galactic value for the hotter temperatures at the
outer radii; (5) the {\sl ASCA} spectra are also consistent with a
bremsstrahlung component arising from discrete sources contributing
$\la 0.1\%$ to the total emission measure.

To escape this version of a Standard Model (at least for these
brightest ellipticals) one has to invoke serious errors in the plasma
codes and/or the calibration of the {\sl ASCA} SIS. (Of course, errors
of such magnitude would render invalid all existing results of
significance for both isothermal and multiphase models.)  The
relative calibration of {\sl ASCA} and the {\sl SAX} satellite has
been examined in detail and no serious discrepancies have been found
apart from the small differences in the energy response below 1 keV
discussed in section \ref{calib} (e.g. Orr et al. 1997).

The accuracy of the Fe L shell emission lines in the plasma codes has
been tested previously by Hwang et al. \shortcite{una} who used {\sl
ASCA} SIS data to show that the Fe abundance of M87 deduced from
analysis of the Fe K shell lines near 6.5 keV is very consistent with
that inferred from broad-band spectral fitting that includes the Fe L
shell emission lines. Since the average temperature of M87 is $\sim 2$
keV, this consistency test may not be entirely appropriate for normal
ellipticals with $T\la 1$ keV. However, BF tested the accuracy of the
Fe L shell lines for normal ellipticals by comparing the temperatures
and abundances derived from fitting MEKAL and RS models over a
restricted energy region where the Fe L lines dominate the
spectrum. These results were then compared to the fits with the Fe L
energy region excluded. They found good agreement between the
temperatures and abundances in both cases, especially for the MEKAL
model.

In our present paper we have found that excellent fits to both the
{\sl ASCA} SIS and GIS data are achieved by multiphase models that use
the MEKAL plasma code. Comparison of fits using the MEKAL and RS
plasma codes shows that, as expected, the MEKAL code is more accurate
and also that remaining inaccuracies in the code are most likely
insufficient to negate the evidence in support of multiphase models
(section \ref{ass}). Hence, we do not find any evidence that errors in
the plasma codes near the Fe L energy region (or calibration errors)
are sufficient to change qualitatively the evidence for multiphase hot
gas with approximately solar abundances presented in this paper.

\subsection{Implications of solar abundances}
\label{abun}

Recently Arimoto et al. \shortcite{arimoto} have shown that the
stellar Fe abundances deduced from optical observations of a large
sample of ellipticals are approximately solar when averaged over an
effective radius for each galaxy. This agrees quite well with the Fe
abundances predicted by the multiphase models computed within the
$r\sim 5\arcmin$ {\sl ASCA} apertures of the ellipticals in our sample
but is inconsistent with the sub-solar Fe abundances predicted by the
isothermal models. Comparable Fe abundances in stars and the hot gas
do not reconcile the standard chemical models where the hot gas is
enriched by Type Ia supernovae (e.g. Ciotti et al. 1993).

There is preliminary evidence that the Fe abundances deduced from
optical data are in fact $\sim 0.5$ solar when averaged over an
effective radius for each galaxy \cite{lconf}. If we take the stellar
Fe abundances to be 0.5 solar and the Fe abundances in the hot gas to
be 1 solar, then using equation (1) of Arimoto et
al. \shortcite{arimoto} we have that the standard enrichment models
require a Type Ia supernovae rate of $\sim 0.05h^{2}_{75}$ SNu. This
is a factor of 2.6 below the Type Ia SN rate of $0.13h^{2}_{75}$ SNu
reported by Capallero et al. \shortcite{cap}, although this
discrepancy can be reduced below a factor of two if an Fe abundance of
1.2 solar for the hot gas is used instead. Such a modest increase is
still consistent with the range of Fe abundances obtained for the
multiphase {\sl ASCA} models.  

Thus, if the stellar Fe abundances are near 0.5 solar then the Fe
abundances of the hot gas predicted by the multiphase models of the
{\sl ASCA} data can be produced to within a factor of two by the
standard chemical enrichment models (e.g. Ciotti et al. 1991).  This
small discrepancy essentially disappears if we increase the Fe
abundances we have obtained by a factor of 1.44 to reflect the
meteoritic solar abundances instead of the photospheric values we have
used -- see Ishimaru \& Arimoto \shortcite{im}.  It should be
mentioned that the 1-2 solar abundances for Fe in the hot gas are also
consistent with other scenarios involving both homogeneous
\cite{bmabun} and inhomogeneous cooling flows \cite{fujita}.

It is possible that the abundances vary within the relatively large
apertures used to analyze the {\sl ASCA} data.  In fact, the cooling
flow model of NGC 5044 indicates that the colder gas, which must be
more centrally concentrated to be consistent with the {\sl ROSAT}
temperature profiles, has higher metallicity than the hotter
gas. Abundance gradients are also inferred from analysis of radially
varying isothermal models of these galaxies with {\sl ROSAT} (see
section \ref{rosat}) and with similar models of the {\sl ASCA} data of
NGC 5846 \cite{fino}. However, these results in themselves are not
compelling since the isothermal approximation need not be valid,
especially for the bins near the centers (see Figure \ref{fig.pspc}).

The 1-2 solar Fe abundances predicted by the multiphase models are
substantially larger than the values of $\sim 0.3$ solar typically
found for the intra-cluster medium (ICM) in a rich cluster (Fukazawa
et al. 1998). If the Fe abundances for these brightest elliptical
galaxies in our sample can be extrapolated to the ellipticals formed
in rich clusters, then simple models of the chemical enrichment of the
ICM by the hot gas expelled from elliptical galaxies in rich clusters
can fit the observations if most of the ICM is primordial
(e.g. Loewenstein \& Mushotzky 1996).

The relative abundances with respect to Fe deduced from the two-temperature
models (section \ref{2t}) also favor near-solar values and are
consistent with the trend discovered by Fukazawa et
al. \shortcite{fuk98} that the Si/Fe abundance is nearly solar for the
least massive systems such as large ellipticals and small groups of
galaxies and increases with the temperature of the system to a value
of $\sim 3$ solar for rich clusters.  However, the solar Si/Fe
abundances indicated by both our two-temperature and cooling flow models
imply nearly solar Si abundances for these galaxies which are similar
to those found in rich clusters by Fukazawa et al.; i.e. the solar Si
abundances predicted by the multiphase models are not consistent with
the trend shown in Figure 2 of Fukazawa et al. wherein the Si
abundance of ellipticals should have very sub-solar values.

\subsection{Two-temperature vs. cooling flow models}
\label{2tvscf}

The two-temperature models tend to provide slightly better broad-band fits
to the {\sl ASCA} spectra than the cooling flow models. The
best-fitting cooling flow models, however, predict Si and S line
fluxes for NGC 1399 and NGC 4472 that agree with the {\sl ASCA} SIS
data slightly better than the best-fitting two-temperature models. Buote et
al. \shortcite{b98} have shown that two-temperature models are very flexible
and, in particular, can approximate to high accuracy a multiphase
cooling flow spectrum throughout the {\sl ASCA} bandpass. With the
{\sl ASCA} data we thus cannot ascertain reliably whether the spectra
consist of only two phases or whether the two-temperature models merely
provide good approximations to spectra which have a continuum of
temperature components.

There are physical models for the hot gas in ellipticals which
accommodate only two phases. Brighenti \& Mathews \shortcite{bm4472}
find that their most successful models of the {\sl ROSAT} temperature
and surface brightness profiles of NGC 4472 require their isothermal
cooling flow model be supplemented with hot gas accreted from the
surrounding medium. In this scenario the spectra observed within the
{\sl ASCA} apertures of the ellipticals in our sample are the
projection of two nearly isothermal components -- a colder component
arising from the isothermal cooling flow and a hotter component from
the accreted hot gas. (A related scenario for galaxy clusters with cD
galaxies is sometimes invoked as an alternative to multiphase cooling
flow models; e.g. Xu et al. 1998.) The approximately solar abundances
deduced from the {\sl ASCA} data can also be reproduced by such
two-temperature models \cite{bmabun}, although the excess absorption (Table
\ref{tab.params}) is not an obvious prediction of the model. (Note,
however, that related models of isothermal cooling flows can produce
material that would give rise to excess absorption -- Pellegrini \&
Ciotti 1998).

Another success of the two-temperature model of Brighenti \& Mathews
\shortcite{bm4472} also highlights a possible problem with the
model. The mass estimated from the stellar dynamics within an
effective radius $(R_e)$ of NGC 4472 is in excellent agreement with
the mass inferred from the X-ray data using the two-temperature
model. However, for $r\la 0.1R_e$ the mass inferred from the X-ray
analysis is less than that indicated by stars. One explanation for
this discrepancy is that magnetic pressure support becomes important
at these small radii which must be accounted for in the X-ray
analysis. Alternatively, mass drop-out from the cooling flow may need
to be considered.

We obtain results for multiphase cooling flow models of the {\sl ASCA}
data of the elliptical galaxies quite analogous to clusters
(e.g. Fabian 1994). In particular, excess absorption on the
cooling-flow component and Galactic absorption on the ambient hot gas
is indicated for NGC 1399, NGC 4472, and NGC 5044.  There is a
discrepancy between the mass implied by this absorbing material and
the amount of cold gas inferred from HI and CO observations (e.g. see
BF and Fabian 1994). This is a critical issue for the simple
constant-pressure cooling flow model because the inferred absorbing
mass is consistent with the mass deposited by the cooling flow over
the galaxy lifetime. It is possible that other versions of multiphase
cooling flows would have different predictions for mass deposition.

Broad-band spectral fitting of future observations of elliptical
galaxies with {\sl Chandra} will be able to distinguish two-temperature and
cooling flow models \cite{b98}. With {\sl ASTRO-E} this distinction
can be made using only individual K shell emission lines (particularly
of oxygen) thus alleviating any remaining concerns at that time about
the accuracy of plasma codes in the Fe L shell energy region.

\section{Conclusions}
\label{conc}

We have examined the {\sl ASCA} and {\sl ROSAT} spectra of NGC 1399,
NGC 4472, NGC 4636, and NGC 5044, which are among the brightest
elliptical galaxies in X-rays, to constrain the temperature structure
and abundances of the hot gas in these systems.  Our principal aim is
to determine whether the {\sl ASCA} spectra are adequately described
by isothermal models of the hot gas, and whether such isothermal
models fit the {\sl ASCA} data as well or better than simple
multiphase models. These ``isothermal'' models actually have two
components (1T+BREM) consisting of one component of isothermal hot gas
(1T) and another component of thermal bremsstrahlung (BREM) presumably
arising from discrete sources.

To begin we simultaneously fit isothermal models to the 0.55-9 keV
{\sl ASCA} SIS and 1-9 keV {\sl ASCA} GIS spectra of each galaxy using
the MEKAL and Raymond-Smith (RS) plasma codes to model the hot gas in
each system. Each {\sl ASCA} spectrum is extracted from a region of
$\sim 5\arcmin$ radius because the large energy-dependent PSF confuses
spatial analysis on scales much smaller than this. As we discuss
below, the spatial variation of the X-ray spectra is examined with the
{\sl ROSAT} PSPC data.

In terms of the $\chi^2$ null hypothesis probability $(P)$ the 1T+BREM
fits to the {\sl ASCA} data are of formally poor quality for each
galaxy if the relative abundances with respect to Fe in the hot gas
are fixed at their solar values. If the relative abundances of the
$\alpha$-process elements (and others for NGC 5044 and NGC 4636) are
allowed to vary for each galaxy, the fits are improved but are still
very poor for all of the galaxies except (marginally) NGC 4636.

The Fe abundances inferred from the isothermal models with variable
relative abundances are $\sim 0.3Z_{\sun}$ for NGC 1399, NGC 4472, and
NGC 5044 consistent with previous studies; for NGC 4636 a very
sub-solar Fe abundance is also obtained with the RS code, though
$Z\sim 0.8Z_{\sun}$ is implied for the MEKAL code when the relative
abundances are allowed to deviate substantially from their solar
values. The best isothermal fits require abundances of O, Ne, and Mg
to be very different from Fe; these abundances are usually required to
be near zero except, e.g., for the Ne abundances of NGC 5044 and NGC
4636 which require values greater than solar for the MEKAL code. Ni
abundances very different from Fe are indicated for NGC 4636 and NGC
5044. Finally, the marginal fit for the isothermal model of NGC 4636
using the MEKAL model requires the Na abundance to be many times the
solar value.

Next we examined whether simple multiphase models fit the {\sl ASCA}
data better than the isothermal models. We studied two-temperature models
(2T or 2T+BREM) consisting of two components of hot gas and (if
desired) a bremsstrahlung component (BREM) for the putative emission
from discrete sources. We also examined a simple multiphase cooling
flow where gas cools at constant pressure and drops out of the flow at
all radii. This model (CF+1T or CF+1T+BREM) consists of one
cooling-flow component (CF), one component of isothermal hot gas (1T)
with temperature tied to the upper temperature of the CF component,
and (if desired) a bremsstrahlung component (BREM).

The two-temperature models provide subtantially better fits to the SIS
and GIS data than the isothermal models. With the exception of NGC
4636, these superior fits are achieved with the relative abundances
with respect to Fe fixed at their solar values. The cooling flow
models fit nearly as well as the two-temperature models for NGC 1399,
NGC 4472, and NGC 5044. The poor fit of the cooling flow model to NGC
4636 is probably due largely to the limitation in our implementation
of the cooling flow model which fixes the relative abundances at their
solar values. The multitemperature fits of NGC 1399 and NGC 4472 are
improved significantly when a BREM component is added (2T+BREM),
though the reduction in $\chi^2$ is not so large as that observed
between the 1T+BREM and 2T cases. For NGC 4636, in contrast, the
addition of the BREM component to the 2T and CF+1T models provides a
much larger improvement to the fit than observed between the 1T+BREM
and 2T cases. The luminosities of these BREM components are consistent
with expectations of emission from discrete sources.

Unlike the isothermal models, the Fe abundances predicted by the
multiphase models indicate values of $\sim 1$-2 solar. For example,
the 2T+BREM models give Fe abundances (90 per cent confidence) of
$1.6^{+0.7}_{-0.6}Z_{\sun}$ for NGC 1399, $2.0^{+2.4}_{-1.0}Z_{\sun}$
for NGC 4472, $0.6^{+0.1}_{-0.1}Z_{\sun}$ for NGC 5044, and
$0.7^{+0.3}_{-0.2}Z_{\sun}$ for NGC 4636. The multiphase cooling flows
also indicate solar Fe abundances: $1.1^{+0.4}_{-0.2}Z_{\sun}$ for NGC
1399, $1.3^{+0.1}_{-0.1}Z_{\sun}$ for NGC 4472,
$0.8^{+0.9}_{-0.2}Z_{\sun}$ for NGC 5044, and
$1.0^{+0.4}_{-0.2}Z_{\sun}$ for NGC 4636.  An abundance gradient is
also implied by the cooling flow model of NGC 5044 since the abundance
determined for the isothermal component ($0.4^{+0.2}_{-0.2}Z_{\sun}$)
is considerably less than the cooling flow component. Both the
two-temperature and cooling flow models for NGC 1399, NGC 4472, and NGC 5044
predict absorption in excess of the Galactic value for the colder gas
components in their respective models while Galactic absorption is
indicated for their hotter components. This is consistent with the
radial variation of column density found in the cores of some cooling
flow clusters (e.g. Fabian 1994).

The key constraints from the broad-band spectral fitting arise from
the SIS data near 1 keV where the Fe L shell emission lines
dominate. Errors in the plasma codes are also potentially most serious
for these Fe L lines (e.g. Fabian et al. 1994; Liedahl et al. 1995).
We find, as expected, that the MEKAL code is much more accurate than
RS for the important energies $\sim 0.7-1.4$ keV, and thus the MEKAL
models should be used in preference to RS especially for elliptical
galaxies.  By comparing fits using the MEKAL and RS plasma codes we
have determined that remaining errors in the modeling of the Fe L
lines by these codes are insufficient to qualitatively change our
results obtained from the broad-band spectral fitting. Previous
studies by Hwang et al. \shortcite{una} and BF also find that the
temperatures and Fe abundances inferred for ellipticals using {\sl
ASCA} data are not overly sensitive to remaining errors in the plasma
codes near 1 keV.

We have measured the emission from the K$\alpha$ line blends of the
H-like and He-like ions of Si and S using a simple model of a
bremsstrahlung continuum with zero-width gaussians for the line
blends. These local measurements provide additional constraints
independent of the Fe L emission on the models obtained from the
broad-band spectral fitting. We find that isothermal models
generally under-predict the equivalent widths of the Si and S lines of
these systems.

The equivalent widths of the multiphase models, in contrast, are much
better matches to the {\sl ASCA} data. The ratios Si XIV/Si XIII and S
XVI/XV are also generally better described by the multiphase
models. Even for NGC 4636, which displays the smallest improvement for
the two-temperature model over the isothermal model for the broad-band
spectral fits (section \ref{broad}), the line ratios and equivalent
widths of Si and S clearly favor the two-temperature model.  Hence, whereas
Loewenstein \& Mushotzky \shortcite{lmushy} and Loewenstein
\shortcite{lconf} found support for isothermal models of ellipticals
using only the Si ratio, we find that when both the ratios and
equivalent widths of the Si and S line blends are considered, the
multiphase models are strongly favored.

We examined whether the temperature profiles inferred from the {\sl
ROSAT} PSPC data of these galaxies can also distinguish between the
isothermal and multiphase models obtained from the single-aperture
analysis of the {\sl ASCA} data. Dividing up the PSPC spectrum for
each galaxy into a series of radial bins within the apertures defined
for the {\sl ASCA} analysis, we fitted single-temperature MEKAL models
to each bin. Each galaxy has a qualitatively similar temperature
gradient which begins with a minimum temperature at the center and
rises to a maximum at the final radial bin which overlaps the {\sl
ASCA} radius in very good agreement with previous studies of these
galaxies (Forman et al. 1993; David et al. 1994; Trinchieri et
al. 1994; Rangarajan et al. 1995; Irwin \& Sarazin 1996; Jones et
al. 1997).

The temperatures of the central and outer bins agree quite well with
the temperatures of the colder and hotter components obtained for the
two-temperature and cooling flow models of the {\sl ASCA} data. The
abundances, though uncertain, tend to have values near solar. Finally,
the central bins of the PSPC data are not satisfactorily described by
the isothermal model and instead suggest multiphase temperature
components and absorption in excess of the Galactic value, especially
for NGC 1399 and NGC 5044.

We also constructed composite {\sl ROSAT} models by summing up the
isothermal models for each radial bin within the {\sl ASCA}
apertures. For each composite PSPC model we simulated an {\sl ASCA}
SIS observation appropriate for each galaxy. We then fitted
isothermal and multiphase models to these simulated SIS data
analogously to our procedure for the real {\sl ASCA} data.

These composite PSPC models are very consistent with the {\sl ASCA}
data of NGC 1399 and NGC 5044 and strongly favor multiphase models of
the hot gas in these systems.  The peculiar abundances required for
the isothermal models of the real {\sl ASCA} data (Table
\ref{tab.params}) of all the galaxies also can be produced by forcing
a isothermal model to fit the multiphase spectrum generated by the
composite PSPC models. Although the PSPC models suggest the spectral
properties of NGC 4636 are quite similar to the other galaxies, strong
conclusions cannot be drawn from the radially varying isothermal
models the PSPC data of NGC 4636 and NGC 4472 since they fail to
produce the {\sl ASCA} emission above $\sim 3$ keV.

Therefore, our analysis of these systems demonstrates that isothermal
models of the {\sl ASCA} data within the $r\sim 5\arcmin$ apertures
are inconsistent with the {\sl ASCA} and {\sl ROSAT} PSPC data, a
conclusion which cannot be attributed to reasonable errors in the
MEKAL plasma code. Simple two-temperature and multiphase cooling flow
models can provide good descriptions of these data sets with Fe
abundances of $\sim 1$-2 solar and (except for NGC 4636) relative
abundances fixed at their solar values (in good agreement with the
results of BF); this situation is very analogous to that found for
poor galaxy groups in a recent study \cite{b99}.  Although the current
data do not clearly distinguish between the two-temperature and
cooling flow models, data will soon be available from the {\sl Chandra},
{\sl XMM}, and {\sl ASTRO-E} satellites which can usefully address
this issue (see section \ref{2tvscf}).

In section \ref{abun} we have discussed the implications of the
approximately solar Fe abundances obtained for the multiphase
models. The stellar abundances averaged over $R_e$ computed by Arimoto
et al. \shortcite{arimoto} imply solar Fe abundances in agreement with
the multiphase models. However, preliminary results reported in
Loewenstein \shortcite{lconf} indicate the stellar Fe abundances are
$\sim 0.5$ solar. Using this smaller value we find that the standard
chemical enrichment models of the hot gas in ellipticals (e.g. Ciotti
et al. 1991) can reproduce the solar Fe abundance in the hot gas to
within a factor of two. (This relatively small discrepancy essentially
disappears if we increase the Fe abundances we have obtained by a
factor of 1.44 to reflect the meteoritic solar abundances instead of
the photospheric values we have used -- see Ishimaru \& Arimoto
\shortcite{im}.) Solar Fe abundances in the hot gas are also
consistent with other enrichment models based on homogeneous
\cite{bmabun} and inhomogeneous cooling flows \cite{fujita}.

Although the ellipticals in our sample are special in that they are
among the brightest galaxies in relatively dense environments and have
among the largest values of $L_{\rm x}/L_{\rm B}$, it is possible the
evidence we have obtained for non-isothermal gas in these systems
could be the result of their {\sl ASCA} data having perhaps the
largest S/N among elliptical galaxies. If this is the case, the better
quality data soon to become available should reveal that the hot gas
in other ellipticals (at least near their centers) also consists of
multiple temperature components with nearly solar Fe abundances.

\section*{Acknowledgments}

It is a pleasure to thank W. Mathews, S. Faber, and G. Blumenthal for
interesting discussions. Helpful comments and suggestions were
provided by the anonymous referee and by the editor, A. Fabian.  We
also gratefully acknowledge K. Arnaud, K. Ebisawa, and others at the
{\sl ascahelp} and {\sl rosathelp} email support services for their
patient responses to questions regarding the {\sl ASCA} and {\sl
ROSAT} data analysis software. The {\sc XSPEC} implementation of the
multiphase cooling flow model was kindly provided by R. Johnstone to
whom we express our gratitude. This research has made use of ASCA data
obtained from the High Energy Astrophysics Science Archive Research
Center (HEASARC), provided by NASA's Goddard Space Flight
Center. Support for this work was provided by NASA through Chandra
Fellowship grant PF8-10001 awarded by the Chandra Science Center, which
is operated by the Smithsonian Astrophysical Observatory for NASA
under contract NAS8-39073.

\end{document}